\documentclass{aa}
\usepackage{graphicx}
\usepackage[varg]{txfonts}
\usepackage{natbib}
\usepackage{subfig}
\usepackage{upgreek}
\usepackage{multirow}
\bibpunct{(}{)}{;}{a}{}{,}
\begin{document}
   \title{Orbital properties of binary post-AGB stars}
   \titlerunning{Post-AGB binaries}
   \author{Glenn-Michael Oomen \inst{1,2} 
           \and 
           Hans Van Winckel \inst{1}
           \and
           Onno Pols \inst{2}
           \and
           Gijs Nelemans \inst{2,1}
           \and
           Ana Escorza \inst{1,3}
           \and
           Rajeev Manick \inst{1}
           \and
           Devika Kamath \inst{4,5}
           \and
           Christoffel Waelkens \inst{1}
         }
   \institute{Instituut voor Sterrenkunde (IvS), KU Leuven,
              Celestijnenlaan 200D, B-3001 Leuven, Belgium\\
              \email{glennmichael.oomen@kuleuven.be}
            \and
            Department of Astrophysics/IMAPP, Radboud University, P.O. Box 9010, 6500 GL Nijmegen, The Netherlands
            \and
            Institut d’Astronomie et d’Astrophysique, Université Libre de Bruxelles, ULB, Campus Plaine C.P. 226, Boulevard du Triomphe, B-1050 Bruxelles, Belgium
            \and
            Department of Physics \& Astronomy, Macquarie University, Sydney, NSW 2109, Australia
             \and
             Astronomy, Astrophysics and Astrophotonics Research Centre, Macquarie University, Sydney, NSW 2109, Australia
   }
   \date{Received ? ? 2018 / Accepted 3 October 2018}
   \authorrunning{Oomen et al.}

  \abstract{
Binary post-asymptotic giant branch (post-AGB) stars are thought to be the products of a strong but poorly-understood interaction during the AGB phase. The aim of this contribution is to update the orbital elements of a sample of galactic post-AGB binaries observed in a long-term radial-velocity monitoring campaign, by analysing these systems in a homogeneous way. Radial velocities are computed from high signal-to-noise spectra by use of a cross-correlation method. The radial-velocity curves are fitted by using both a least-squares algorithm and a Nelder-Mead simplex algorithm. We use a Monte Carlo method to compute uncertainties on the orbital elements. The resulting mass functions are used to derive a companion mass distribution by optimising the predicted to the observed cumulative mass-function distributions, after correcting for observational bias. As a result, we derive and update orbital elements for 33 galactic post-AGB binaries, among which 3 are new orbits. The orbital periods of the systems range from 100 to about 3000~days. Over 70 percent (23 out of 33) of our binaries have significant non-zero eccentricities ranging over all periods. Their orbits are non-circular despite the fact that the Roche-lobe radii are smaller than the maximum size of a typical AGB star and tidal circularisation should have been strong when the objects were on the AGB. We derive a distribution of companion masses that is peaked around 1.09~$M_\sun$ with a standard deviation of 0.62~$M_\sun$. The large spread in companion masses highlights the diversity of post-AGB binary systems. Post-AGB binaries are often chemically peculiar, showing in their photospheres the result of an accretion process of circumstellar gas devoid of refractory elements. We find that only post-AGB stars with high effective temperatures (> 5500~K) in wide orbits are depleted in refractory elements, suggesting that re-accretion of material from a circumbinary disc is an ongoing process. It appears, however, that depletion is inefficient for the closest orbits irrespective of the actual surface temperature.

}
 
   \maketitle
%
%________________________________________________________________

\section{Introduction}
The final evolution of low- and intermediate-mass stars is still
veiled by many uncertainties. One of the important research
questions is the impact of binarity on the ultimate fate of such stars \citep[see recent reviews by][and references therein]{demarco17,jones17}. Binary interaction can have an impact on the intrinsic properties
of an evolved star. For example, it can alter the pulsations, the mass-loss
efficiency and geometry, the dust-formation processes, and the
circumstellar envelope morphology. Binary interaction can
even play a dominant role in determining the ultimate fate of the
object. Moreover, a plethora of peculiar objects and violent phenomena result from mass transfer in binary stars, ranging from the
spectacular thermonuclear novae, supernovae type Ia, sub-luminous
supernovae, gravitational wave sources, etc., to less energetic systems
such as sub-dwarf B stars, barium stars, cataclysmic variables,
bipolar planetary nebulae (PNe).

The final phase of evolution of low- and intermediate-mass stars (0.8 -- 8~$M_{\sun}$) is a rapid transition from the asymptotic giant branch
(AGB) via the post-AGB phase towards the PN
stage. Post-AGB stars are considered transition objects which have left the
AGB but are not yet hot enough to ionise the circumstellar
environment \citep{vanwinckel03}. When the envelope mass of the AGB
star is reduced to $\sim$ 0.02~$M_{\sun}$, the post-AGB phase starts which
coincides with the end of dust-driven mass loss. The speed of evolution of
a post-AGB star is mainly determined by how fast its remaining envelope mass
is reduced even further
\citep{blocker95,vassiliadis94,millerbertolami16}.

\begin{figure}
\includegraphics[width=\columnwidth]{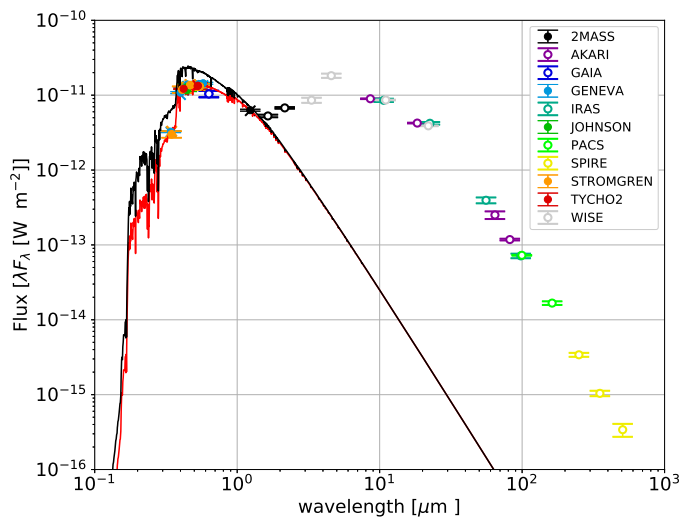}
\caption{SED of post-AGB star HD~108015. The black solid line denotes the atmospheric model, the red line is the reddened model with E(B-V) = 0.15. The different coloured symbols are the photometric data points of different passband filters.}
\label{Fig:sed}
\end{figure}

When the first binary post-AGB stars were serendipitously discovered
\citep[e.g.][]{waelkens91,pollard95,waelkens96,vanwinckel95}, it
turned out that their spectral energy distributions (SEDs) had several
common, but distinct properties. This is illustrated in
Fig.~\ref{Fig:sed} where a typical SED of a post-AGB binary is
displayed. The SED displays the photospheric contribution of the
central luminous post-AGB star with a clear excess due to thermal
emission of dust in the infrared. The distinct property is that the
excess starts in the near-infrared (near-IR), often at wavelengths shorter than 2~$\mu$m, indicating that this circumstellar dust must be close to the
central star, near sublimation radius.  The peak of the dust
excess is typically around 10~$\mu$m and in the long-wavelength tail,
the spectral index follows the Rayleigh-Jeans slope up to sub-millimetre
wavelengths. It is now well established that these specific features
in the SED indicate the presence of a stable compact disc in the
system \citep[e.g.][and references therein]{hillen17, kluska18}. This
type of SEDs are called {\sl disc-type SEDs}.

The disc-type SEDs are common and in the Galaxy alone, some 80
post-AGB stars with such an SED were identified
\citep{deruyter06,gezer15,vanwinckel17}. Moreover, in systematic
surveys of the Large and Small Magellanic Clouds in search of post-AGB
candidates, the evolved objects with these disc-type SEDs were found
to outnumber the optically bright post-AGB stars which likely have expanding
dust shells \citep{kamath14,kamath15}.

Prior to this work, the orbits of 30 galactic post-AGB binaries were known (references are given in Table~\ref{tableorbitalelements}). The periods were found to be in the range 100--2500~days. Furthermore, the observed orbits show an excess of non-zero eccentricities even though the orbital sizes are too small to accommodate a full-grown AGB star \citep{vanwinckel03, vanwinckel07, vanwinckel09}. 

%Post-AGB stars are chemically very diverse \citep{vanwinckel97, gielen09, gezer15}. Most, but not all disc-type post-AGB stars are depleted in refractory elements, i.e. elements that are easily captured in dust grains have low photospheric abundances. There is a clear correlation between the condensation temperature of a particular element and its abundance in the post-AGB photosphere \citep{maas05}.  The proposed mechanism that produces this phenomenon is re-accretion of gas from a circumbinary disc \citep{waters92}. Because dust is subject to a much larger radiation pressure than gas, the gas separates from the dust and can be accreted back onto the post-AGB star. This gas, which is poor in refractory elements, will deplete the stellar photosphere of these refractory elements, while the volatile elements are less affected.

In this contribution we present the result of a systematic and
long-term effort to find and investigate the orbital properties of
binary post-AGB stars selected on the basis of their SED properties \citep[see][]{deruyter06,gezer15}. In Sect.~\ref{sec:data}, we state the origin of the data used in this work. In Sect.~\ref{sec:dataanalysis}, we present the results of the orbital and SED fitting. We compare depletion in post-AGB stars with the stellar and orbital properties in Sect.~\ref{sec:depletion}. The distribution of companion masses is derived in Sect.~\ref{sec:companionmasses} and the orbital properties of the post-AGB binaries are presented in Sect.~\ref{sec:orbitalproperties}. We discuss the results in Sect.~\ref{sec:discussion}.

\section{Observational data} \label{sec:data}
\subsection{Older data}
Prior to 2009, we accumulated data via specific observing runs at the 1.2-m Swiss telescope and the CORALIE spectrograph \citep{queloz99}. These radial velocity data were supplemented with radial velocity determinations based on high-resolution high signal-to-noise spectra obtained for our abundance determinations. These spectra were obtained with the 1.4-m CAT telescope and the CES spectrograph \citep{enard82}, the FEROS spectrograph on the 1.5-m ESO telescope \citep{kaufer99}, or the EMMI spectrograph of the 3.5-m NTT ESO telescope \citep{dekker86}. Additionally, for some objects we obtained radial velocities with the CORAVEL instrument \citep{baranne79} mounted on the 1.5-m Danish telescope also at the ESO La Silla observatory. In Table~\ref{tableorbitalelements}, we refer to the specific original papers. In addition, we collected radial velocity data from the literature at large. 
\subsection{HERMES data}
A very systematic long-term monitoring started in 2009 when our HERMES spectrograph \citep{raskin11} on the 1.2-m Mercator telescope at the Roque de los Muchachos observatory became operational. This efficient fibre-fed spectrograph is available during the whole year and allows for an extensive observing campaign \citep{vanwinckel10,gorlova13,vanwinckel15}. Most of the data analysed here is from this effort. Radial velocities (RVs) from all the new HERMES spectra have been computed by use of a cross-correlation method. This method fits a line mask to the observed spectrum and maximises the cross-correlation function. This consequently returns the radial velocity of the star, along with the 1$\sigma$-uncertainty. This method is described in more detail by \citet{vanwinckel14}. HERMES is calibrated against the IAU radial velocity zeropoint. We did not find systematic offsets between the different spectrographs used so we accumulated all radial velocity information per star irrespective of the source.

\subsection{SEDs} \label{sec:seds} 
In order to construct the disc-type SEDs in Appendix~\ref{appendix:seds}, we have collected photometric data by making use of the Vizier database \citep{ochsenbein00} and the general catalogue of photometric data \citep[GCPD,][]{gcpd}. Photometry at blue, optical, and near-infrared wavelengths characterise the photospheric emission of the post-AGB stars. Consequently, these data points are used to fit the post-AGB photosphere (see Sect. \ref{sec:sedfitting}). Emission at longer wavelengths is dominated by lower-temperature emission from the dusty disc, hence photometry at mid- and far-infrared wavelengths characterise the circumstellar environment. The origin of the photometric data used to construct the SEDs is given in Appendix~\ref{appendix:seds}.

\section{Data Analysis} \label{sec:dataanalysis}

\subsection{Orbital fitting}
\begin{table*}[]
\centering
\caption{Updated orbital elements}
\label{tableorbitalelements}
\begin{tabular}{ll|ccccccc} 
\hline
\hline
$\#$ & Star name & Period (days) & Eccentricity & T$_0$ (days) & $\omega$ $(^{\circ})$ & K$_1$ (km/s) & $\gamma$ (km/s) & ref. \\
\hline
1  & 89~Her         & 289.1$\pm$0.2    & 0.29$\pm$0.07    & 2447832$\pm$12   & 68.4$\pm$15  & 4.2$\pm$0.3    & -27.0$\pm$0.2  &  1 \\
2  & AC~Her         & 1188.9$\pm$1.2   & 0.0$+$0.05    & /         & /    & 10.8$\pm$0.7   & -28.8$\pm$0.5  &  2\\
3  & BD+39~4926     & 871.7$\pm$0.4    & 0.024$\pm$0.006 & 2451040$\pm$36  & 150.8$\pm$15 & 16.06$\pm$0.08 & -30.45$\pm$0.07 & 3 \\
4  & BD+46~442      & 140.82$\pm$0.02  & 0.085$\pm$0.005 & 2455233.9$\pm$1.3   & 275.8$\pm$3.3  & 23.8$\pm$0.1   & -98.13$\pm$0.08 & 4 \\
5  & DY~Ori         & 1248$\pm$36    & 0.22$\pm$0.08   & 2455990$\pm$56  & 87.2$\pm$19  & 12.4$\pm$1.0   & -0.11$\pm$0.04 &  5 \\
6  & EP~Lyr         & 1151$\pm$14    & 0.39$\pm$0.09   & 2455029.3$\pm$8.7  & 61.1$\pm$7.6  & 13.4$\pm$1.3   & 15.9$\pm$1.0   &  5\\
7  & HD~44179        & 317.6$\pm$1.1    & 0.27$\pm$0.03   & 2408735.9$\pm$5.3   & 346.6$\pm$4.8  & 12.1$\pm$0.3   & 20.0$\pm$0.2   &  6, 7\\
8  & HD~46703        & 597.4$\pm$0.2    & 0.30$\pm$0.02   & 2443519.6$\pm$7.6   & 241.9$\pm$4.2   & 16.0$\pm$0.3   & -93.3$\pm$0.2  & 8 \\
9  & HD~52961        & 1288.6$\pm$0.3  & 0.23$\pm$0.01   & 2407308$\pm$24  & 297.4$\pm$5.9  & 13.1$\pm$0.3   & 6.2$\pm$0.2     &   7 \\
10 & HD~95767        & 1989$\pm$61   & 0.25$\pm$0.05   & 2449500$\pm$95 & 197.7$\pm$19 & 12.1$\pm$0.8   & -20.3$\pm$0.8   & 9 \\
11 & HD~108015       & 906.3$\pm$5.9    & 0.0$+$0.03    & /         & /    & 3.4$\pm$0.3    & 4.0$\pm$0.2     & 9 \\
12 & HD~131356       & 1488.0$\pm$8.7   & 0.32$\pm$0.04   & 2449398$\pm$32  & 162.7$\pm$6.9  & 16.3$\pm$0.7   & -6.7$\pm$0.4    & 9 \\
13 & HD~158616       & 363.3$\pm$1.0    & 0.0$+$0.1     & /         & /    & 8.4$\pm$1.0    & 56.1$\pm$0.7    & 10 \\
14 & HD~213985       & 259.6$\pm$0.7    & 0.21$\pm$0.05    & 2407110$\pm$16   & 104.8$\pm$26  & 31.4$\pm$1.0   & -42.0$\pm$0.9  & 9 \\
15 & HP~Lyr         & 1818$\pm$80    & 0.20$\pm$0.04    & 2456175$\pm$61  & 14.2$\pm$13  & 7.8$\pm$0.2    & -115.6$\pm$0.2 & 5 \\
16 & HR~4049         & 430.6$\pm$0.1    & 0.30$\pm$0.01   & 2447176.6$\pm$3.8   & 236.5$\pm$3.5  & 16.6$\pm$0.2   & -31.9$\pm$0.2  & 11 \\
17 & IRAS~05208-2035 & 234.38$\pm$0.04    & 0.0$+$0.02    & /         & /    & 18.4$\pm$0.2   & 35.6$\pm$0.1    & 17 \\
18 & IRAS~06165+3158 & 262.6$\pm$0.7    & 0.0$+$0.05    & /  		  & /  & 15.5$\pm$0.5   & -16.4$\pm$0.3   & \\
19 & IRAS~06452-3456 & 215.4$\pm$0.4    & 0.0$+$0.03    & /         & /    & 36.9$\pm$0.6   & 45.9$\pm$0.3 &    \\
20 & IRAS~08544-4431 & 501.1$\pm$1.0    & 0.20$\pm$0.02   & 2451499.6$\pm$7.8  & 230.1$\pm$5.9  & 8.8$\pm$0.2    & 62.4$\pm$0.1 &  12  \\
21 & IRAS~09144-4933 & 1762$\pm$27    & 0.30$\pm$0.04    & 2451302$\pm$39  & 145.7$\pm$7.7  & 14.5$\pm$0.6   & 29.6$\pm$0.5  &  5 \\
22 & IRAS~15469-5311 & 390.2$\pm$0.7    & 0.08$\pm$0.02   & 2451530$\pm$13  & 114.0$\pm$16 & 12.3$\pm$0.4   & -13.9$\pm$0.3  & 12 \\
23 & IRAS~16230-3410 & 649.8$\pm$3.5    & 0.0$+$0.13    & /         & /    & 3.9$\pm$0.3    & -154.3$\pm$0.2 & \\
24 & IRAS~17038-4815 & 1394$\pm$12    & 0.63$\pm$0.06   & 2451694$\pm$15  & 124.5$\pm$5.5  & 15.2$\pm$1.5   & -25.6$\pm$0.5  &  5 \\
25 & IRAS~19125+0343 & 519.7$\pm$0.7    & 0.24$\pm$0.03   & 2451503$\pm$11  & 243.0$\pm$8.1  & 12.0$\pm$0.5   & 67.3$\pm$0.3 &  12  \\
26 & IRAS~19135+3937 & 126.97$\pm$0.08  & 0.13$\pm$0.03   & 2454997.7$\pm$1.0   & 66.0$\pm$4.4   & 18.0$\pm$0.6   & 2.1$\pm$0.4 &  13  \\
27 & IRAS~19157-0247 & 119.6$\pm$0.1    & 0.34$\pm$0.04   & 2451366.5$\pm$2.9   & 72.1$\pm$8.0  & 8.0$\pm$0.4    & 31.7$\pm$0.3  & 12 \\
28 & RU~Cen         & 1489$\pm$10    & 0.62$\pm$0.07   & 2449885$\pm$25  & 315.2$\pm$12 & 22.1$\pm$1.9   & -25.9$\pm$0.8  & 14 \\
29 & SAO~173329      & 115.951$\pm$0.002  & 0.0$+$0.04   & /       & /    & 12.4$\pm$0.3   & 73.3$\pm$0.2    & 9 \\
30 & ST~Pup         & 406.0$\pm$2.2    & 0.0$+$0.04    & /         & /    & 17.9$\pm$0.7   & 0.1$\pm$0.2    & 15 \\
31 & SX~Cen         & 564.3$\pm$7.6    & 0.0$+$0.06    & /         & /    & 21.5$\pm$0.9   & 24.3$\pm$1.0   & 14 \\
32 & TW~Cam         & 662.2$\pm$5.3    & 0.25$\pm$0.04   & 2455111$\pm$18  & 144.4$\pm$10 & 14.1$\pm$0.6   & -49.8$\pm$0.5  & 5 \\
33 & U~Mon          & 2550$\pm$143   & 0.25$\pm$0.06    & 2451988$\pm$316 & 87.3$\pm$15 & 14.9$\pm$1.1   & 24.1$\pm$1.0   & 16 \\
\hline
\end{tabular}
\tablefoot{References point to previously published orbits of stars for which additional RV data was collected in order to update the available orbital elements.}
\tablebib{(1)~\citet{waters93}; (2)~\citet{vanwinckel98}; (3)~\citet{kodaira70}; (4)~\citet{gorlova12}; (5)~\citet{manick17}; (6)~\citet{waelkens96}; (7)~\citet{vanwinckel95}; (8)~\citet{hrivnak08}; (9)~\citet{vanwinckel00}; (10)~\citet{desmedt16}; (11)~\citet{waelkens91};  (12)~\citet{vanwinckel09}; (13)~\citet{gorlova15}; (14)~\citet{maas02}; (15)~\citet{gonzalez96};  (16)~\citet{pollard06}; (17)~\citet{gielen08}}
\end{table*}

The orbital analysis is performed for all objects in a homogeneous way by first making an initial guess for the orbit. For the binaries that already have published orbits, the initial guess is simply the set of orbital elements available for that star. In the case of unpublished orbits, a Lomb-Scargle periodogram has been used to estimate the period. The next step involves an optimisation of the orbit to the observed radial velocities. Since the orbit is Keplerian, this requires a nonlinear optimisation algorithm. In this work, we used two separate optimisation routines: a least-squares algorithm and a Nelder-Mead simplex algorithm \citep{neldmead65}.

The use of two separate routines was required since the optimisation algorithms do not always converge to a good solution. This is mainly because of the large scatter in the RV curve due to pulsations. Many post-AGB stars in our sample are variable, and some of them populate the population II Cepheid instability strip in the HR diagram. The latter are known as RV~Tauri stars, and their pulsation periods are of the order of 10--100 days. The amplitude of the pulsations in the radial-velocity data can be of the same magnitude as the orbital motion. This makes it difficult to fit the orbit of the binary, especially if the orbit is wide, the companion is of low-mass, and/or if the inclination is small. In some cases, the optimisation fails for one algorithm, but works for the other. In cases where the optimisation fails for both algorithms, a better initial guess is required.

In order to improve the orbital fitting, we have obtained pulsationally cleaned data from \citet{manick17} for the RV~Tauri stars DY~Ori, EP~Lyr, HP~Lyr, IRAS~09144-4933, IRAS~17038-4815, and TW~Cam. For strong pulsators without cleaned data available, we removed pulsations using a similar approach as \citet{manick17}. This was done using an iterative procedure involving a Lomb-Scargle periodogram to estimate the pulsation frequencies in RV data after the orbital motion is subtracted, and then fitting a sine curve to the pulsations and subtracting these from the residuals of the orbit. This procedure is repeated until the peak of the pulsational frequency in the periodogram is less than three times the noise level. The pulsations found are subtracted from the original RV data, and the orbit is re-fit. This iterative procedure can be repeated until no improvement on the orbital fitting is obtained. The pulsational cleaning was applied to AC~Her, IRAS~08544-4431, IRAS~19157-0247, and U~Mon (see Appendix~\ref{appendix:orbits}).

\begin{table*}[] 
\centering
\caption{Projected semi-major axis, mass functions, and minimum masses}
\label{massfunctions}
\begin{tabular}{llccc}
\hline
\hline
\multirow{2}{*}{$\#$} & \multirow{2}{*}{Star name} & \multirow{2}{*}{$a_1 \sin i$ (AU)} & \multicolumn{1}{p{2cm}}{\centering Mass function\\($M_\sun$)} & \multicolumn{1}{p{3cm}}{\centering Minimum mass ($M_\sun$)\\($M_1 = 0.6$, $i=75^\circ$)} \\
\hline
1  & 89~Her         & 0.106$\pm$0.007   & 0.0019$\pm$0.0004 & 0.10\\
2  & AC~Her         & 1.176$\pm$0.080     & 0.153$\pm$0.032 & 0.64  \\
3  & BD+39~4926     & 1.286$\pm$0.007   & 0.373$\pm$0.006 & 1.03  \\
4  & BD+46~442      & 0.3074$\pm$0.0014 & 0.195$\pm$0.003 & 0.72  \\
5  & DY~Ori         & 1.39$\pm$0.11     & 0.23$\pm$0.05  & 0.79    \\
6  & EP~Lyr         & 1.30$\pm$0.12      & 0.22$\pm$0.06  & 0.77    \\
7  & HD~44179        & 0.342$\pm$0.008   & 0.053$\pm$0.003 & 0.38  \\
8  & HD~46703        & 0.839$\pm$0.015   & 0.220$\pm$0.012 & 0.77  \\
9  & HD~52961        & 1.507$\pm$0.034     & 0.274$\pm$0.019 & 0.87  \\
10 & HD~95767        & 2.14$\pm$0.16     & 0.33$\pm$0.07 & 0.96     \\
11 & HD~108015       & 0.28$\pm$0.02     & 0.0036$\pm$0.0009 & 0.13\\
12 & HD~131356       & 2.11$\pm$0.09     & 0.57$\pm$0.07  & 1.33     \\
13 & HD~158616       & 0.28$\pm$0.03     & 0.022$\pm$0.008 & 0.26  \\
14 & HD~213985       & 0.733$\pm$0.025   & 0.777$\pm$0.079 & 1.62  \\
15 & HP~Lyr         & 1.27$\pm$0.06     & 0.083$\pm$0.007 & 0.47  \\
16 & HR~4049         & 0.627$\pm$0.010   & 0.177$\pm$0.008 & 0.69    \\
17 & IRAS~05208-2035 & 0.396$\pm$0.004   & 0.150$\pm$0.005 & 0.64   \\
18 & IRAS~06165+3158 & 0.374$\pm$0.011   & 0.10$\pm$0.01  & 0.52      \\
19 & IRAS~06452-3456 & 0.73$\pm$0.01     & 1.12$\pm$0.05  & 2.07     \\
20 & IRAS~08544-4431 & 0.398$\pm$0.008     & 0.033$\pm$0.002 & 0.31    \\
21 & IRAS~09144-4933 & 2.25$\pm$0.11     & 0.49$\pm$0.07  & 1.21     \\
22 & IRAS~15469-5311 & 0.438$\pm$0.015   & 0.074$\pm$0.008  & 0.45 \\
23 & IRAS~16230-3410 & 0.232$\pm$0.021    & 0.004$\pm$0.001  & 0.13\\
24 & IRAS~17038-4815 & 1.52$\pm$0.08     & 0.24$\pm$0.04  & 0.81     \\
25 & IRAS~19125+0343 & 0.56$\pm$0.02     & 0.086$\pm$0.010 & 0.48  \\
26 & IRAS~19135+3937 & 0.209$\pm$0.008   & 0.075$\pm$0.008 & 0.45  \\
27 & IRAS~19157-0247 & 0.083$\pm$0.003   & 0.0053$\pm$0.0007 & 0.15\\
28 & RU~Cen         & 2.38$\pm$0.15     & 0.81$\pm$0.17  & 1.66    \\
29 & SAO~173329      & 0.132$\pm$0.003   & 0.023$\pm$0.001 & 0.27\\
30 & ST~Pup         & 0.67$\pm$0.02     & 0.241$\pm$0.026  & 0.81 \\
31 & SX~Cen         & 1.12$\pm$0.05     & 0.58$\pm$0.07  & 1.35    \\
32 & TW~Cam         & 0.83$\pm$0.04     & 0.174$\pm$0.022  & 0.68 \\
33 & U~Mon          & 3.38$\pm$0.31       & 0.79$\pm$0.18   & 1.64  \\
\hline
\end{tabular}
\end{table*}

Because the scatter in the RV curve due to the residuals of pulsations is inherently larger than the uncertainty on the RV measurements, the best-fitting model is selected based on the variance of the residuals of the fit. The optimisation algorithms cannot handle circular orbits very well and will almost always find a low-eccentricity orbit that is better-fitting, even though the orbit might be (very close to) circular in reality. To handle this, we use the Lucy and Sweeney test for circular orbits \citep{lucy71}. The null hypothesis is that the eccentricity is zero and a 5$\%$ significance level is adopted. Only if the null hypothesis is rejected, the eccentric orbit is taken as best-fitting model.

In this work, we adopted a Monte Carlo method to compute the uncertainty intervals on the parameters from the fitting process. Since the main source of noise in the data is due to unknown stellar pulsations rather than the intrinsic measurement error, we apply the following method. First, the best-fitting orbit was subtracted from the observed radial velocities. The standard deviation on the residuals is then used to generate 1000 RV curves for each of the binaries by randomly sampling new RV data normally distributed around the best-fitting model. This assumes that the scatter in the radial velocities also is normally distributed. These synthetic RV curves are fitted again and the 1$\sigma$-uncertainties on the parameters are computed from the standard deviation of the resulting distribution of orbital elements. The list of orbital elements for each of the binaries, along with their uncertainties, is given in Table~\ref{tableorbitalelements}. Note that the uncertainty on the eccentricity for circular orbits is in fact a one-sided confidence interval, such that the 1$\sigma$-uncertainty quoted in Table~\ref{tableorbitalelements} corresponds to the 68$\%$ confidence level. Furthermore, circular orbits do not have a time of periastron passage ($T_0$) nor an argument of periastron ($\omega$), hence these values are omitted for these objects in Table~\ref{tableorbitalelements}.

The orbital elements describe the orbit of the post-AGB star around the centre of mass. Consequently, we can derive useful quantities, such as the projected distance of the post-AGB star to the centre of mass, $a_1 \sin i$. Another interesting quantity is the mass function $f(m)$ which is defined as
\begin{equation}
f(m) = \frac{P K_1^3}{2\pi G}(1-e^2)^{\frac{3}{2}} = \frac{M_2^3}{(M_1+M_2)^2} \sin^3i,
\label{Eq:massfct}
\end{equation}
where the first equation consists solely of measured orbital elements and physical constants, while the second equation relates to the masses of both stars and the (unknown) inclination angle. The projected semi-major axis and mass function for each of the post-AGB binaries are given in Table~\ref{massfunctions}.

\subsection{SED fitting} \label{sec:sedfitting}
In this work, we perform a preliminary analysis of the SEDs of all the targets in our sample. In future work, we will provide the SED fits of all known galactic disc-type post-AGB stars with sufficient photometry. All the SEDs in our binary sample (see Appendix~\ref{appendix:seds}) have been fitted with a MARCS model atmosphere \citep{marcspaper}. The fit was performed with a grid search over all free parameters using a $\chi^2$-minimalisation method \citep{degroote11}. The fit was only performed at optical and near-IR wavelengths, since longer wavelengths are affected by disc emission. The IR-excess caused by the disc starts at different wavelengths, depending on both the temperature at the inner rim of the disc as well as the inclination angle, but always starts between 1--3~$\mu$m. Consequently, for each star individually we determine from which wavelength disc emission starts to significantly contribute to the observed flux. In Fig.~\ref{fig:discseds}, the photometry included in the fitting procedure is shown as filled circles, while photometry excluded from the fit are shown as empty circles. To fit the model atmospheres, the effective temperature, surface gravity, and metallicity were constrained with spectroscopic data from literature (see Table~\ref{specdata}) within their uncertainty, although we allowed for a larger range in temperatures ($\pm500$~K) for the large-amplitude pulsators in our sample. We assume in this process that the total extinction in the line-of-sight has the wavelength dependency of the interstellar-medium (ISM) extinction law \citep{cardelli89}. 

After the model atmosphere was fitted, the infrared excess of the star was integrated. The fraction of the infrared luminosity with respect to the post-AGB luminosity is a useful quantity that provides information on how much energy of the star is reprocessed by dust in the disc. Note that the $L_\mathrm{IR}/L_*$ ratio can be larger than 1 in cases where the disc partly obscures the post-AGB star, as in the case of HD~44179 (also known as the Red Rectangle) where we only see the post-AGB star via scattered light. The results of the SED fitting (E(B-V) and $L_\mathrm{IR}/L_*$) are presented in Table~\ref{specdata}.

\begin{table*}[]
\centering
\caption{Spectroscopic data and results of SED fitting for post-AGB stars in the sample.}
\label{specdata}
\begin{tabular}{llcccccccccc}
\hline
\hline
$\#$ & Star name & $T_\mathrm{eff}$ (K) & $\log g$ & E(B-V)  &  $L_\mathrm{IR}/L_*$ & [Fe/H] & [Zn/Fe] & [Zn/Ti] & [S/Ti] & Depletion & Ref.\\
\hline
1  & 89~Her        & 6600 & 0.8   & 0.02 & 0.38  & -0.5  & 0.1   & 0.6   & 0.7 & mild & 1  \\
2  & AC~Her        & 5800 & 1.0   & 0.46 & 0.24  & -1.4  & 0.5   & 0.7   & 1.2 & mild & 2  \\
3  & BD+39~4926    & 7750 & 1.0   & 0.23 & 0.0   & -2.4  & 1.7   & 2.0   & 3.2 & strong & 3  \\
4  & BD+46~442     & 6250 & 1.5   & 0.23 & 0.19  & -0.8 & -0.1 & -0.2   & -0.4 & no & 4 \\
5  & DY~Ori        & 5900 & 1.5   & 0.90 & 0.74  & -2.3  & 2.1   & 2.1   & 2.5 & strong & 5  \\
6  & EP~Lyr        & 6200 & 1.5   & 0.48 & 0.04  & -1.8  & 1.1   & 1.3   & 1.4 & moderate & 5  \\
7  & HD~44179        & 7500 & 0.8   & 0.15 & 18.1  & -3.3  & 2.7   & / & / & strong & 6, 7 \\
8  & HD~46703        & 6250 & 1.0   & 0.23 & 0.02  & -1.7  & 0.8   & 0.9   & 1.1 & mild & 8  \\
9  & HD~52961        & 6000 & 0.5   & 0.04 & 0.13  & -4.8  & 3.4   & 3.0   & 3.4 & strong & 9, 10 \\
10 & HD~95767        & 7500 & 2.0   & 0.58 & 0.55  & 0.1   & -0.2  & 0.0   & 0.1 & no & 11  \\
11 & HD~108015       & 7000 & 1.5   & 0.15 & 1.04  & -0.1 & -0.1 & 0.1   & 0.1 & no & 11 \\
12 & HD~131356       & 6000 & 1.0   & 0.15 & 0.65  & 0.0   & 0.2   & 0.6  & 0.5 & mild & 11  \\
13 & HD~158616       & 7250 & 1.25   & 0.51 & 0.23  & -0.6  & 0.2   & 0.0   & 0.1 & no & 12  \\
14 & HD~213985       & 8250 & 1.5   & 0.12 & 0.35  & -0.9  & / & / & 1.9 & strong & 13  \\
15 & HP~Lyr        & 6300 & 1.0   & 0.39 & 0.56   & -1.0  & 0.6  & 2.6  & 3.0 & strong &  14 \\
16 & HR~4049         & 7600 & 1.1    & 0.20 & 0.12  & -4.8  & 3.5   & / & / & strong & 6 \\
17 & IRAS~05208-2035 & 4250 & 0.75    & 0.01 & 0.43 & -0.7 & / & / & / & no & 3 \\
18 & IRAS~06165+3158 & 4250 & 1.5 & 0.53 & 0.39 & -0.9 & -0.1 & 0.0 & / & no & 10 \\
19 & IRAS~06452-3456 & /  & /   & 0.94 &  0.11  & / & / & / & / & / & /\\
20 & IRAS~08544-4431 & 7250 & 1.5   & 1.32  & 0.49  & -0.3  & 0.4   & 0.9   & 1.0 & mild & 15  \\
21 & IRAS~09144-4933 & 5750 & 0.5   & 1.78  & 0.81  & -0.3  & / & / & 1.3 & moderate & 15  \\
22 & IRAS~15469-5311 & 7500 & 1.5   & 1.27 & 0.74   & 0.0   & 0.3   & 1.8   & 2.1  & strong & 15  \\
23 & IRAS~16230-3410 & 6250 & 1.0     & 0.72 & 0.46   & -0.7  & 0.3   & 1.0   & 1.1 & moderate & 15  \\
24 & IRAS~17038-4815 & 4750 & 0.5 & 0.57 & 0.79   & -1.5  & 0.3   & / & / & no & 15 \\
25 & IRAS~19125+0343 & 7750 & 1.0   & 0.94 & 0.90  & -0.3  & 0.4   & 2.3   & 2.6 & strong & 15  \\
26 & IRAS~19135+3937 & 6000 & 0.5   & 0.28 & 0.26  & -1.0 & 0.0 & / & / & no & 10\\
27 & IRAS~19157-0247 & 7750 & 1.0    & 0.66 & 0.79  & 0.1   & / & / & 0.4 & no & 15  \\
28 & RU~Cen        & 6000 & 1.5   & 0.18 & 0.39    & -1.9  & 0.9   & 1.0   & 1.3 & moderate & 16  \\
29 & SAO~173329      & 7000 & 1.0   & 0.31 & 0.35  & -0.9 & 0.1  & -0.1 & 0.4 & no & 3 \\
30 & ST~Pup        & 5500 & 1.0   & 0.06 & 1.32 & -1.5  & 1.4   & 2.1   & 2.0 & strong & 17  \\
31 & SX~Cen        & 6250 & 1.5   & 0.17 & 0.40  & -1.1  & 0.6   & 1.5   & 1.9 & strong & 16  \\
32 & TW~Cam        & 4800 & 0.0  & 0.42 & 0.43  & -0.5  & 0.1   & 0.3   & 0.7 & no & 18  \\
33 & U~Mon         & 5000 & 0.0   & 0.34 & 0.28   & -0.8  & 0.2   & 0.0   & 0.5 & no & 18 \\
\hline
\end{tabular}
\tablefoot{
Values for metallicity and surface gravity are in dex. Uncertainties are not quoted, but formal errors for temperature are $\pm$ 250~K, for surface gravity $\pm$ 0.5~dex, and for abundances $\pm$ 0.3~dex. Effective temperature, surface gravity, and abundances come from literature, quoted in the rightmost column. The E(B-V) values come from SED fitting (see Sect. \ref{sec:sedfitting}), and the infrared luminosities are computed by integrating the infrared excess in SEDs in Appendix~\ref{appendix:seds}.}
\tablebib{
(1)~\citet{kipper11}; (2) \citet{giridhar98}; (3) \citet{rao12}; (4) \citet{gorlova12}; (5) \citet{gonzalez97}; (6) \citet{vanwinckel95}; (7) \citet{waelkens96}; (8) \citet{hrivnak08}; (9) \citet{waelkens91}; (10) \citet{rao14}; (11) \citet{vanwinckel97}; (12) \citet{desmedt16}; (13) \citet{deruyter06}; (14) \citet{giridhar05}; (15) \citet{maas05}; (16) \citet{maas02}; (17) \citet{gonzalez96}; (18) \citet{giridhar00}.}
\end{table*}

\section{Depletion} \label{sec:depletion}
Post-AGB stars are chemically very diverse \citep{vanwinckel97, gielen09, gezer15, kamath18}. Most, but not all disc-type post-AGB stars are depleted in refractory elements, i.e. elements that are easily captured in dust grains have low photospheric abundances. There is a clear inverse correlation between the condensation temperature of a particular element and its abundance in the post-AGB photosphere \citep{maas05}.  The proposed mechanism that produces this phenomenon is re-accretion of gas from a circumbinary disc \citep{waters92}. Because dust is subject to a much larger radiation pressure than gas, the gas separates from the dust and can be accreted back onto the post-AGB star. This gas, which is poor in refractory elements, will deplete the stellar photosphere of these refractory elements, while the volatile elements are less affected. Consequently, a good measure for the `depletion' of a post-AGB star is the ratio of S or Zn, which have low condensation temperatures, against Fe or Ti, which have high condensation temperatures. These ratios are listed in Table~\ref{specdata}, relative to the solar abundance ratios. In this work, we consider a post-AGB star not to be depleted if the [S/Ti] or [Zn/Ti] ratios are smaller than 0.5~dex. If [S/Ti] and [Zn/Ti] are in the range $0.5-1.0$~dex, the post-AGB star is mildly depleted. The post-AGB stars are moderately depleted if the ratios are in the range $1.0-1.5$~dex, and strongly depleted if the ratios are larger than 1.5~dex. In the absence of a [S/Ti] and [Zn/Ti] tracer, we use the [Zn/Fe] ratio to quantify depletion.

\begin{figure}
\resizebox{\hsize}{!}{\includegraphics{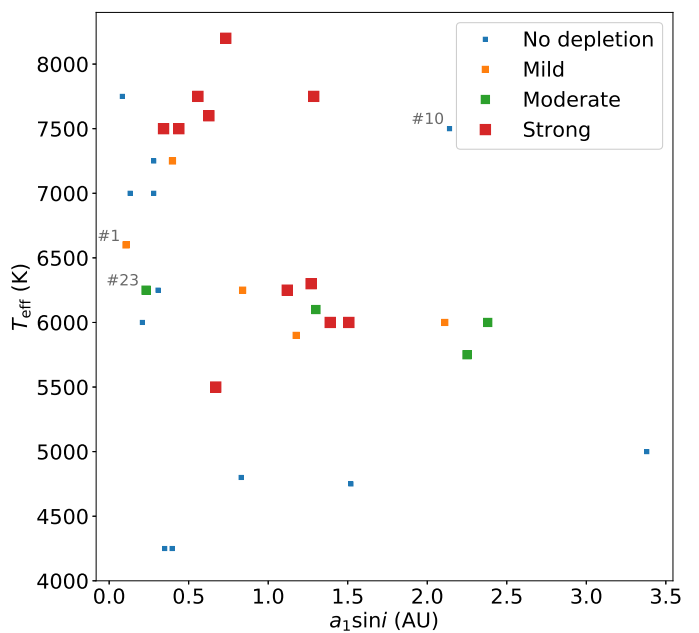}}
\caption{Effective temperature of the post-AGB stars versus the projected separation of the post-AGB orbits. Different levels of depletion are depicted by different colours and increasing symbol sizes. Objects that are numbered are discussed in Sect.~\ref{sect:depletion discussion}.}
\label{deplfig}
\end{figure}

All post-AGB stars that are depleted harbour a circumbinary disc (as seen from the SED), but not all post-AGB stars with discs are depleted \citep[e.g.][]{gezer15}. Consequently, having a circumbinary disc is a necessary condition for depletion, but not sufficient. In order to investigate why some post-AGB stars in our sample are (not) depleted, we plot the level of depletion in a diagram of effective temperature versus projected semi-major axis of the post-AGB orbit around the centre of mass in Fig.~\ref{deplfig}. The different symbol sizes and colours correspond to different levels of depletion. 

Figure~\ref{deplfig} shows that the depleted objects are clearly separated from the non-depleted objects. On one hand, post-AGB stars show no depletion if their effective temperature is smaller than about 5000~K. On the other hand, the high-temperature objects are depleted only if the distance of the post-AGB star to the centre of mass is large enough. Since we only see the projected distance, this boundary is not clearly defined, but can be estimated to be around 0.3~AU.

The correlation between depletion and orbital size for high-temperature post-AGB stars is also present if one compares depletion with the orbital period instead of $a_1\sin i$. The orbital period is related to the semi-major axis of the binary via the total mass of the system (and hence the mass of the companion star). Since the mass of the companion is unknown for the individual systems, the threshold for an object to be depleted or non-depleted is also not clear in this case and lies around $P \approx 300$~days.

\section{Companion masses} \label{sec:companionmasses}
\subsection{Mass function distribution}
The mass functions from Table~\ref{massfunctions} are shown as a cumulative distribution in Fig.~\ref{fm_dist_fit}. The peak of the mass-function distribution is around 0.2~$M_\sun$. The sample contains 4 systems that have very low mass functions $\lesssim 0.005$~$M_\sun$ (i.e., 89~Her, HD~108015, IRAS~19157-0247, and IRAS~16230-3410). A small mass function can be the result of a very low inclination angle and/or a small companion mass (see Eq.~\ref{Eq:massfct}). For the well-studied case 89~Her, the geometry and kinematics of the CO outflow favours an inclination of approximately 12 degrees \citep{bujarrabal07, hillen13}. Assuming a post-AGB star of mass 0.6~$M_\sun$, the corresponding companion mass is between 0.5$-$1.2~$M_\sun$.  For the other three systems, no additional data is available to constrain the inclination.

%\begin{figure}
%\resizebox{\hsize}{!}{\includegraphics{images/hist_log_massfunctions.eps}}
%\caption{Histogram of the mass functions in the sample of post-AGB binaries.}
%\label{hist_massfct}
%\end{figure}

\subsection{Companion mass distribution}

Although companion masses of individual systems are unknown, we can derive the distribution of masses of companions in a statistical way by making several assumptions. First of all, we must assume a mass for the post-AGB star. Post-AGB masses are expected to be limited to quite a narrow range similar to that of white dwarfs, because these stars have already lost their entire envelope. Post-AGB stars with an initial mass between 1 and 3~$M_\sun$ are expected to have masses in the range $0.5-0.7$~$M_\sun$ \citep{millerbertolami16}. However, this is influenced by the physics in previous evolutionary stages. In the main sequence, core convective overshooting can act to increase the mass of the core. Increased wind mass loss and/or common-envelope evolution will act to decrease the final mass of the core, hence decrease the mass of the post-AGB star. Furthermore, the core mass -- luminosity relation cannot help us as the distances to most post-AGB stars are very uncertain. The recent GAIA DR2 release is still based on single-star fits to the data to obtain the parallaxes and hence distances. As the orbits are in the range of one
to several AUs, the orbital motion projected on the sky will be of similar
amplitude as the parallax. We therefore do not include distance estimates of
our binaries in this paper. In what follows, we assume that the post-AGB mass distribution is similar to the observed distribution of white dwarf masses. The GAIA DR2 catalogue shows that this distribution peaks at 0.6~$M_\sun$ and 0.8~$M_\sun$ \citep{jimenezesteban18,kilic18}. However, the peak around 0.8~$M_\sun$ is suggested to be resulting from mergers. Given that our stars are in binaries which avoided merging, we assume a Gaussian distribution for the mass of the post-AGB star centred around 0.6~$M_\sun$ with a standard deviation of 0.05~$M_\sun$, as given by \citet{kilic18}. Even though this white-dwarf mass distribution is based on a population of single white dwarfs and binary interactions can result in lower-mass remnant cores, we argue from Eq.~\ref{Eq:massfct} that a small change in post-AGB mass $M_1$ does not significantly impact the measured mass functions. 

It is also required to impose a distribution for the inclination angle. For completely random orientations of the orbital plane in space, high inclinations (edge-on) are more frequently observed than low inclinations (face-on). This projection effect is properly described by
\begin{equation}
i = \arccos(z),
\label{Eq. incl}
\end{equation}
where $z$ is the length of the projection of the unit normal vector along the line-of-sight and is uniformly distributed in $[-1,1]$ for random orientations of the orbital plane in space. However, we do not expect to observe the post-AGB binaries edge-on, as the disc would obscure the post-AGB photosphere from our field-of-view. Since the target selection of our optically bright post-AGB stars is based on the SED properties, we have no targets that are observed edge-on, except for HD~44179. However, for the case of HD~44179, the post-AGB star is observed via scattered light which itself is inclined with respect to the orbit. This means that we observe this binary as if it was inclined and hence its measured mass function is also attenuated by the scattering angle. 

In order to take into account the observational bias because of disc obscuration in our post-AGB sample, we limit the inclination angle distribution to 75$^\circ$ at maximum, which accounts for the typical scale heights of the circumbinary discs \citep[e.g.,][]{hillen16}.

\begin{figure}
\resizebox{\hsize}{!}{\includegraphics{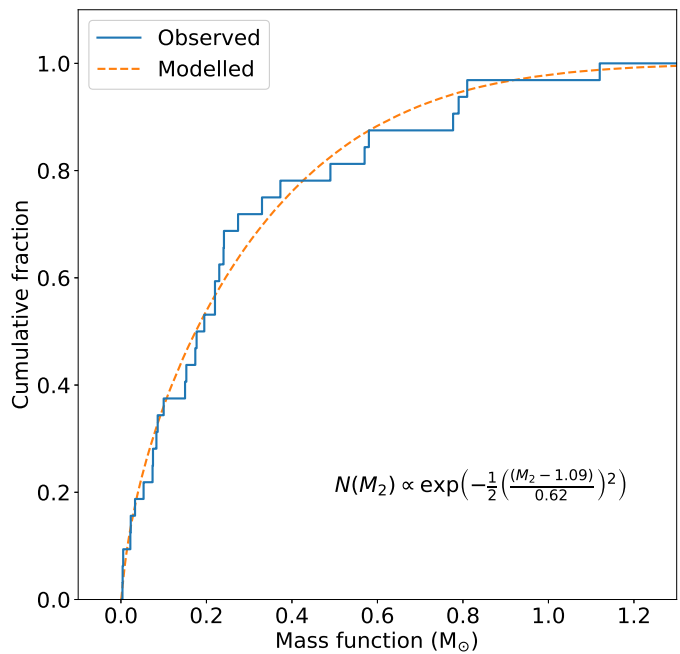}}
\caption{Observed and modelled cumulative mass-function distribution. The distribution for the companion masses is a Gaussian profile centred around 1.1~$M_\sun$ with standard deviation of 0.62~$M_\sun$.}
\label{fm_dist_fit}
\end{figure}

\begin{figure}
\resizebox{\hsize}{!}{\includegraphics{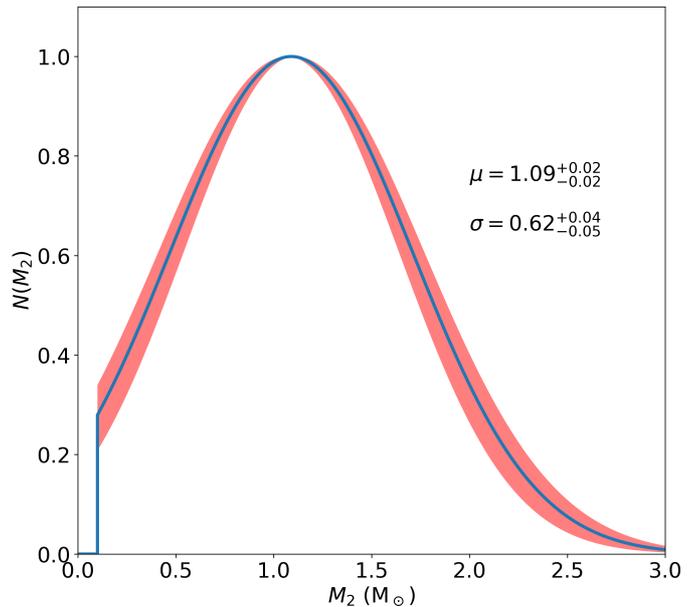}}
\caption{Mass distribution of secondary stars in post-AGB binary systems in our sample. The best-fit values for the Gaussian profile are $\mu = 1.09^{+0.02}_{-0.02}$ and $\sigma = 0.62^{+0.04}_{-0.05}$ in units in solar masses. The red-shaded region denotes the uncertainty range on the mass distribution.}
\label{M2_dist}
\end{figure}

We derive the distribution of companion masses from the observed cumulative mass-function distribution (Fig.~\ref{fm_dist_fit}) by first assuming a shape for the distribution as a prior. For simplicity, we assume a Gaussian distribution for the companion masses, $\mathrm{d}N/\mathrm{d}M_2 \propto \exp\left(-\frac{1}{2}\left(\frac{M_2-\mu}{\sigma}\right)^2\right)$, which has two free parameters, the average $\mu$ and the standard deviation $\sigma$. Next, a large number of random values is generated within the distributions of the post-AGB mass, the inclination, and the companion mass, which in turn gives us a distribution for the mass functions according to Eq.~\ref{Eq:massfct}. By optimising $\mu$ and $\sigma$, a companion mass distribution can be found that, together with distributions of $M_1$ and $\sin i$, results in a distribution of mass functions that best fits the observed cumulative mass-function distribution. The optimisation was performed using the Nelder-Mead simplex algorithm, while the uncertainties on $\mu$ and $\sigma$ were computed with a Monte Carlo method.

In this procedure, we have imposed an additional constraint on the semi-amplitude of the velocity of the post-AGB star. Since the orbit of a post-AGB star becomes very hard to detect when the radial-velocity shift is low, there is an observational bias towards binaries with high velocity semi-amplitude $K_1$. In order to account for this effect, we take a distribution for the orbital period which we assume to be logarithmically distributed (similar to what we observe). For each generated value of the mass function, we draw a value for the period from a logarithmic distribution in the range 100--3000~days such that a velocity semi-amplitude can be computed according to Eq.~\ref{Eq:massfct}. We do not take eccentricity into account here since its effect on the velocity semi-amplitude is relatively small. If this semi-amplitude of the radial velocity is below 3~km/s, we remove this combination from the fitting procedure as we probably would not have detected this binary. The main effect of this constraint is to reject low-mass companion stars in an orbit with low inclination from the fit, as this results in a low value for the mass function. This shifts the companion-mass distribution towards slightly lower masses, as there is a small population of low-mass stars that we would not observe.

The result of the fitting procedure is shown in Fig.~\ref{fm_dist_fit}. The best-fit values from the optimisation are $\mu = 1.09^{+0.02}_{-0.02}$ and $\sigma = 0.62^{+0.04}_{-0.05}$. This distribution is shown in Fig.~\ref{M2_dist}. According to this simple model, the peak of the distribution of companion masses is around 1~$M_\sun$. The distribution is bound from below at 0.1~$M_\sun$, as this is the minimum mass ($i = 75^\circ$) of the lowest mass function target in our sample (see Table~\ref{massfunctions}). With a standard deviation of 0.62~$M_\sun$, the distribution of masses ranges from 0.1~$M_\sun$ up to 2.8~$M_\sun$ at the 3$\sigma$-level. 

We tested a more complex mass distribution by assuming that there are also white-dwarf companions in the sample, which have a very peaked mass distribution. We do not expect this to be dominant in our sample as there is no detected symbiotic activity, nor is there evidence in any of the systems for the thermal contribution of the white-dwarf photospheres. We tried fitting the distribution with a double Gaussian profile in order to take into account a white-dwarf companion population. By including this additional population, the mass distribution of the main-sequence companions shifts towards higher masses. However, this does not give a significantly better fit, so we refrained from adding this population.

Finally, a Kolmogorov-Smirnov goodness-of-fit test (K-S test) was applied to test whether the Gaussian model is adequate for the observed mass-function distribution. The null hypothesis states that the two distributions are the same. The K-S test returned a p-value of 0.9, which indicates a good fit and hence we cannot reject the null hypothesis. This Gaussian model provides a good estimate of the most probable masses of the secondary stars. Moreover, the large standard deviation of 0.6~$M_\sun$ shows that a wide range of masses is needed for the companion population.

\section{Orbital properties} \label{sec:orbitalproperties}

\subsection{Period--eccentricity diagram} \label{sec:elogp}

\begin{figure}
\resizebox{\hsize}{!}{\includegraphics{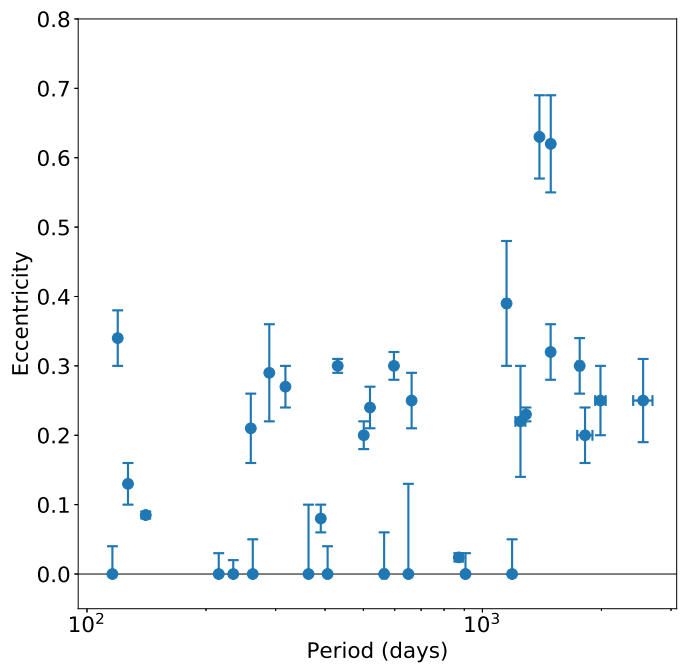}}
\caption{$e\log P$-diagram of the sample of binary post-AGB stars.}
\label{elogP}
\end{figure}

The period and eccentricity are arguably the most important observable characteristics of an orbit as they are directly related to the angular momentum of the system. Consequently, the period vs eccentricity diagram (also known as $e\log P$-diagram) is an important tool to understand the properties of a population of binaries. The $e\log P$-diagram of the sample of post-AGB binaries is presented in Fig.~\ref{elogP}. We find that 10 out of 33 binaries have orbits indistinguishable from circular, while the other 23 have eccentricities ranging up to 0.7. The uncertainty on the period of the orbits is relatively small, as this is quite easy to determine. This is especially the case for the shorter-period orbits, where a large number of cycles were observed. The eccentricity on the other hand is much harder to determine precisely. This is visible in Fig.~\ref{elogP}, where the error bars on the eccentricity are in some cases quite large. This is most often a consequence of the large scatter due to pulsations.

We tested for correlations in the orbital elements by performing a Spearman rank correlation test \citep{spearman04}, which is suitable for small data sets and is robust against outliers. We find that the correlation between period and eccentricity is small but significant. The r-statistic from the Spearman correlation test is 0.35, which results in a p-value of 0.045. In a similar fashion, there is a correlation between eccentricity and $a_1 \sin i$  ($r=0.39$, $p=0.025$). It is clear that larger eccentricities occur, on average, in wider orbits. There are some interesting exceptions, however, as some very close orbits appear to be more eccentric. Moreover, at periods below 1000~days, there are both eccentric and non-eccentric orbits in the sample. Only the very wide orbits seem to be exclusively eccentric.

Figure~\ref{elogP} shows an apparent lack of orbits with an eccentricity between 0.1 and 0.2. However, given the limited number of binaries in our sample and the relatively large uncertainties on some of the eccentricities, we cannot confirm whether this gap in the eccentricity distribution is real.

\subsection{Size of the orbits}
\begin{figure}
\resizebox{\hsize}{!}{\includegraphics{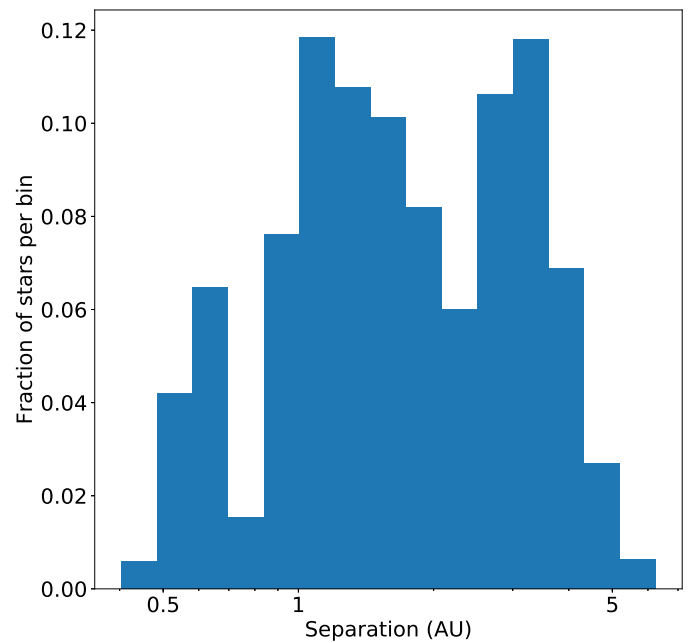}}
\caption{Histogram of the distribution of separations in astronomical units for the binaries in our sample. The height of the bins corresponds to the fraction of binaries in that bin with a total number of 15 bins.}
\label{hist_separations}
\end{figure}

The semi-major axes of the post-AGB binary systems are related to the orbital periods via Kepler's third law. Since we now have knowledge of the companion masses derived in Sect.~\ref{sec:companionmasses}, we can compute the distribution of orbital separations for the binaries in our sample by using the measured orbital periods. In order to do so, we take the post-AGB mass distribution as before to be a Gaussian distribution around 0.6~$M_\sun$ with standard deviation of 0.05~$M_\sun$. A semi-major-axis distribution is derived for each binary individually by randomly drawing 10\,000 values from all distributions, while imposing a minimum mass for the companion distribution (see Table~\ref{massfunctions}). The histogram constructed from the sum of all semi-major-axis distributions is shown in Fig.~\ref{hist_separations}. The majority of the binaries have a separation smaller than approximately 4~AU. The semi-major-axis distribution goes up to 6~AU, which corresponds to the longest-period binary (U~Mon) in the sample with a high companion mass. The distribution is bound from below at 0.4~AU.

%\subsection{Radius of the Roche lobes} \label{sec:rochelobes}
\begin{figure}
\resizebox{\hsize}{!}{\includegraphics{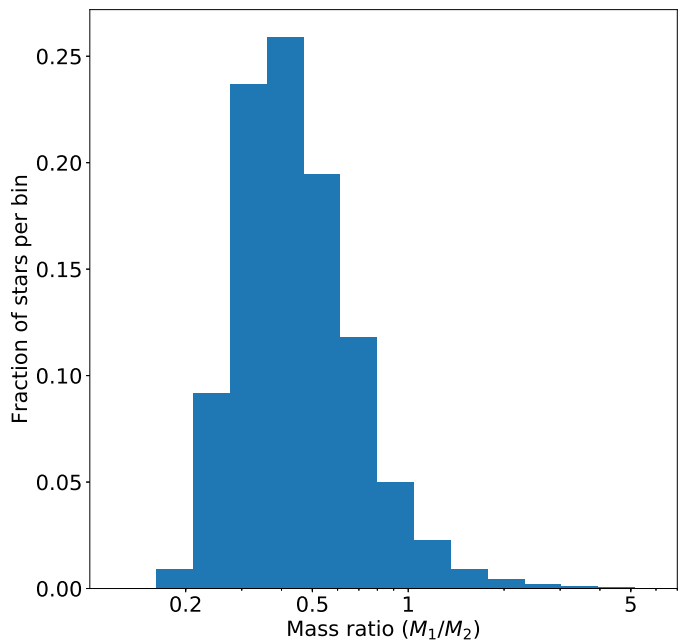}}
\caption{Histogram of the distribution of mass ratios. The height of the bins corresponds to the fraction of binaries in that bin with a total number of 15 bins.}
\label{q_distribution}
\end{figure}

The companion mass distribution can also be used to derive a distribution for the mass ratios, which are defined as $M_1/M_2$. We again draw 10\,000 values from the post-AGB mass distribution around 0.6~$M_\sun$ and the companion mass distribution for each binary individually, taking into account the minimum mass from Table~\ref{massfunctions}. The result is shown in Fig.~\ref{q_distribution}. For most binaries, the companion is more massive than the post-AGB star, as is expected from the companion mass distribution. As the companion mass distribution extends to masses as low as 0.1~$M_\sun$, the mass ratio can reach a maximum value of 5, but this is the case for only a small number of stars.
 
\begin{figure}
\resizebox{\hsize}{!}{\includegraphics{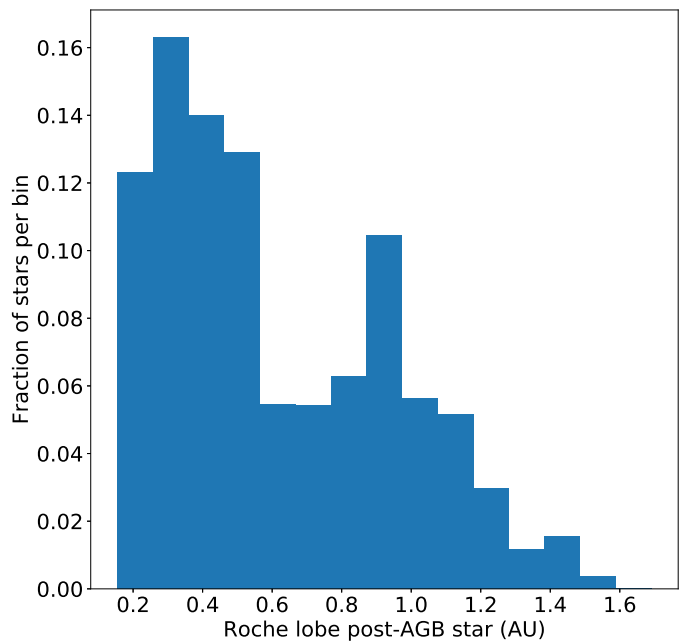}}
\caption{Same as Fig.~\ref{hist_separations}, but for the distribution of Roche-lobe radii.}
\label{RL_distribution}
\end{figure}

The radius of the Roche lobe of a star in a binary is only a function of the semi-major axis $a$ of the binary system and the mass ratio of the two stars. From \citet{eggleton06}, we have the relation
\begin{equation}
R_{L,1} = \frac{0.44q^{0.33}}{(1+q)^{0.2}}a,
\label{eq:RLformula}
\end{equation}
where $R_{L,1}$ is the radius of a sphere with the same volume as the Roche lobe, $q$ is the mass ratio, and $a$ is the separation of the binary system. This equation is valid for $0.1 \leq q \leq 10$, which is always satisfied in our sample of post-AGB binaries (see Fig.~\ref{q_distribution}). We apply the same procedure as before, i.e. we randomly draw 10\,000 values out of all distributions ($M_1, M_2, P_\mathrm{orb}$) for each star, which results in a distribution for the radii of the Roche lobes of the post-AGB stars. The result is shown in Fig.~\ref{RL_distribution}. We find that the Roche-lobe radii range from 0.2~AU to 1.6~AU, with the majority of the Roche lobes smaller than 1~AU. Consequently, the current sizes of most post-AGB Roche lobes are smaller than the maximum size of a typical AGB star ($\sim$ 1--2~AU).

Since the effect of an unknown companion mass partly cancels in Eq.~\ref{eq:RLformula}, the Roche-lobe radii are well-defined for the individual post-AGB stars. As a result, the 3$\sigma$-uncertainty on the Roche-lobe radii of the stars in our sample is on average 0.07~AU, despite the relatively large uncertainty on the companion masses.

\section{Discussion} \label{sec:discussion}

\subsection{Depletion in binary post-AGB stars} \label{sect:depletion discussion}
As discussed in Sect.~\ref{sec:depletion} and shown in Fig.~\ref{deplfig}, post-AGB stars with low effective temperatures ($T_\mathrm{eff}< 5500$~K) are not depleted. This can be understood from the structure of a low-temperature post-AGB star. To quantify this, we computed stellar evolution models with the \texttt{MESA} code \citep{MESApaper1,MESApaper2,MESApaper3,MESApaper4}. Our models show that in the post-AGB phase, the effective temperature is strongly dependent on the mass in the envelope of the star (see Appendix~\ref{appendix:MESA}). For an 0.6~$M_\sun$ post-AGB star of solar metallicity, the mass in the envelope is of the order of $10^{-2}$~$M_\sun$ at 3500~K, while at 5500~K only $\sim 10^{-3}$~$M_\sun$ remains. Therefore, a higher temperature post-AGB star needs to accrete much less metal-poor gas in order to show depletion patterns, assuming mixing in the envelope is efficient. Moreover, since a post-AGB star evolves towards higher temperatures in the HR diagram, a hotter post-AGB star can be regarded as `more evolved' compared to its lower-temperature analogues. One can argue that the hotter post-AGB star had more time to accrete material from a disc compared to a cooler post-AGB star. The increase of depletion signatures with effective temperature is therefore consistent with the idea that the accretion of gas from a circumbinary disc is an ongoing process.

Another remarkable feature in Fig.~\ref{deplfig} is that the stars with the lowest $a_1\sin i$ show no signature of depletion, regardless of their effective temperature. The absence of the depletion phenomenon in relatively compact systems can have two possible reasons. Either the post-AGB star in a close orbit does not efficiently accrete gas from the circumbinary disc, or the gas that is accreted from the disc is not depleted. The latter could be the result of a too high temperature at the inner rim of the disc such that the dust sublimates and the refractory elements remain in the gas phase.
%We propose that this is due to the fact that the circumbinary disk is closer to the post-AGB star in the more compact systems, as the position of the disk is thought to scale with the semi-major axis of the system. Consequently, the inner-rim of the disk in these systems might be too hot to form dust, even for elements of high condensation temperature. The material accreted back onto the post-AGB star then contains both volatile and refractory elements, which causes the system not to show depletion.

Examining Fig.~\ref{deplfig} in more detail, there appears to be one object that does not follow the trend discussed above. In the top right corner, HD~95767 ($\# 10$) has a high effective temperature and a large projected semi-major axis, yet shows no sign of depletion. However, \citet{vanwinckel97} reported that this object has a suprasolar sodium abundance, typical for massive F-type supergiants. Added to the fact that this is the only post-AGB star in our sample that lies in the galactic plane and its surface gravity (see Table~\ref{specdata}) is higher than that of post-AGB stars of similar temperature, \citet{vanwinckel97} concludes that ``it probably is a genuinely massive supergiant, and its infrared excess is then not the result of post-AGB evolution''.

The two objects on the left of Fig.~\ref{deplfig} are 89~Her and IRAS~16230-3410 ($\# 1$ and $\# 23$, respectively). These are two objects that show depletion, yet have a small projected semi-major axis of less than 0.25~AU. The word `projected' is key here, since we know that 89~Her has a small inclination angle ($a_1\approx$ 0.4--0.7~AU), and also IRAS~16230-3410 has a very small mass function which could be the result of a near-face-on orbit. For these two objects, the average distance of the post-AGB star to the centre of mass can be much larger in reality.

\subsection{Companion stars}
The mass-function distribution for our sample of post-AGB binaries can be modelled by a Gaussian distribution of companion masses centred around 1.09~$M_\sun$ with a standard deviation of 0.62~$M_\sun$ depicted in Fig.~\ref{M2_dist}.  The mass distribution of the companions depends on star formation history, initial mass function, mass transfer history, and could suffer from observational bias, so the real companion mass distribution may be more complicated. The lack of constraints on the inclination prevents us from determining it accurately. However, the Gaussian model provides good insight into the general properties of the companion stars. 

The large standard deviation is the sign of a large range of companion masses, ranging from 0.1~$M_\sun$ to 2.5~$M_\sun$. This is quite remarkable, since this implies that the mass ratios of the post-AGB binaries also have a large range, while the outcome of binary interaction is thought to depend to a large extent on the mass ratio of a binary system. As the objects were selected by their IR excess, circumbinary discs are present in all the post-AGB binaries of our sample. Consequently, the mass of the companion, and hence also the mass ratio, apparently do not impact the formation of a disc significantly.

The mass distribution of the companions is not strongly peaked, as we would expect for white dwarfs, and our cumulative mass-function distribution fit does not convincingly show that there is a significant population of white dwarfs among the companions. Additionally, we do not observe symbiotic activity typical for accreting white dwarfs, which suggests that almost all companions are of main-sequence nature.

Additional arguments for the companions to be main-sequence stars come from some binaries where the H$_{\alpha}$ profiles were found to turn into P-Cygni profiles at superior conjunction, when the companion is in front of the luminous primary \citep{gorlova15, bollen17}. The interpretation is that continuum photons are scattered out of the line-of-sight by atoms in a jet created by an accretion disc around the companion. This accretion disc was even resolved in one of post-AGB binaries by \cite{hillen16}. 
 The jets originating from these companion stars show escape velocities that are lower than the escape velocity of a typical white dwarf. We therefore conclude that there is good observational evidence that most, if not all, companion stars are unevolved main-sequence stars.

\subsection{Orbits}

\subsubsection{Periods}
A remarkable feature in the orbits of our sample of post-AGB binaries is the period distribution, with periods ranging from 100--3000~days. Given the mass distribution of the companion stars, we can relate the period of the binaries to a distribution of post-AGB Roche-lobe radii (see Fig.~\ref{RL_distribution}). The radii of the Roche lobes range up to only about 1.5~AU, which is similar to the radius of a low-mass AGB star near the end of the AGB phase. Moreover, the wide binary systems have eccentric orbits, giving rise to a smaller separation at periastron. Consequently, if we assume that the radius of the Roche lobe did not change significantly since the end of the AGB, we expect the progenitors of all our post-AGB stars to have filled their Roche lobes at some point in their evolution.

%\begin{figure}
%\resizebox{\hsize}{!}{\includegraphics[angle=-90]{images/period-histogram.eps}}
%\caption{Period distribution of post-AGB binaries in our sample. Periods range from 100 to 3000~days.}
%\label{period-distribution}
%\end{figure}

As a star in a binary fills its Roche lobe, it will initiate mass transfer to the companion star. Since mass transfer is unstable during the AGB for many configurations \citep[][however, see \citealp{chen08, woods11, pavlovskii15}]{hjellming87}, it is expected that a common envelope forms around the binary, which should cause the system to spiral in considerably with resulting periods of the order of days \citep{izzard12}. On the other hand, the longer period systems that do not interact significantly are expected to widen because of the mass lost by an isotropic stellar wind from the AGB star \citep{nie12}. So a bimodal distribution can be expected with a population of close orbits (< 100~d), as products of the common envelope channel, and a population of very wide systems (> 3000~d) which did not come into contact.  Figure~\ref{elogP} shows a very different picture \citep[see also][]{vanwinckel03, vanwinckel09, manick17}: all periods fall into a regime least expected by standard population synthesis predictions \citep{izzard10,nie12}. Producing the period distribution observed in our sample requires additional physical processes \citep{pols14}, such as high accretion rates from wind-Roche-lobe overflow \citep{abate13}, efficient loss of angular momentum due to mass loss \citep{chen18, saladino18}, and/or efficient common-envelope ejection \citep{soker15}. 

Note that post-AGB stars in binaries with periods less than 100~days would currently be filling their Roche lobes (see below). Consequently, these systems would most likely consist of a hotter post-AGB star, or evolve straight to the PN phase to become a close binary central star of a planetary nebula \citep{miszalski09}. On the other hand, wide systems with periods over 3000~days are much harder to detect. Firstly, the HERMES monitoring campaign has been operational since 2009. Consequently, very wide orbits cannot be resolved yet. Secondly, as mentioned earlier in this work, the orbital analysis is limited by the pulsations of the post-AGB stars. Cleaning these pulsations requires a large amount of observations over a relatively short time span. This is observationally challenging to apply to a large sample of stars. The lack of these long-period systems could therefore be the result of observational bias.

The distribution of Roche-lobe radii in Fig.~\ref{RL_distribution} is bound from below at 0.2~AU. Our \texttt{MESA}-models of the post-AGB evolution show that this corresponds to the size of a typical (0.6~$M_\sun$) post-AGB star at 8000~K. This means that the short-period systems are currently close to filling their Roche lobes. However, the post-AGB phase is a transition phase from the AGB to the PN. This means that post-AGB stars are in a state of rapid contraction, in which the radius decreases from the order of an astronomical unit to the size of a white dwarf on a timescale of several times $10^3$--$10^4$ years. Consequently, post-AGB stars that currently nearly fill their Roche lobe either formed (very) recently from some interaction, or are currently undergoing an interaction in which the orbit shrinks at a similar pace as the radius of the star.

\subsubsection{Eccentricities}
From tidal evolution theory, it is expected that binaries in which one of the stars fills a substantial fraction of its Roche lobe become circularised in a relatively short time scale, since tidal forces depend sensitively on the Roche-lobe filling factor \citep[$\dot{e}\propto\left(r/a\right)^8$, see][]{zahn77}. It is a long-standing problem that the detected large eccentricities in evolved binaries are in contradiction to tidal theory. This is the case for systems in which the former AGB star is now a white dwarf, which includes barium and CH stars \citep{jorissen98,vanderswaelmen17}, carbon-enhanced metal-poor (CEMP) stars \citep{jorissen16,hansen16}, and S-type symbiotics \citep{fekel00}, but also other binaries that have undergone an interaction with a red-giant star, such as blue stragglers in old stellar clusters \citep{mathieu15} and wide sdB binaries \citep{vos17}. 

Post-AGB binaries are no exception to the eccentricity problem, as over 70$\%$ of the post-AGB binaries in our sample are significantly eccentric. Not only are the closest orbits eccentric (see Fig.~\ref{elogP}), but we would expect \textit{all} of the post-AGB binaries to have circularised by now, since all the current Roche-lobe radii are too small to fit an AGB star.

To produce the observed eccentric orbits in all these different evolved binary systems, either the tidal forces should be much weaker than predicted, or an extra eccentricity-pumping mechanism is required, or a combination of both. A study by \citet{nie17} on ellipsoidal variables in the Large Magellanic Cloud suggests tidal circularisation rates are a factor $\sim$100 lower than predicted. Additionally, there are several mechanisms that could increase the eccentricity. Phase-dependent mass loss on the AGB can pump the eccentricity by tidally enhancing the mass-loss rate at the periastron of the orbit \citep{bonacicmarinovic08}. Once a circumbinary disc forms, likely due to preferential mass loss via the Lagrangian point L$_2$, resonances between material in the disc and the stars can transfer angular momentum from the binary to the disc \citep{dermine13}. This in turn could increase the eccentricity of the orbit \citep{vos15}, but this is criticised by \citet{rafikov16} who concludes that the eccentricity pumping capacity of the circumbinary discs must be limited. Alternatively, \citet{kashi18} suggest that increased mass loss at periastron due to a grazing envelope evolution can maintain the eccentricity during the interaction. Finally, the presence of a third component in the system could cause a dynamical interaction leading to an eccentric orbit \citep{perets12}.

\subsection{Comparison to other evolved binaries}

Even though the previously mentioned classes of evolved binaries show similar orbital characteristics and their evolution can be linked to the post-AGB phase via the white dwarf, there are some distinct differences. First of all, there is no signature of \textit{s}-process enrichment in the photospheres of any of the binary post-AGB stars in our sample, with one noticeable exception being HD\,158616 \citep{desmedt16}. If anything, the binary post-AGB photospheres are generally poor in \textit{s}-process elements. The \textit{s}-process elements are refractory so their abundances are severely affected by the depletion process. Whether the depletion masks the former enrichment is currently unknown. The \textit{s}-process elements do not deviate from the trend seen in depleted photospheres, where the underabundance scales with the condensation temperature \citep{maas05,giridhar05,reyniers07,gielen09,rao12,venn14}, so there is no observational indication that \textit{s}-process enrichment was present on the AGB.  Furthermore, all the post-AGB stars in our sample harbour an O-rich disc \citep{vanwinckel09,gielen11}. This would imply that the envelope of the former AGB star also was O-rich, suggesting that the third dredge-up in these stars was inefficient and did not enrich the object enough to become a carbon star. If the \textit{s}-process abundances in the photosphere of the AGB-star are not enhanced, the companion would not be polluted and hence this would lead to a normal main sequence + white dwarf binary.

In a large fraction of the low-mass binaries, the stars can interact on the red giant branch (RGB) as well \citep{kamath16}. If the red giant loses enough mass during this interaction, the core will not be massive enough to ignite helium and the star will evolve as a `post-RGB' star. These stars are very similar to the post-AGB stars, except that they are less luminous (1000 -- 3000~$L_\sun$). Since we have no information on the distance and hence on the luminosity, we cannot easily differentiate between the two. But it is likely that a fraction of the stars in our sample are post-RGB stars. These stars would of course not pollute the companion as the \textit{s}-process nucleosynthesis does not take place inside an RGB star.

\section{Conclusions} \label{sec:conclusions}

We present the orbital elements of 33 galactic post-AGB binaries, of which three orbits are new (IRAS~06165+3158, IRAS~06452-3456, and IRAS~16230-3410). The orbital elements are given in Table~\ref{tableorbitalelements}. The orbital periods of the binaries lie in the range 100--3000 days, while over 70 percent of the binaries have a significant non-zero eccentricity.

The $e\log P$-diagram (Fig.~\ref{elogP}) shows many similarities to other types of binaries that have undergone an interaction with a red-giant star. The distribution of Roche-lobe radii of the binary post-AGB stars are small enough that the progenitors of the post-AGB stars must have filled their Roche lobes. From tidal evolution theory one would expect these systems to circularise on a timescale much shorter than the interaction timescale. Clearly, the physics of the interaction of stars on the AGB is still poorly understood. We conclude that extra physics is needed in binary evolution models in order to produce the observed orbital periods and eccentricities, such as enhanced angular-momentum loss, eccentricity pumping by circumbinary discs, phase-dependent mass loss, etc.

By assuming a distribution for the masses of the post-AGB stars and a distribution for the inclination, we find that the distribution of the masses of the companion stars is peaked around 1.09~$M_\sun$ with a large standard deviation of 0.62~$M_\sun$. This large spread in companion masses ($0.1-2.5~M_\sun$) shows the large diversity of post-AGB binaries. This implies that the mass ratio plays only a marginal role in the binary interactions producing the orbits we see today. The observed spread in companion masses, together with the lack of symbiotic activity and the low escape velocities of jets observed in post-AGB binaries, shows that the post-AGB binaries contain mainly main-sequence companions. The companion masses cover a wide distribution, which also means that binaries with a wide range of masses can evolve into post-AGB binaries.

Finally, we found a correlation between the effective temperature, the size of the post-AGB orbit, and the phenomenon of depletion of refractory elements in the post-AGB photosphere. Post-AGB stars with high effective temperatures in a wide orbit are depleted, contrary to post-AGB stars with either a low temperature or a close orbit. Since depletion exclusively occurs in post-AGB stars with disc-type SEDs, we conclude that re-accretion of material from a circumbinary disc is an ongoing process in the post-AGB phase, and that this process is inefficient in more compact post-AGB binary systems.

\begin{acknowledgements}
Based on observations made with the Mercator Telescope, operated on the island of La Palma by the Flemmish Community, at the Spanish Observatorio del Roque de los Muchachos of the Instituto de Astrofísica de Canarias. We used data obtained with the HERMES spectrograph, which is supported by the Research Foundation - Flanders (FWO), Belgium, the Research Council of KU Leuven, Belgium, the Fonds National de la Recherche Scientifique (F.R.S.-FNRS), Belgium, the Royal Observatory of Belgium, the Observatoire de Genève, Switzerland, and the Thüringer Landessternwarte Tautenburg, Germany. The authors thank the many observers of the HERMES consortium. GMO acknowledges support of the Research Foundation - Flanders under contract G075916N. HVW acknowledges support of the research council of the KU Leuven under contract C14/17/082. This research has made use of the SIMBAD database and the Vizier online catalogue operated at CDS, Strasbourg, France. This research has made use of NASA's Astrophysics Data System Bibliographic Services. The authors thank the anonymous referee for the useful suggestions and valuable comments that have improved this paper.

\end{acknowledgements}

\bibliographystyle{aa}
\bibliography{references.bib}

%\Online
\begin{appendix}

\section{Spectral energy distributions} \label{appendix:seds}
 \captionsetup[subfigure]{labelformat=empty}

In this appendix we show the SEDs of the post-AGB stars in the sample. We used a grid of MARCS model atmospheres \citep{marcspaper} to fit the SEDs. Filled circles in Fig.~\ref{fig:discseds} show data points that have been used in the fitting routine, while empty circles are excluded from the fit. This can be either due to a contribution of flux from the disc, or because the photometry is of bad quality or unreliable. All the SEDs show an infrared excess starting at near-infrared wavelengths. This points towards hot dust in the system in the shape of a circumbinary disc. 

A variety of surveys provide the photometric data points with which we construct all the disc-type SEDs in Appendix~\ref{appendix:seds}. Photometry at blue, optical, and near-infrared wavelengths was used to fit the post-AGB photosphere. The most important photometry bands at these wavelengths are the UBVRI Johnson-Cousins bands \citep{johnson,anderson12,morel78,nascimbeni16, kharchenko01, richmond07, ofek08, henden16, lasker08, girard11}, Geneva photometry \citep{gcpd}, and Str\"omgren photometry \citep{hauck98}. In some cases, we used photometry from the Hipparcos \citep{hipparcos} and Tycho-2 \citep{tycho2} catalogues. Other sources include optical photometry from the USNO-B catalogue \citep{usno-b}, ANS \citep{wesselius82}, OAO2 \citep{code80}, TD1 \citep{humphries76}, the Vilnius bands \citep{vilnius}, and SDSS \citep{sdss6, sdss7}.\footnote{See the relevant instrument papers for the expansion of the acronyms used in this paragraph.} At near-infrared wavelengths, the most important survey is 2MASS \citep{2mass}, while at mid- to far-infrared wavelengths, the WISE bands \citep{wise}, Akari photometry \citep{akari}, and the IRAS mission \citep{iras} give us most coverage. For some sources, we have also used MSX photometry \citep{egan03}. Beyond 100~$\mu$m, we have photometry from PACS \citep{PACS} and SPIRE \citep{SPIRE} on the Herschel space observatory, and in few cases SCUBA photometry \citep{difrancesco08}.

 \begin{figure*}
 \centering
   \subfloat[\#1: 89~Her]{%
     \includegraphics[width=0.33\textwidth]{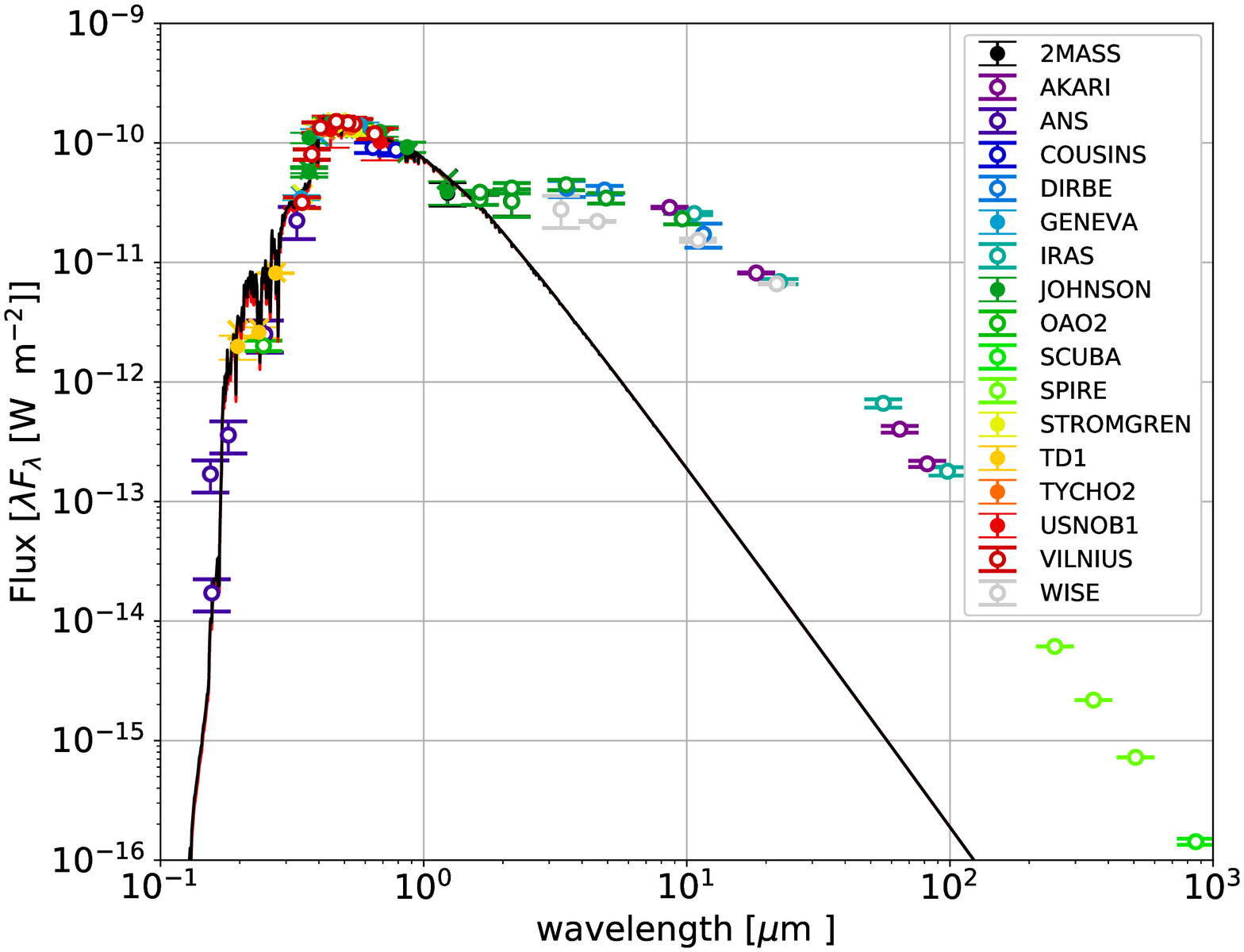}          \label{figure:sed1}
   }
   \subfloat[\#2: AC~Her]{%
     \includegraphics[width=0.33\textwidth]{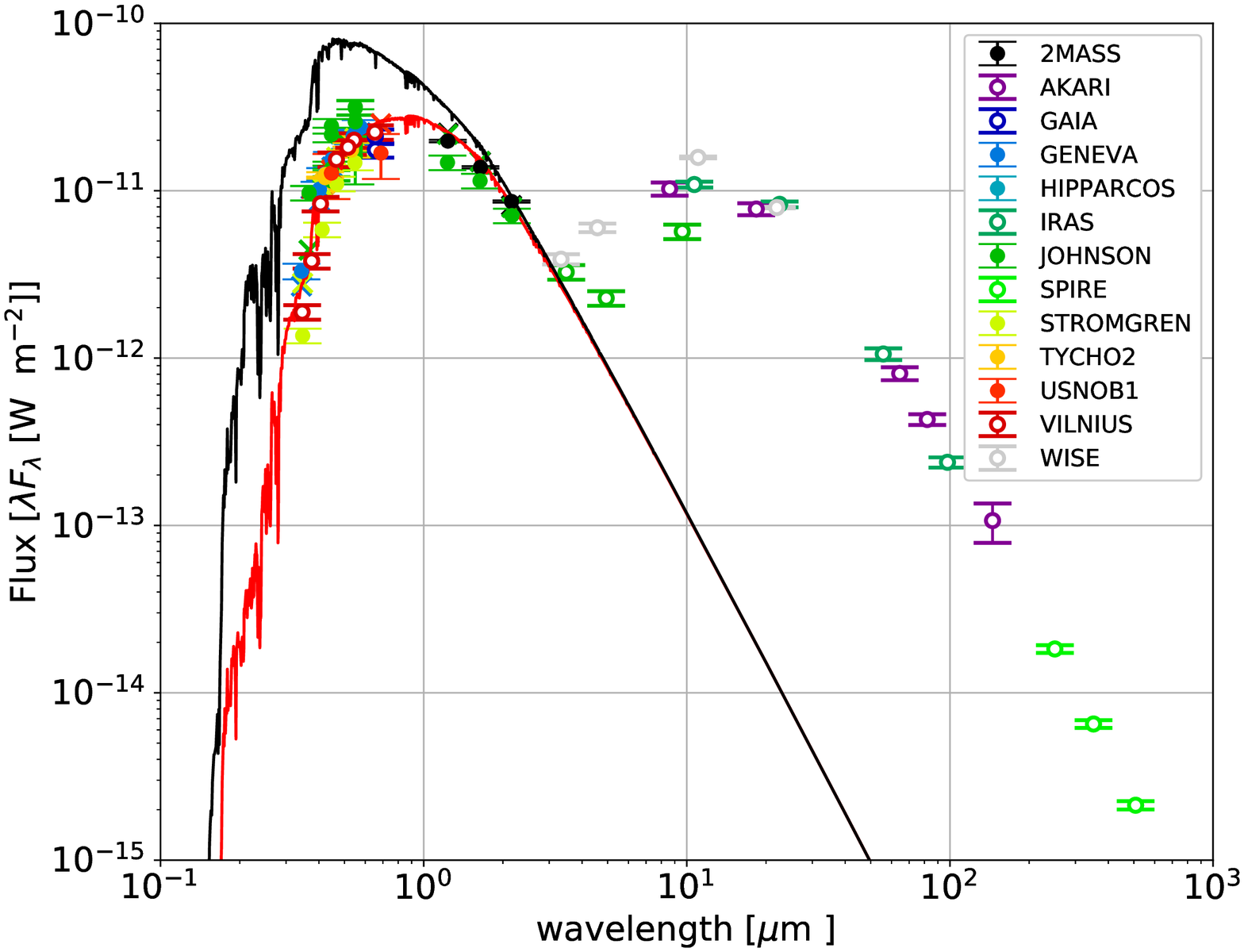}        \label{figure:sed2}
   }
   \subfloat[\#3: BD+39~4926]{%
     \includegraphics[width=0.33\textwidth]{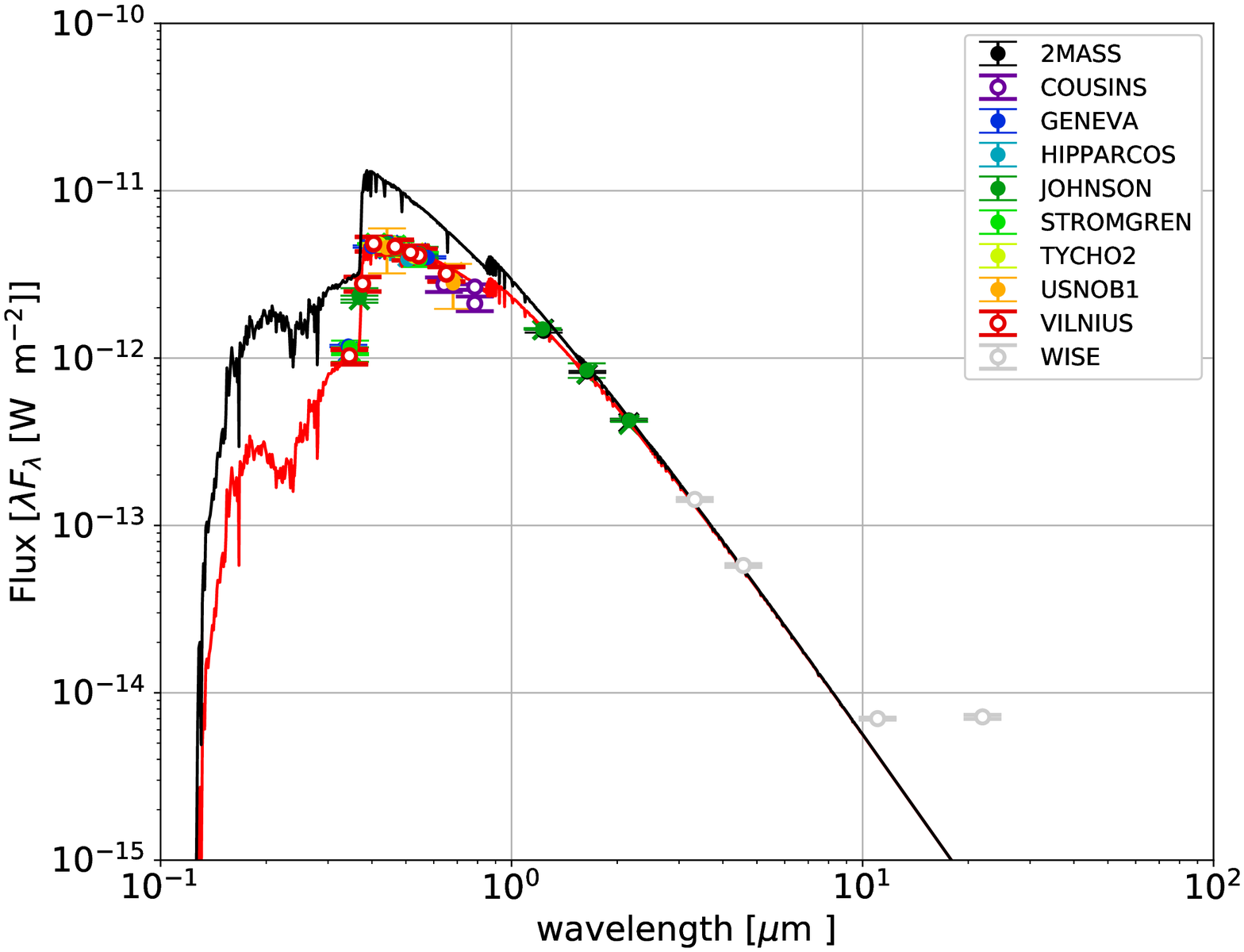}       \label{figure:sed3}
   }\\

   \subfloat[\#4: BD+46~442]{%
     \includegraphics[width=0.33\textwidth]{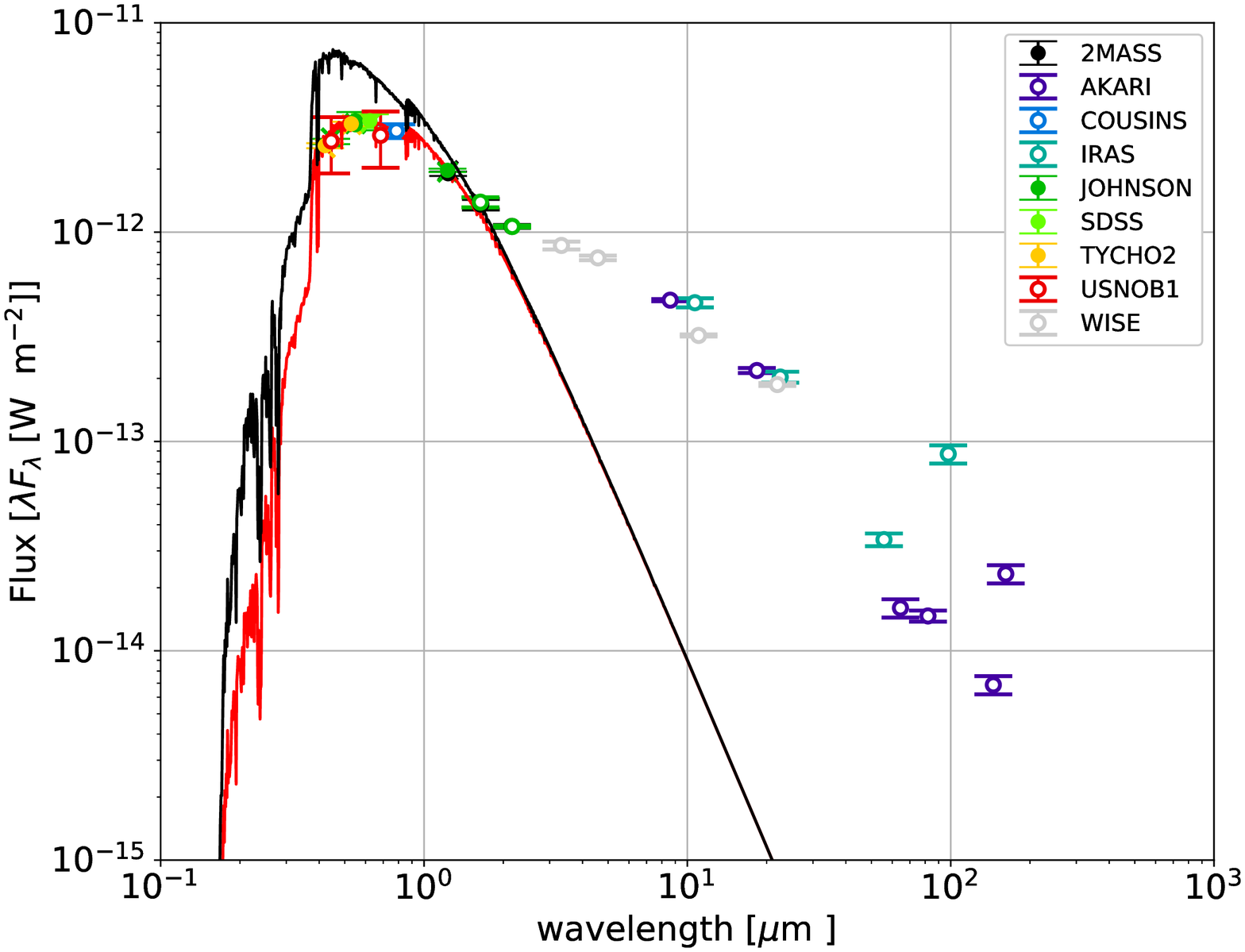}        \label{figure:sed4}
   }
   \subfloat[\#5: DY~Ori]{%
     \includegraphics[width=0.33\textwidth]{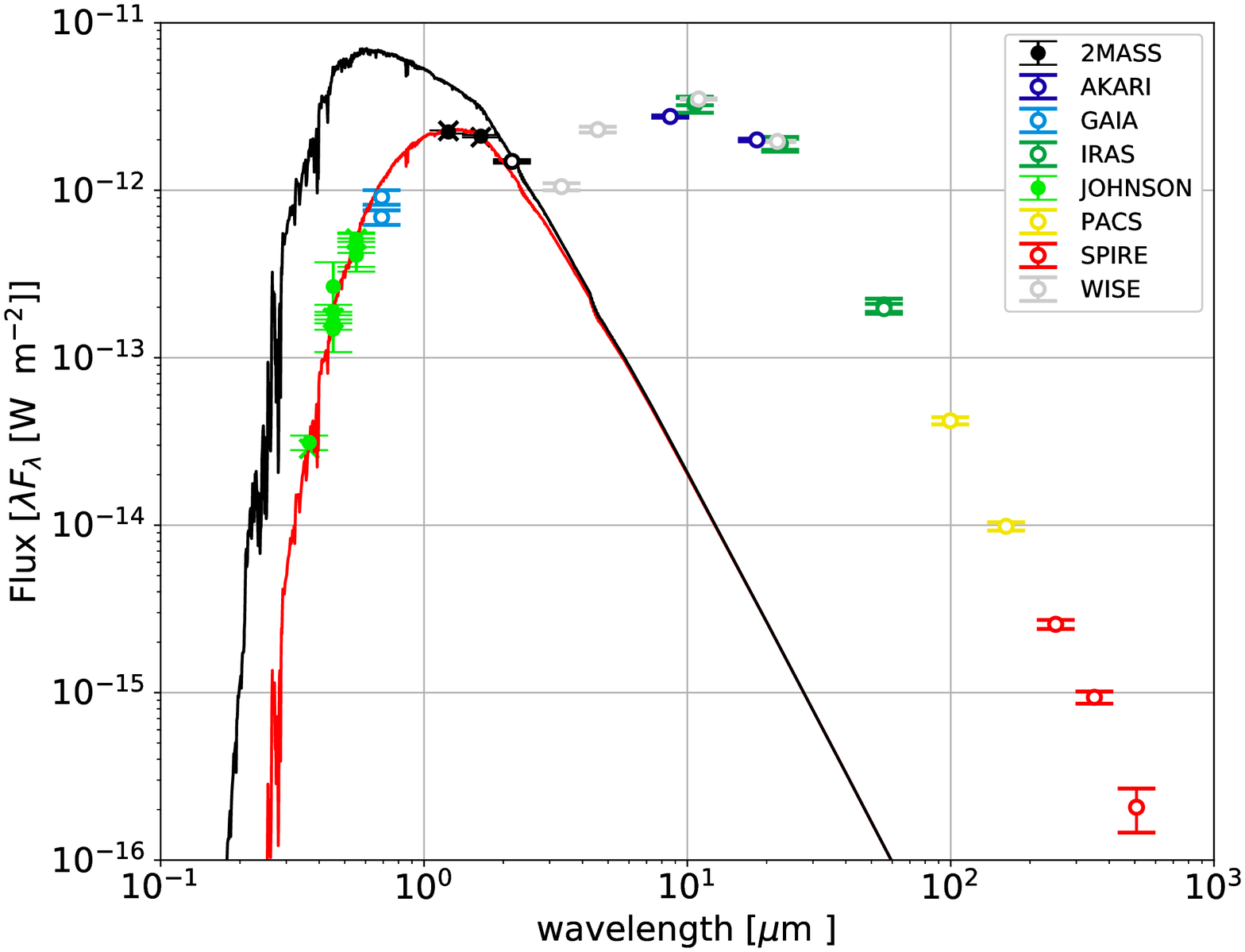}       \label{figure:sed5}
   }
   \subfloat[\#6: EP~Lyr]{%
     \includegraphics[width=0.33\textwidth]{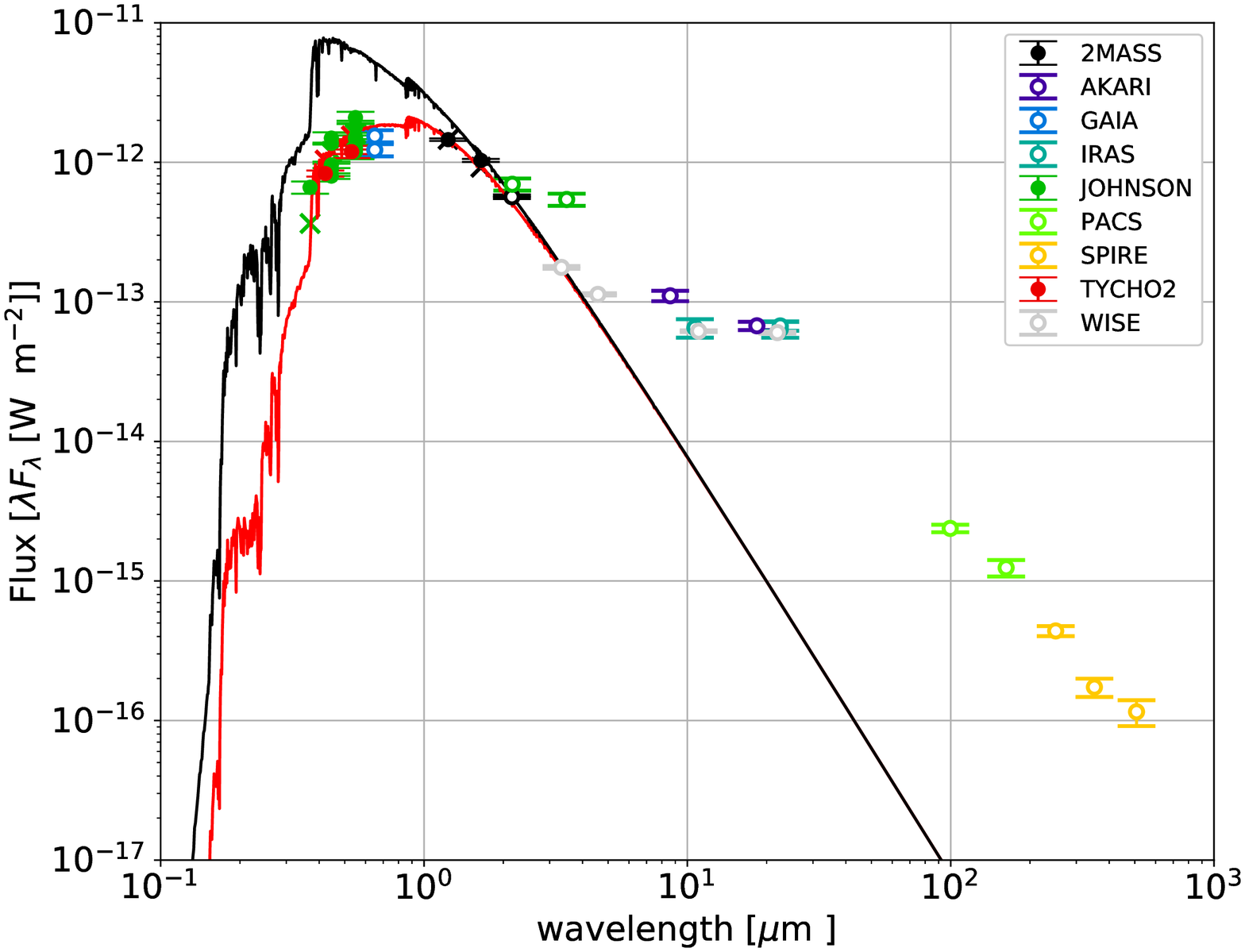}        \label{figure:sed6}
   }\\

   \subfloat[\#7: HD~44179]{%
     \includegraphics[width=0.33\textwidth]{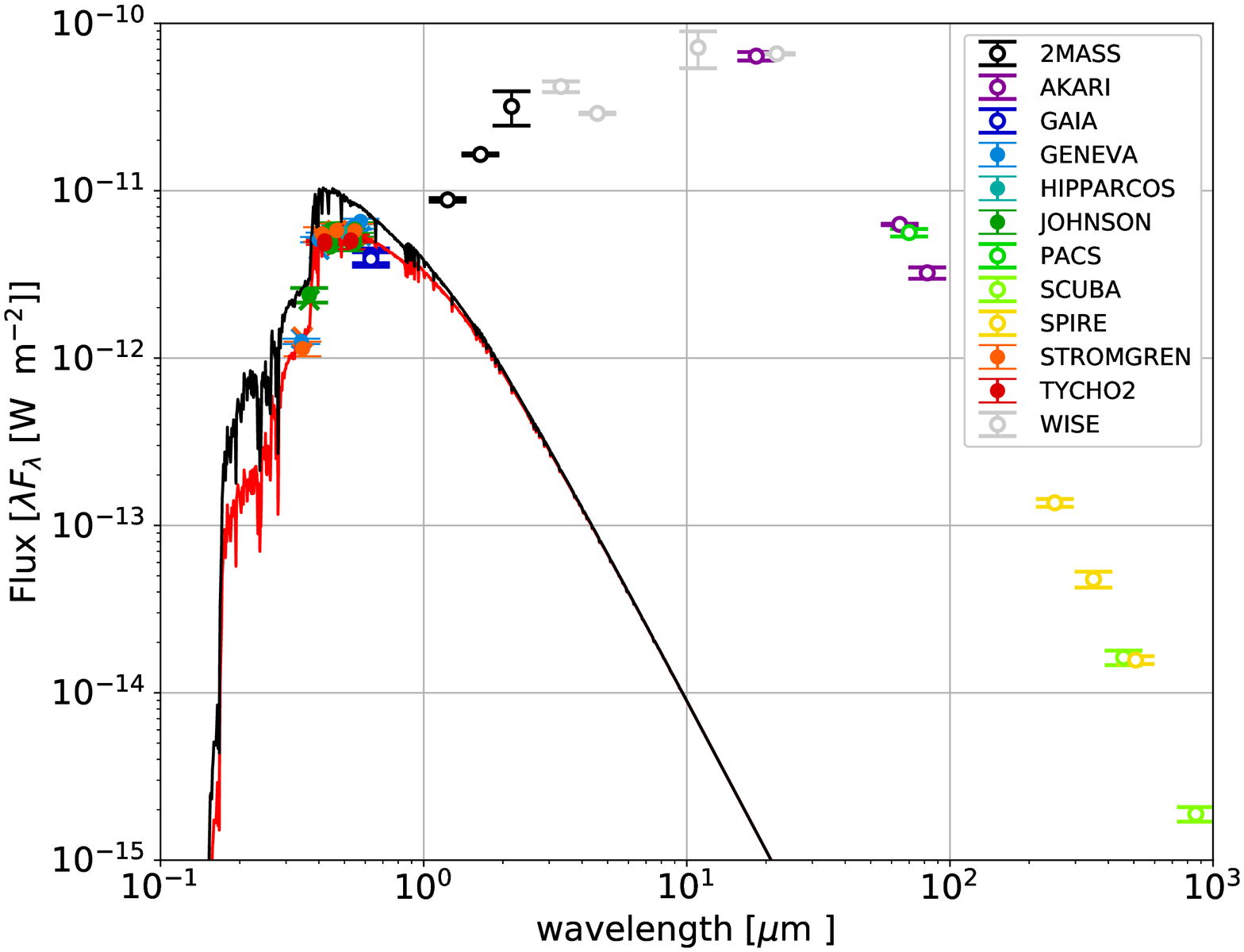}      \label{figure:sed7}
   }
   \subfloat[\#8: HD~46703]{%
     \includegraphics[width=0.33\textwidth]{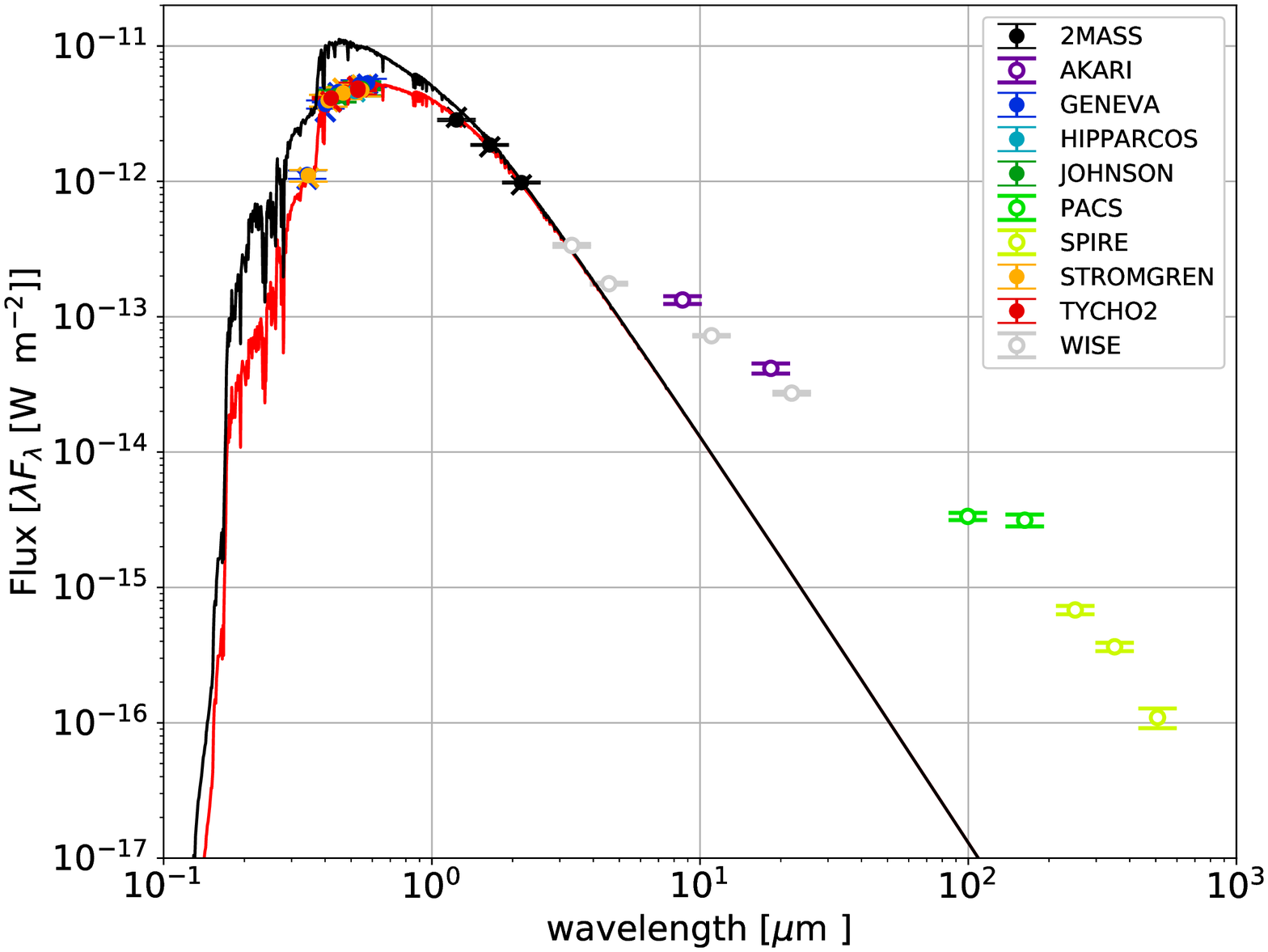}      \label{figure:sed8}
   }
   \subfloat[\#9: HD~52961]{%
     \includegraphics[width=0.33\textwidth]{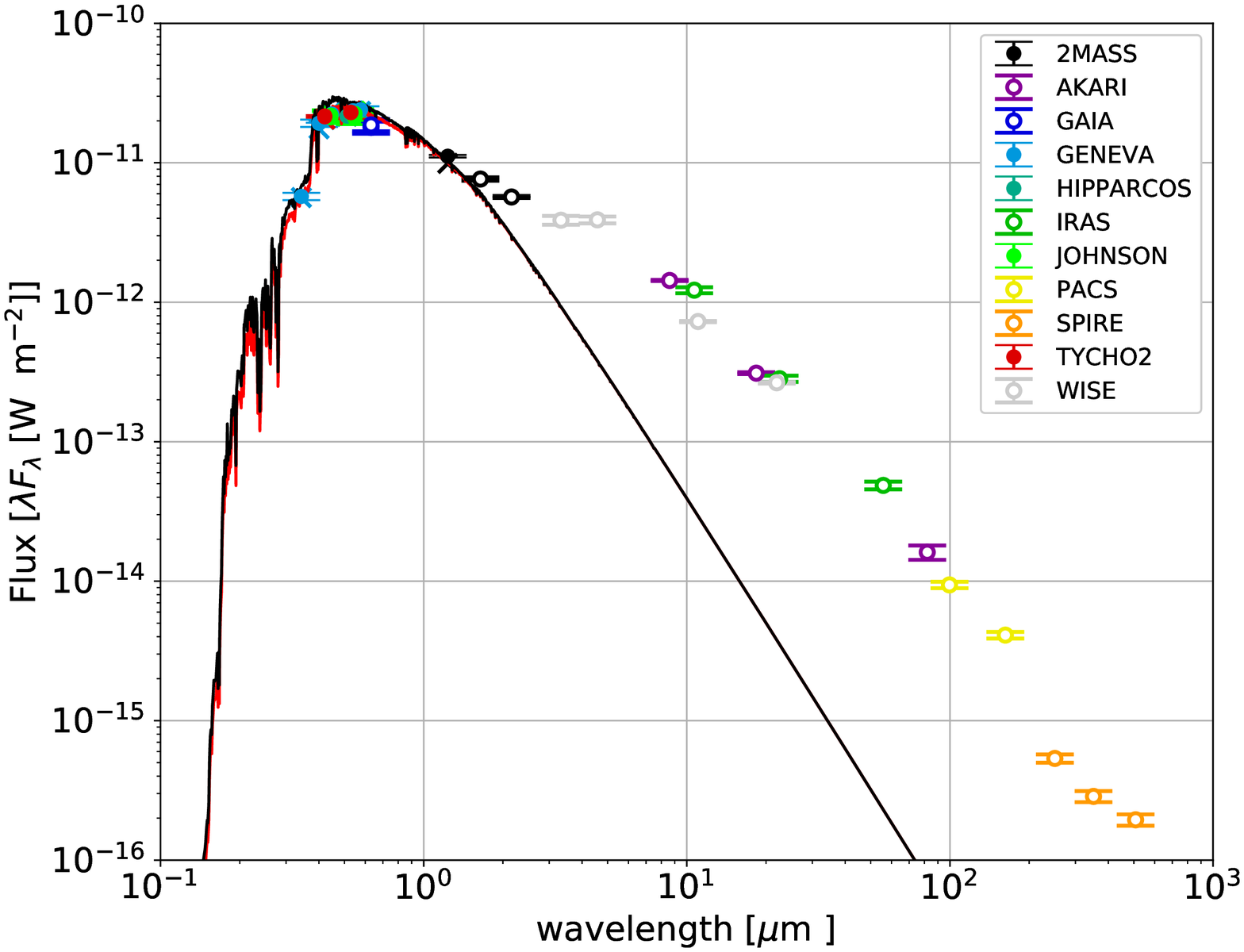}     \label{figure:sed9}
   }\\

   \subfloat[\#10: HD~95767]{%
     \includegraphics[width=0.33\textwidth]{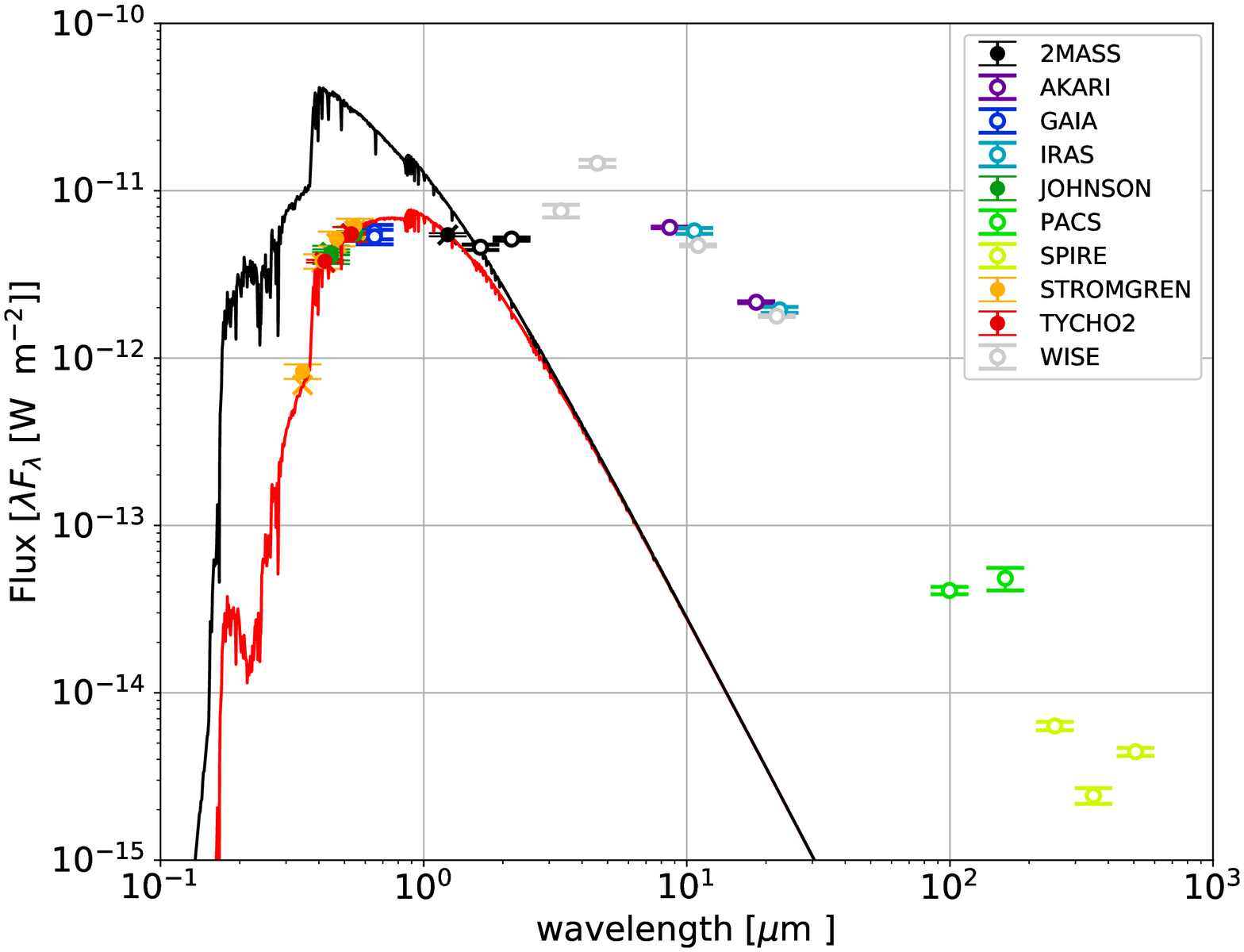}      \label{figure:sed10}
   }
   \subfloat[\#11: HD~108015]{%
     \includegraphics[width=0.33\textwidth]{HD108015_SED.eps}      \label{figure:sed11}
   }
   \subfloat[\#12: HD~131356]{%
     \includegraphics[width=0.33\textwidth]{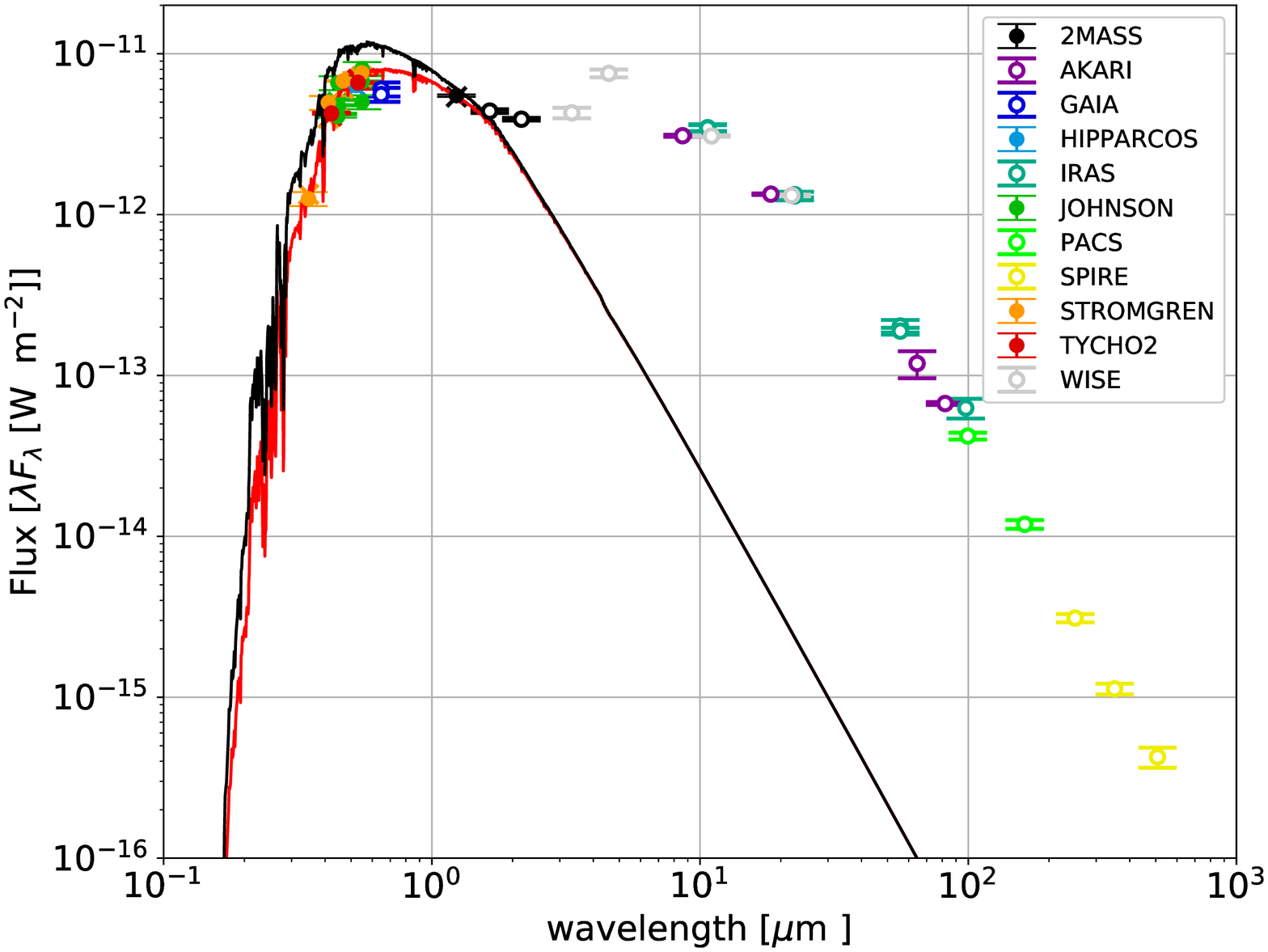}     \label{figure:sed12}
   }\\
   \caption{List of all SEDs of post-AGB stars in our sample. The black curve is the atmospheric model, while the red curve represents the reddened model. The symbols correspond to the observed photometry where different colours denote different surveys.}
   \label{fig:discseds}
 \end{figure*}
 
 \begin{figure*}
 \ContinuedFloat
 \centering
   \subfloat[\#13: HD~158616]{%
     \includegraphics[width=0.33\textwidth]{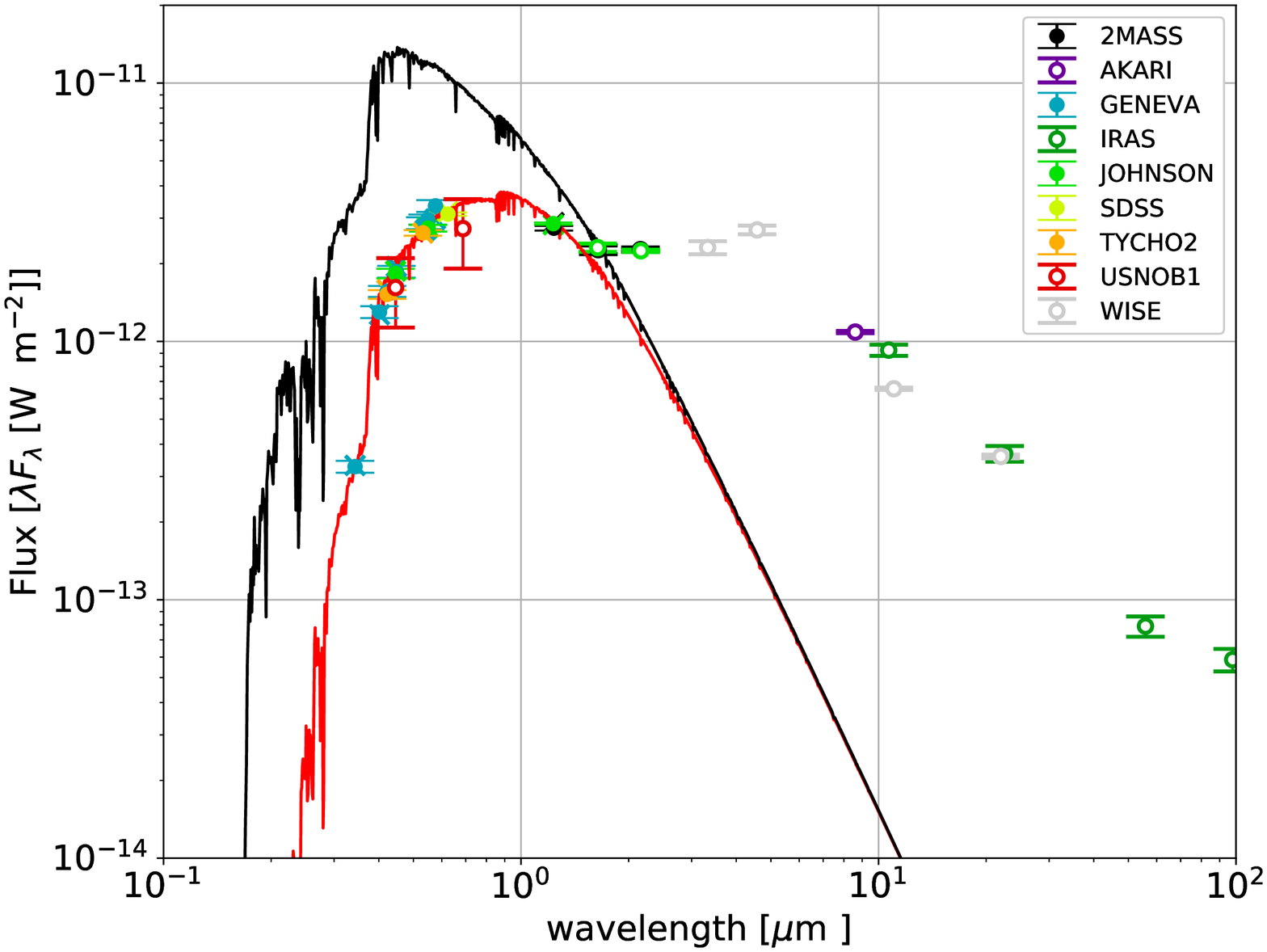}          \label{figure:sed13}
   }
   \subfloat[\#14: HD~213985]{%
     \includegraphics[width=0.33\textwidth]{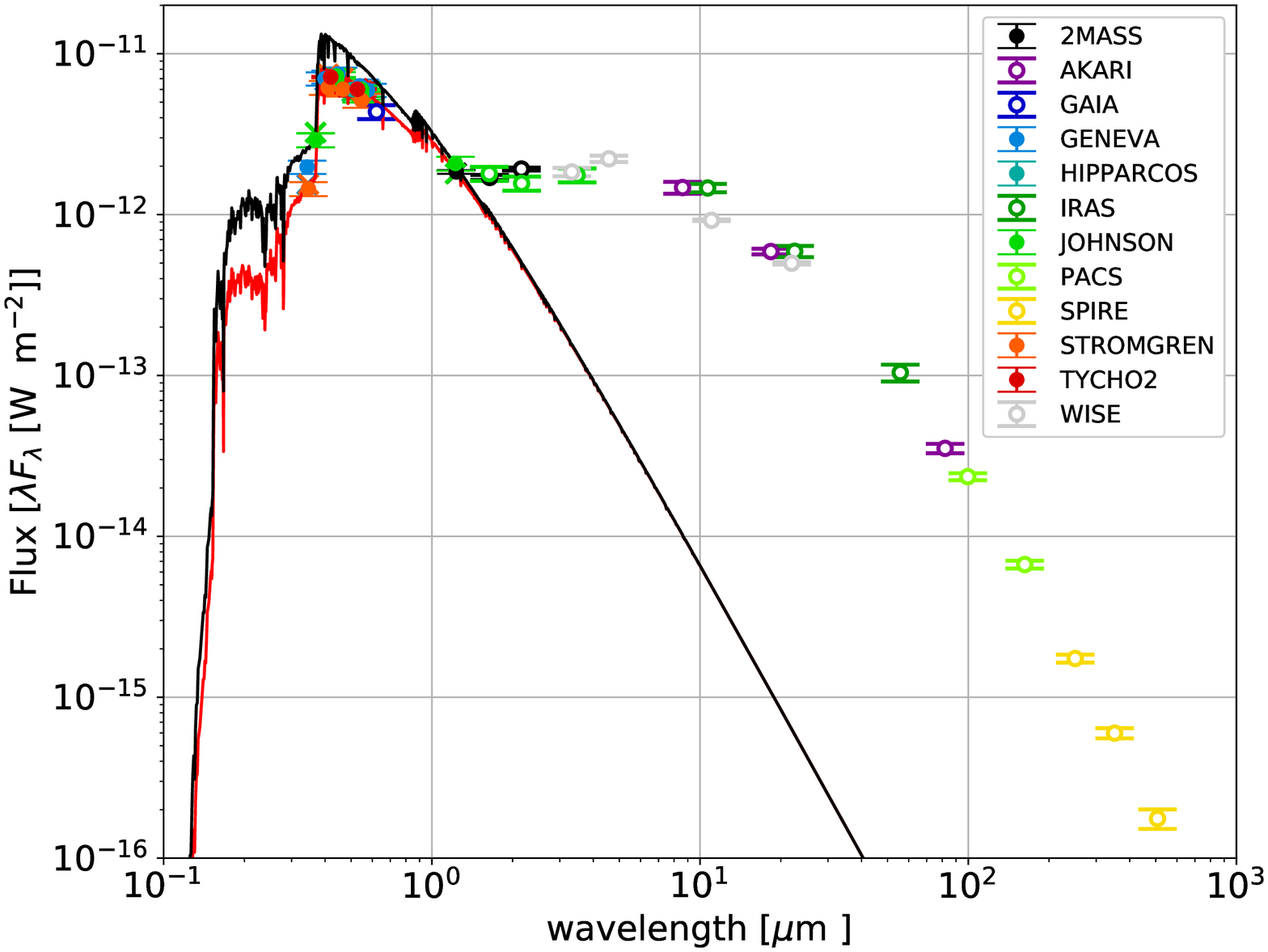}        \label{figure:sed14}
   }
   \subfloat[\#15: HP~Lyr]{%
     \includegraphics[width=0.33\textwidth]{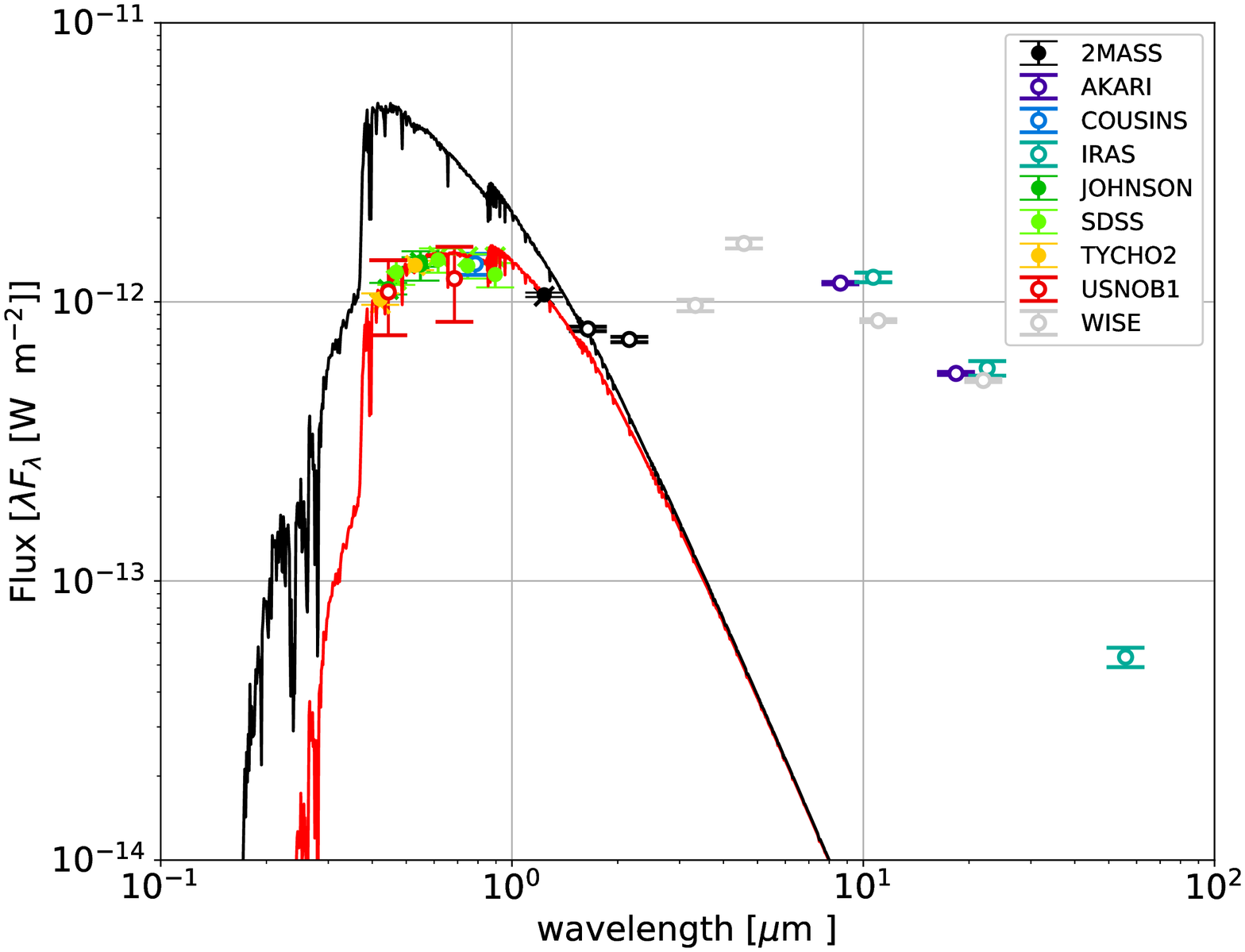}       \label{figure:sed15}
   }\\

   \subfloat[\#16: HR~4049]{%
     \includegraphics[width=0.33\textwidth]{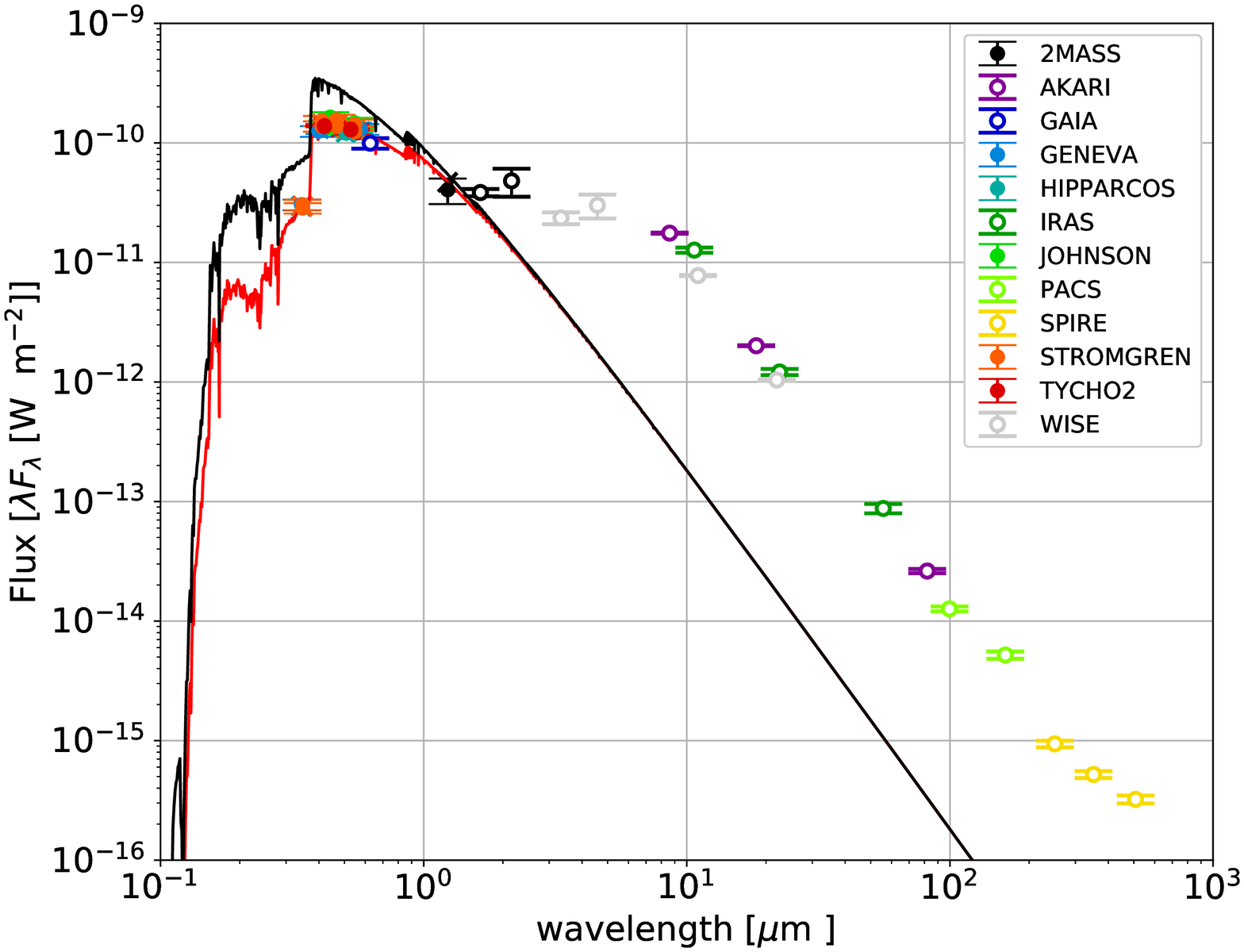}        \label{figure:sed16}
   }
   \subfloat[\#17: IRAS~05208-2035]{%
     \includegraphics[width=0.33\textwidth]{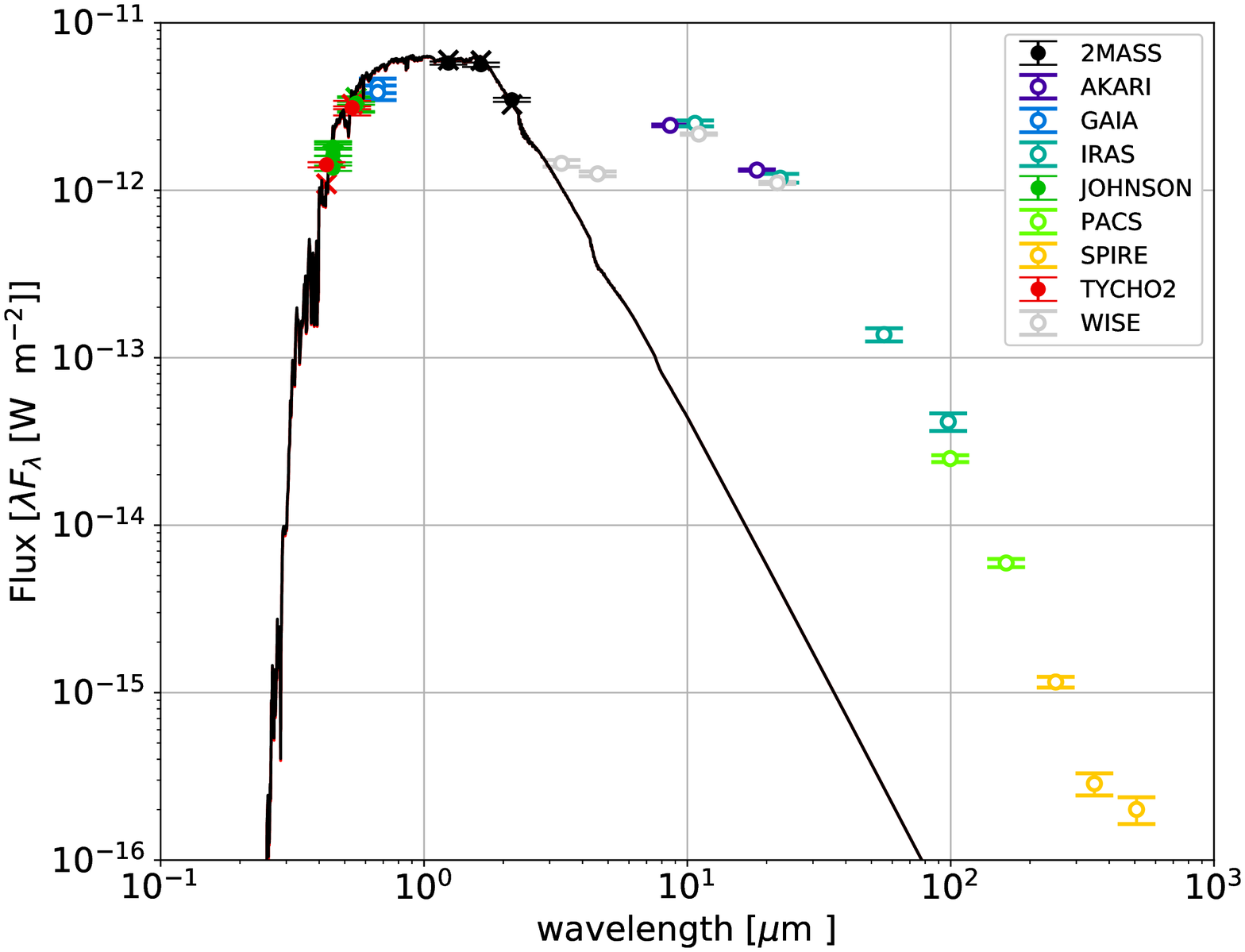}       \label{figure:sed17}
   }
   \subfloat[\#18: IRAS~06165+3158]{%
     \includegraphics[width=0.33\textwidth]{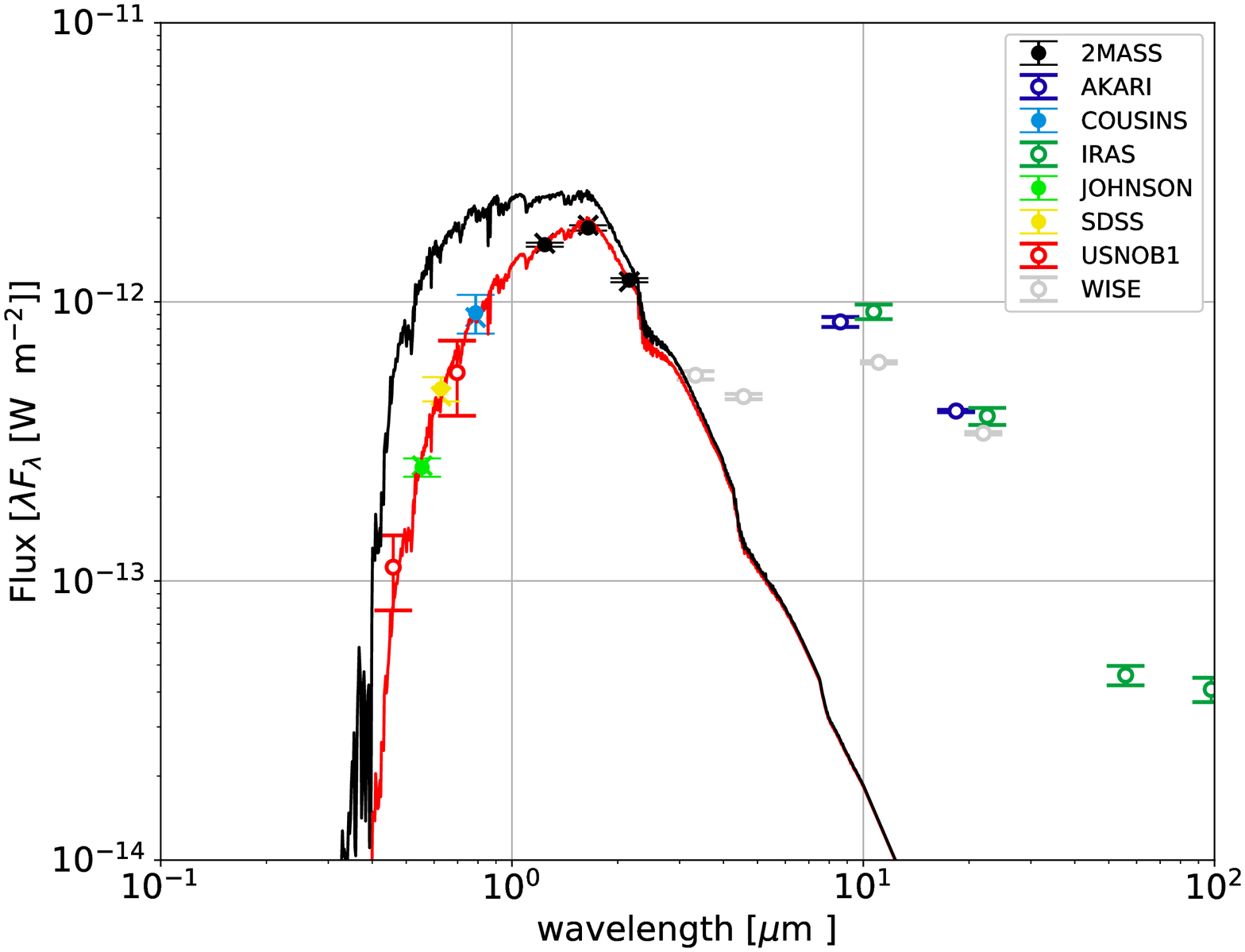}        \label{figure:sed18}
   }\\
   
   \subfloat[\#19: IRAS~06452-3456]{%
     \includegraphics[width=0.33\textwidth]{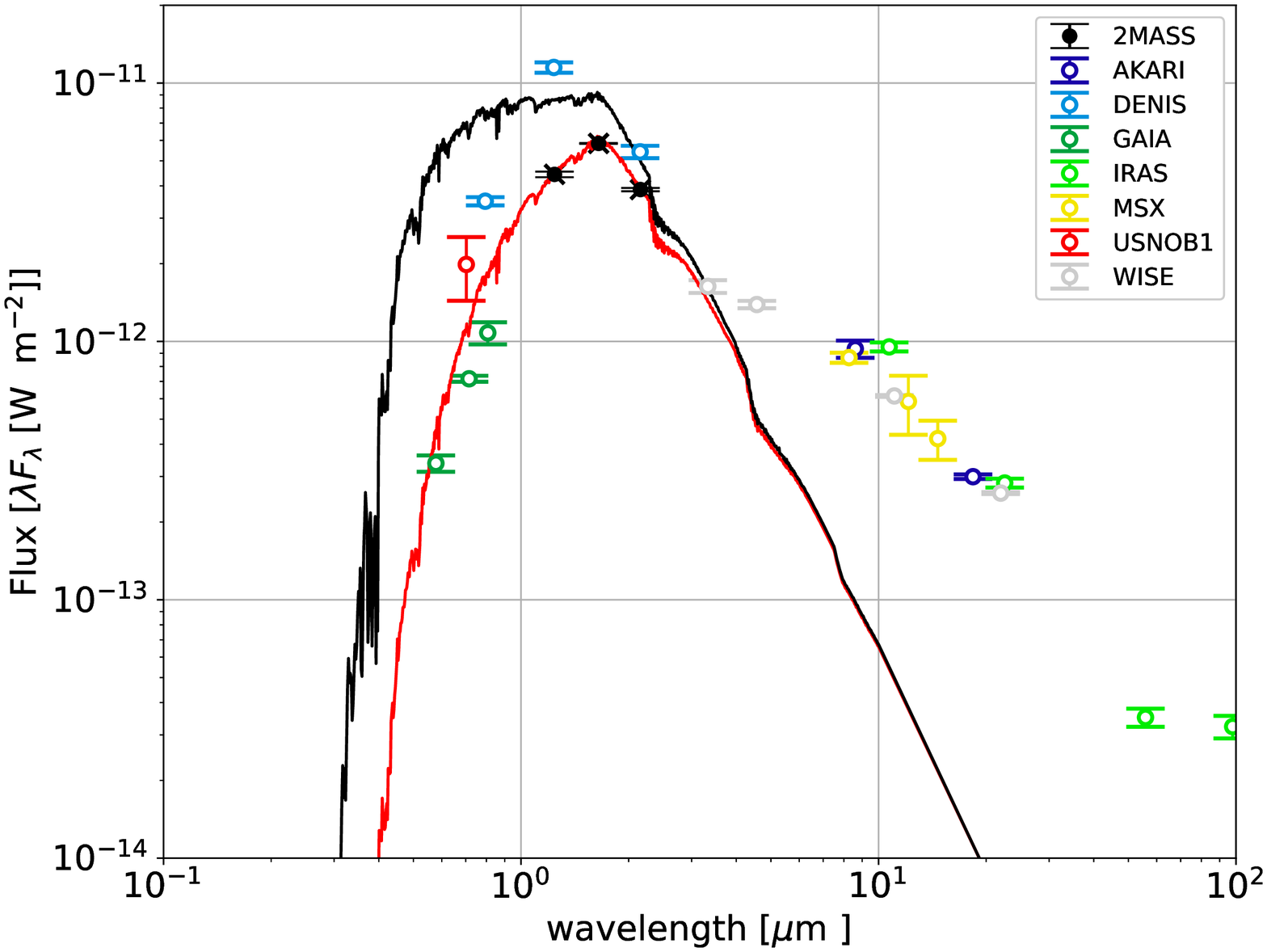}      \label{figure:sed19}
   }
   \subfloat[\#20: IRAS~08544-4431]{%
     \includegraphics[width=0.33\textwidth]{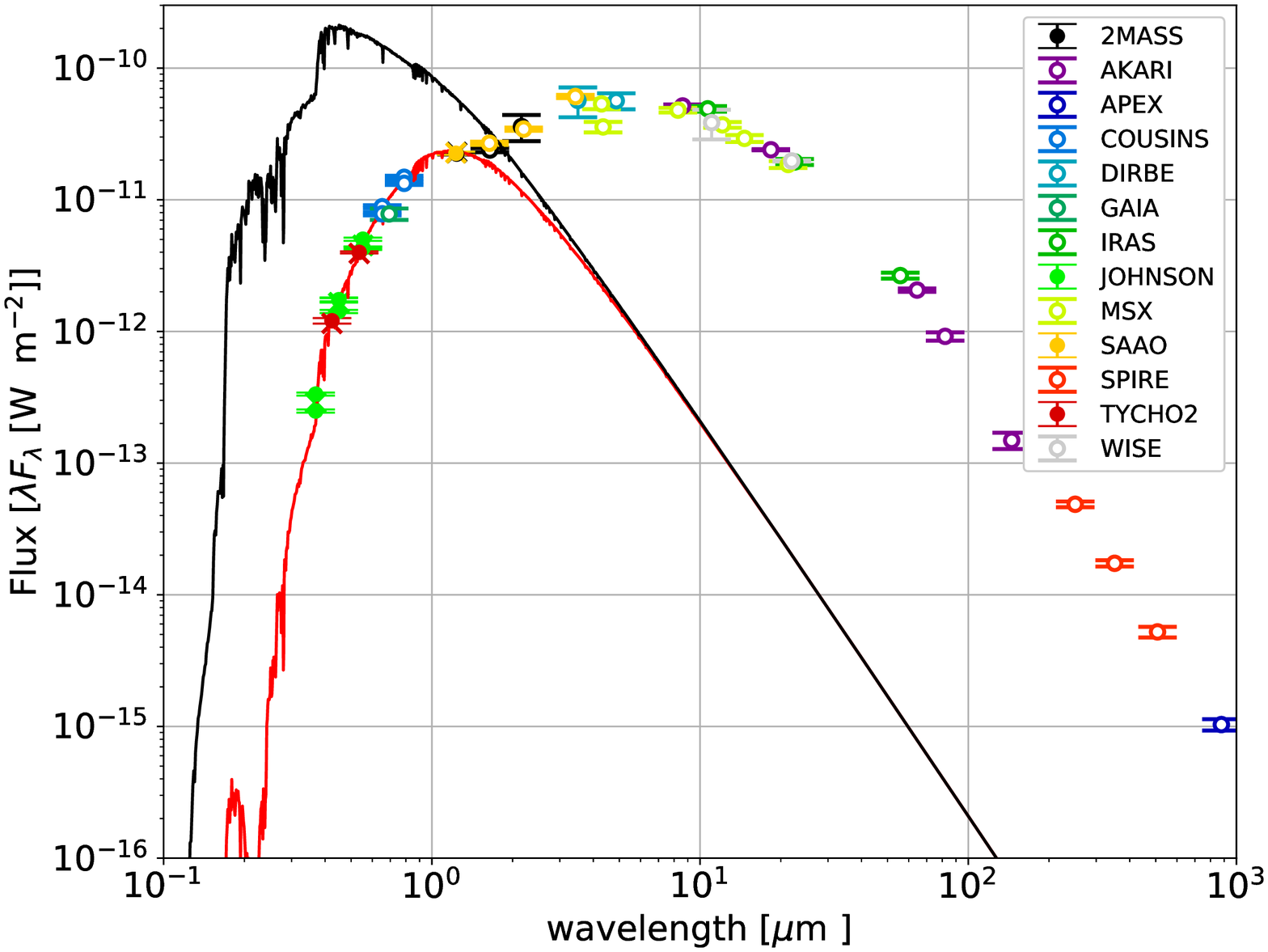}      \label{figure:sed20}
   }
   \subfloat[\#21: IRAS~09144-4933]{%
     \includegraphics[width=0.33\textwidth]{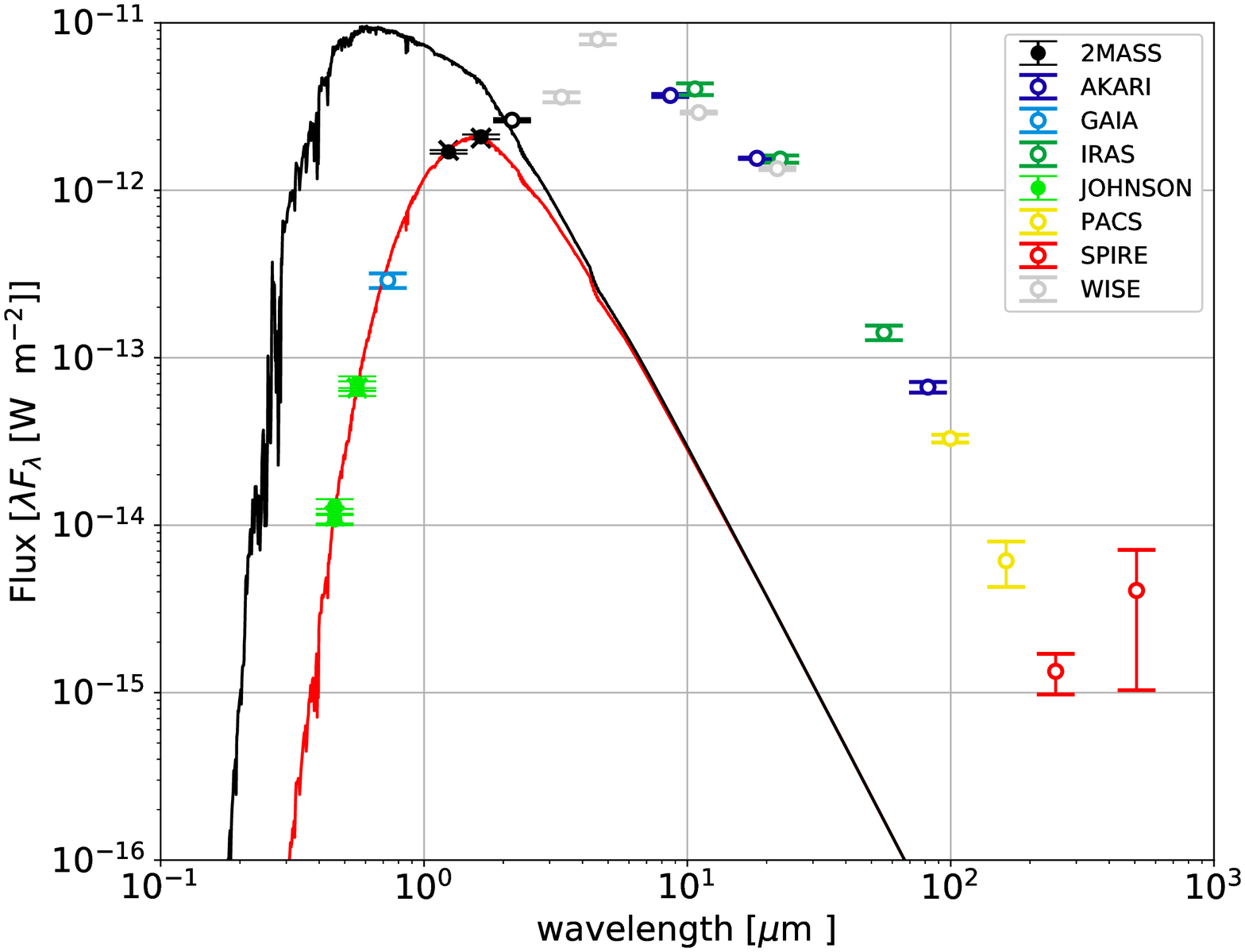}      \label{figure:sed21}
   }\\
   
   \subfloat[\#22: IRAS~15469-5311]{%
     \includegraphics[width=0.33\textwidth]{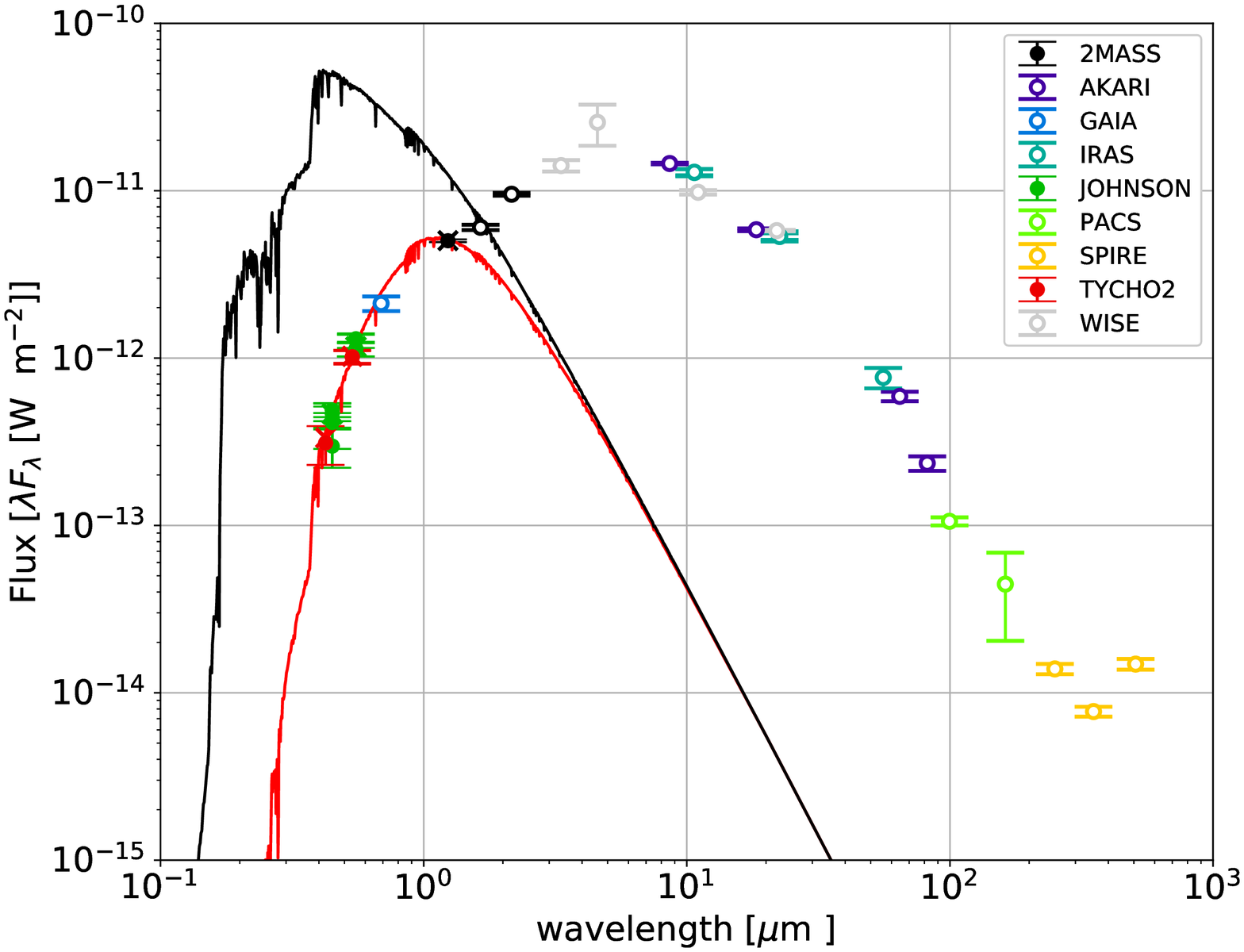}      \label{figure:sed22}
   }
   \subfloat[\#23: IRAS~16230-3410]{%
     \includegraphics[width=0.33\textwidth]{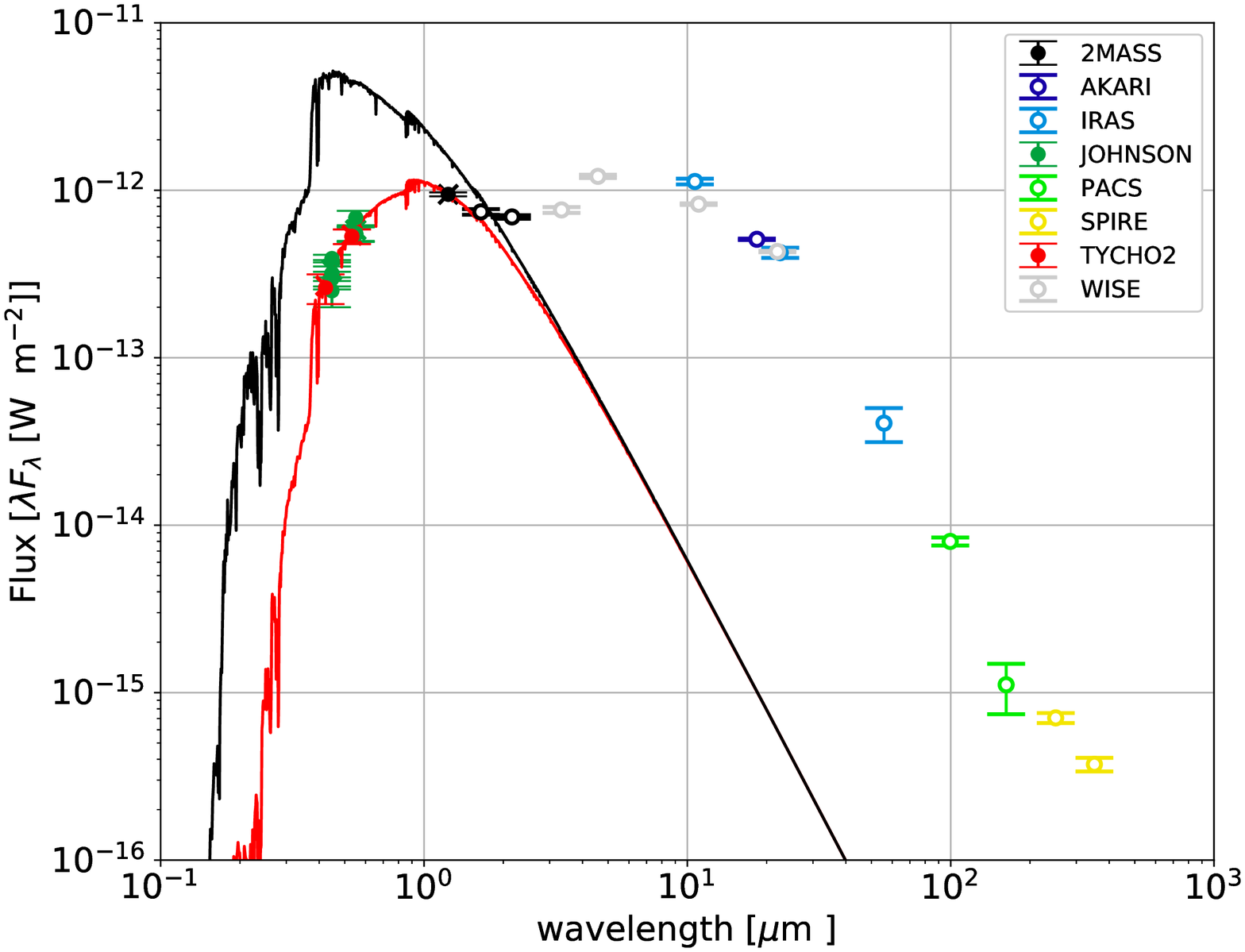}     \label{figure:sed23}
   }
   \subfloat[\#24: IRAS~17038-4815]{%
     \includegraphics[width=0.33\textwidth]{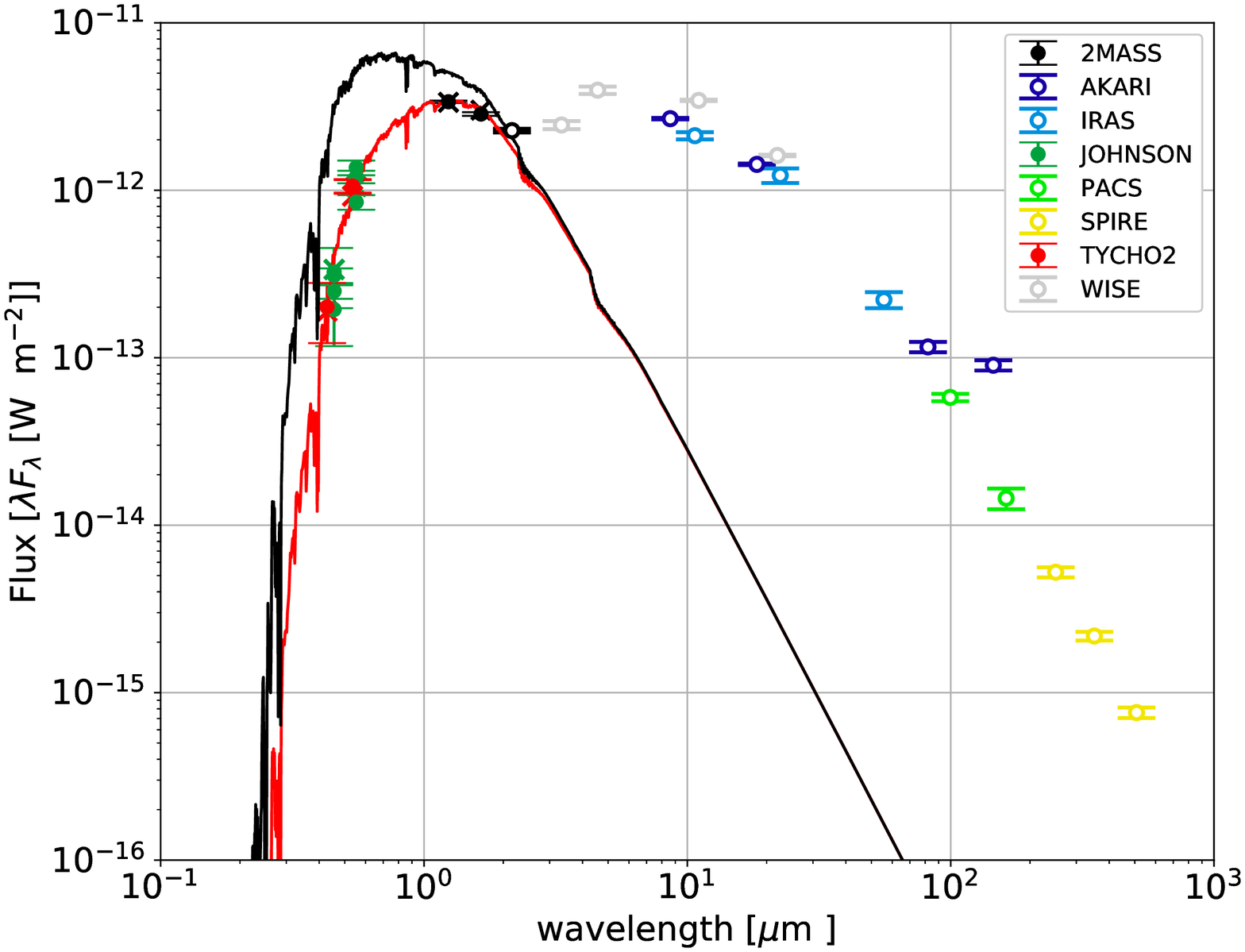}      \label{figure:sed24}
   }\\
   \caption{\textit{continued}}
   \label{fig:sedsappendix}
 \end{figure*}
 
 \begin{figure*}
 \ContinuedFloat
 \centering
   \subfloat[\#25: IRAS~19125+0343]{%
     \includegraphics[width=0.33\textwidth]{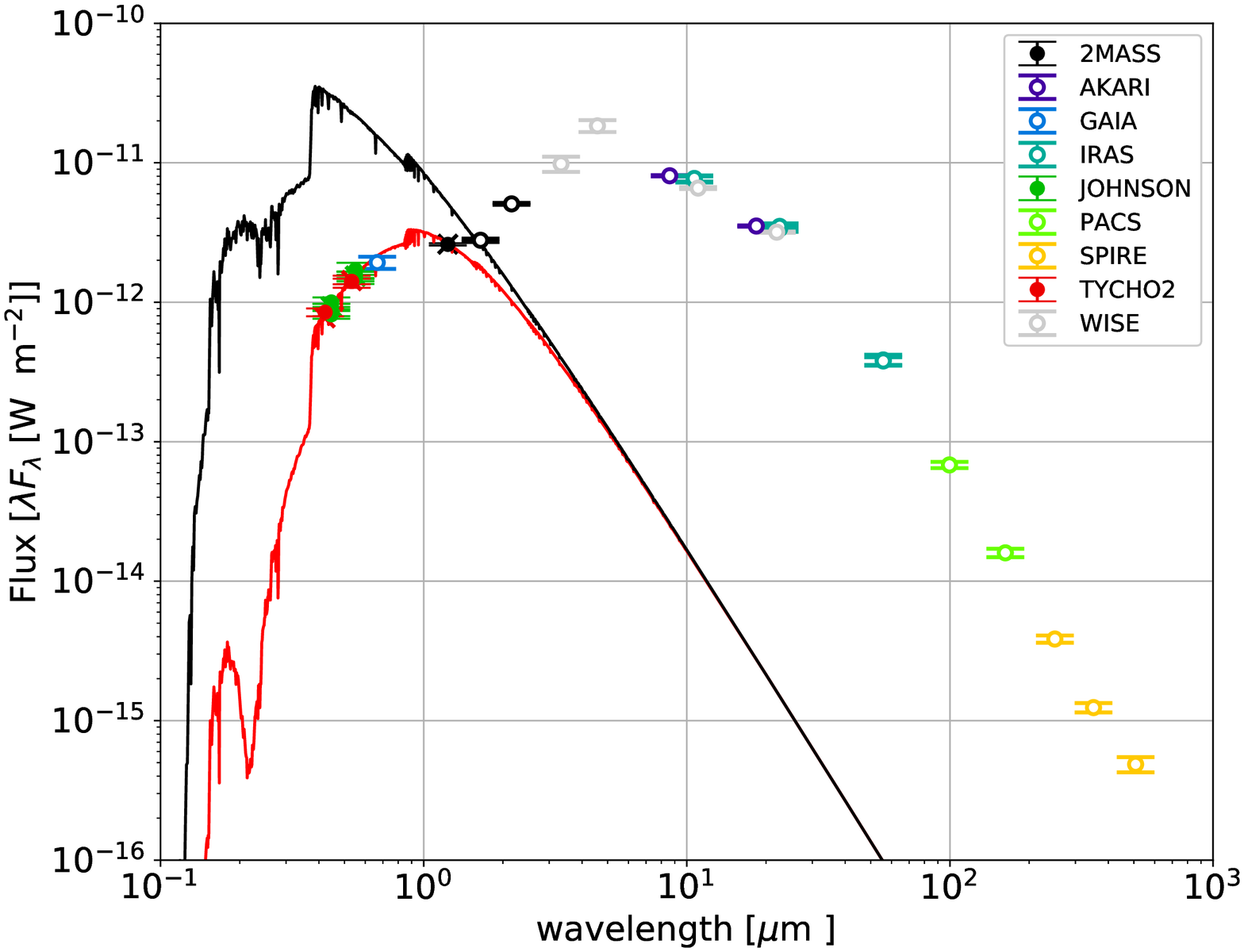}      \label{figure:sed25}
   }
   \subfloat[\#26: IRAS~19135+3937]{%
     \includegraphics[width=0.33\textwidth]{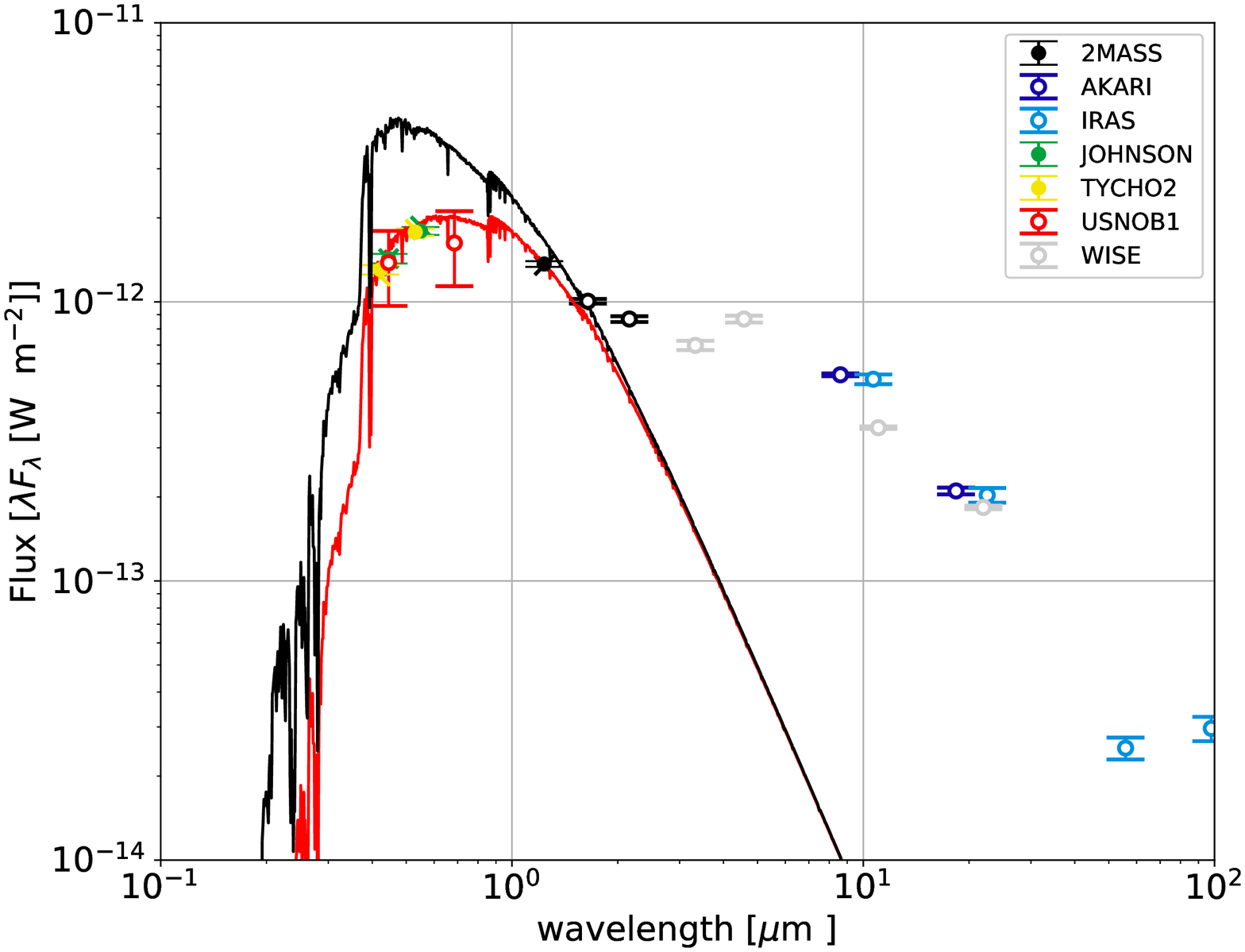}     \label{figure:sed26}
   }
   \subfloat[\#27: IRAS~19157-0247]{%
     \includegraphics[width=0.33\textwidth]{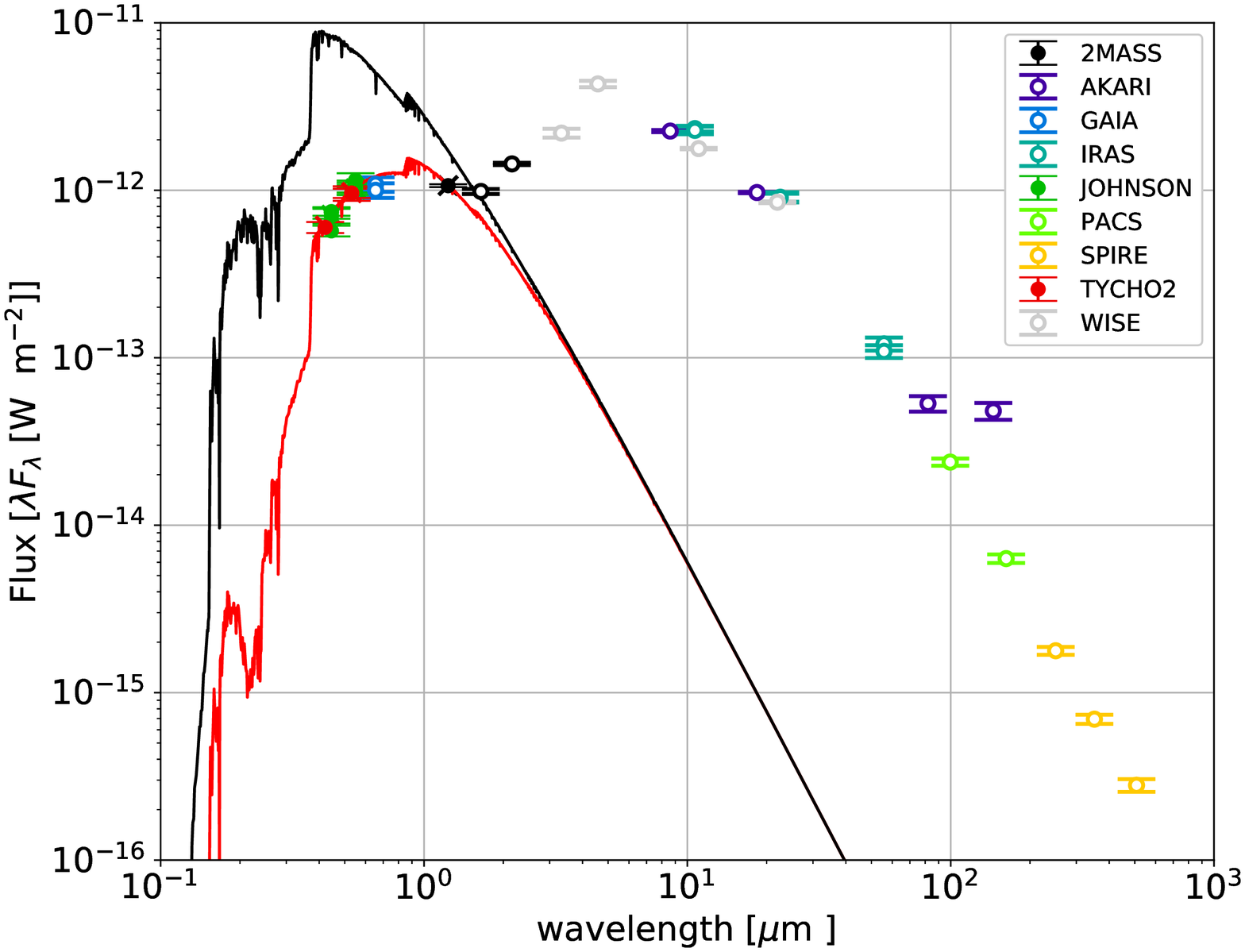}          \label{figure:sed27}
   }\\
   
   \subfloat[\#28: RU~Cen]{%
     \includegraphics[width=0.33\textwidth]{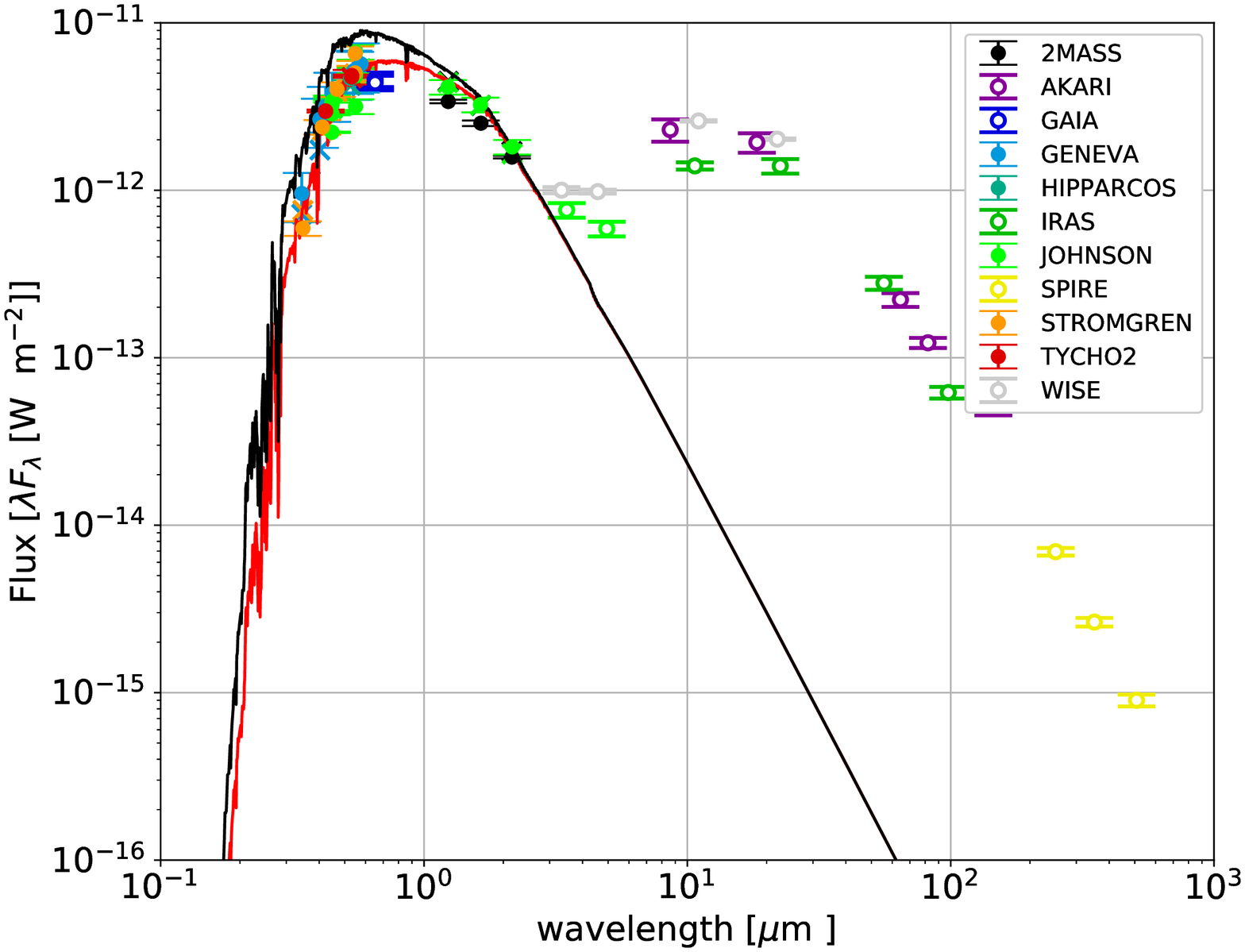}        \label{figure:sed28}
   }
   \subfloat[\#29: SAO~173329]{%
     \includegraphics[width=0.33\textwidth]{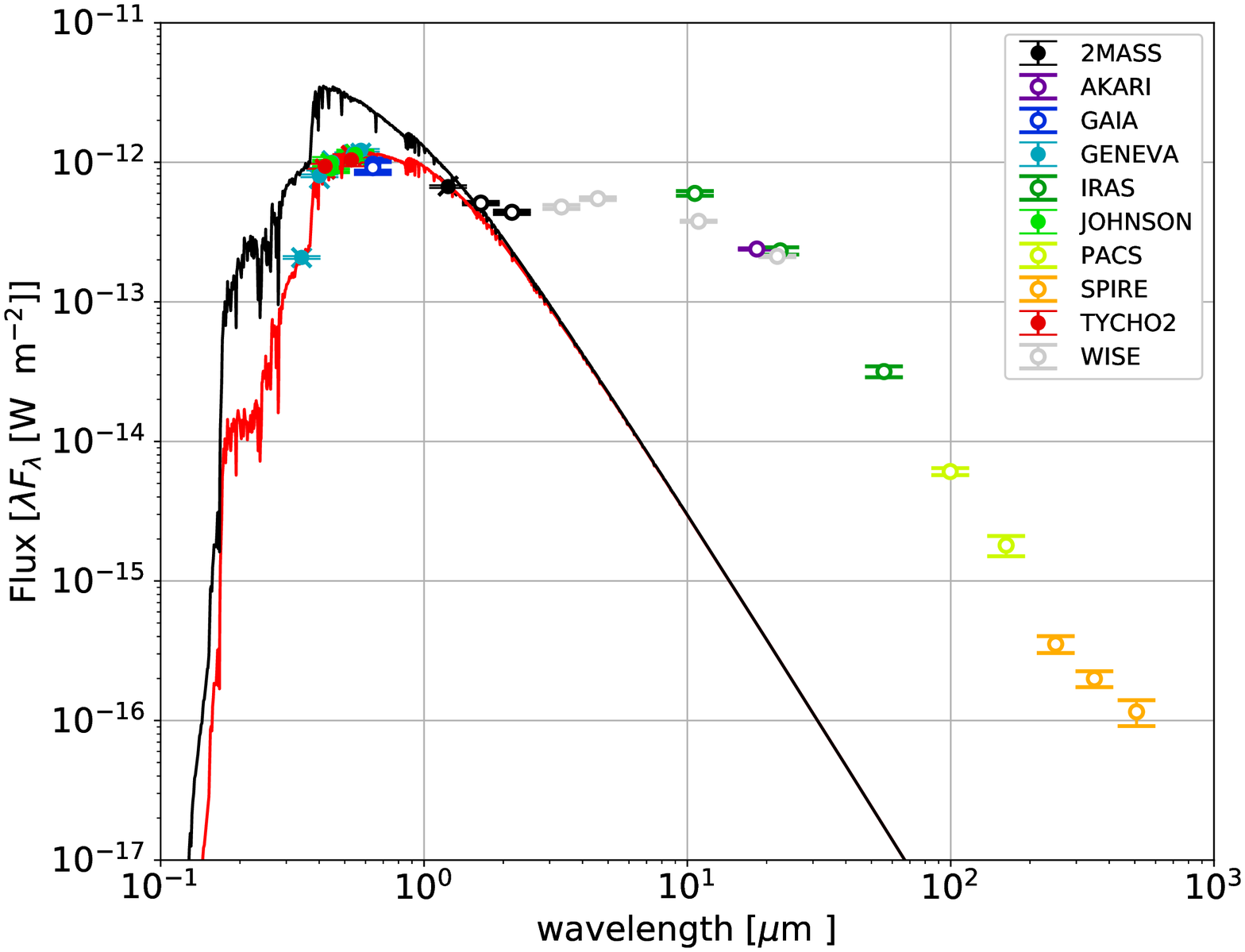}       \label{figure:sed29}
   }
   \subfloat[\#30: ST~Pup]{%
     \includegraphics[width=0.33\textwidth]{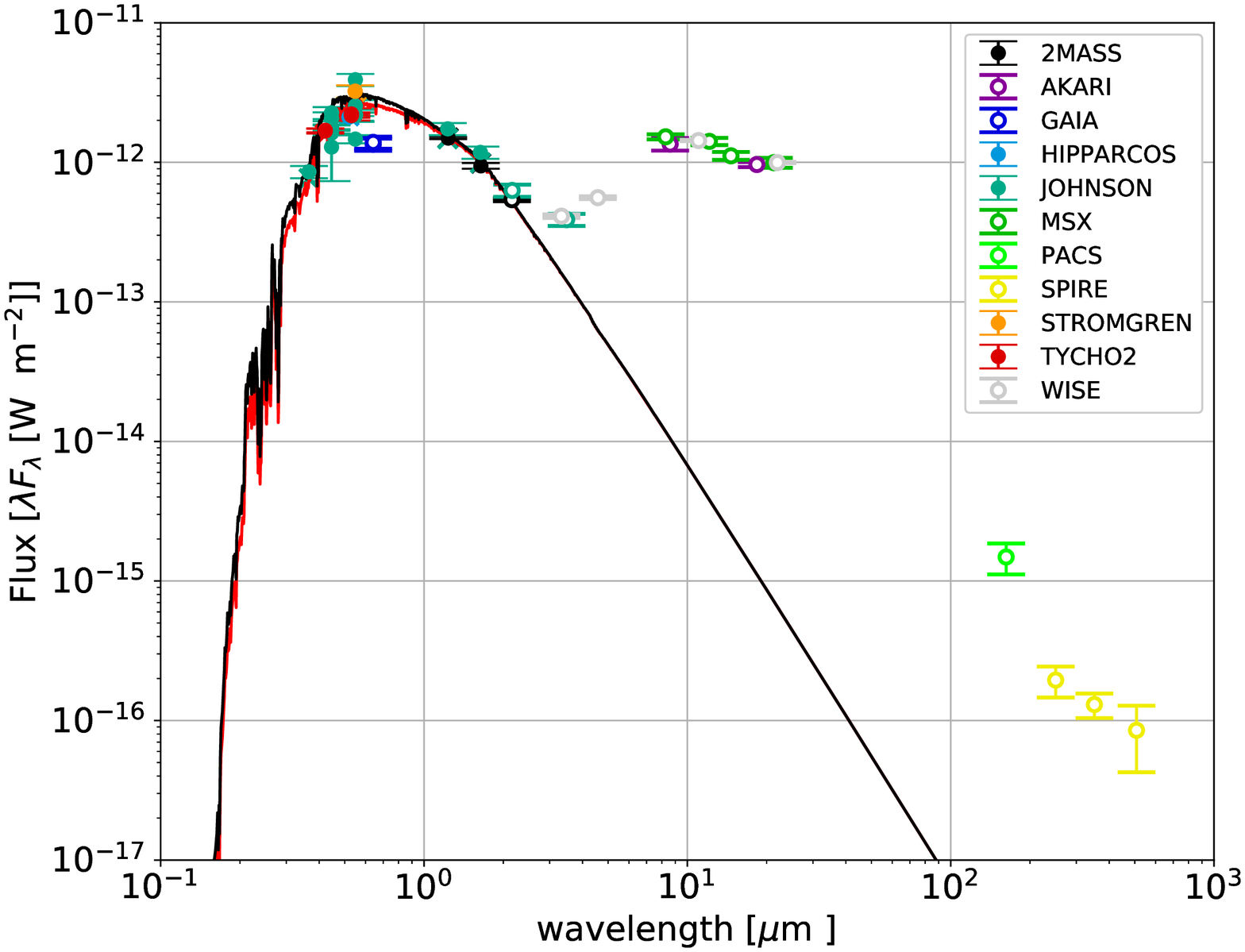}        \label{figure:sed30}
   }\\
   
   \subfloat[\#31: SX~Cen]{%
     \includegraphics[width=0.33\textwidth]{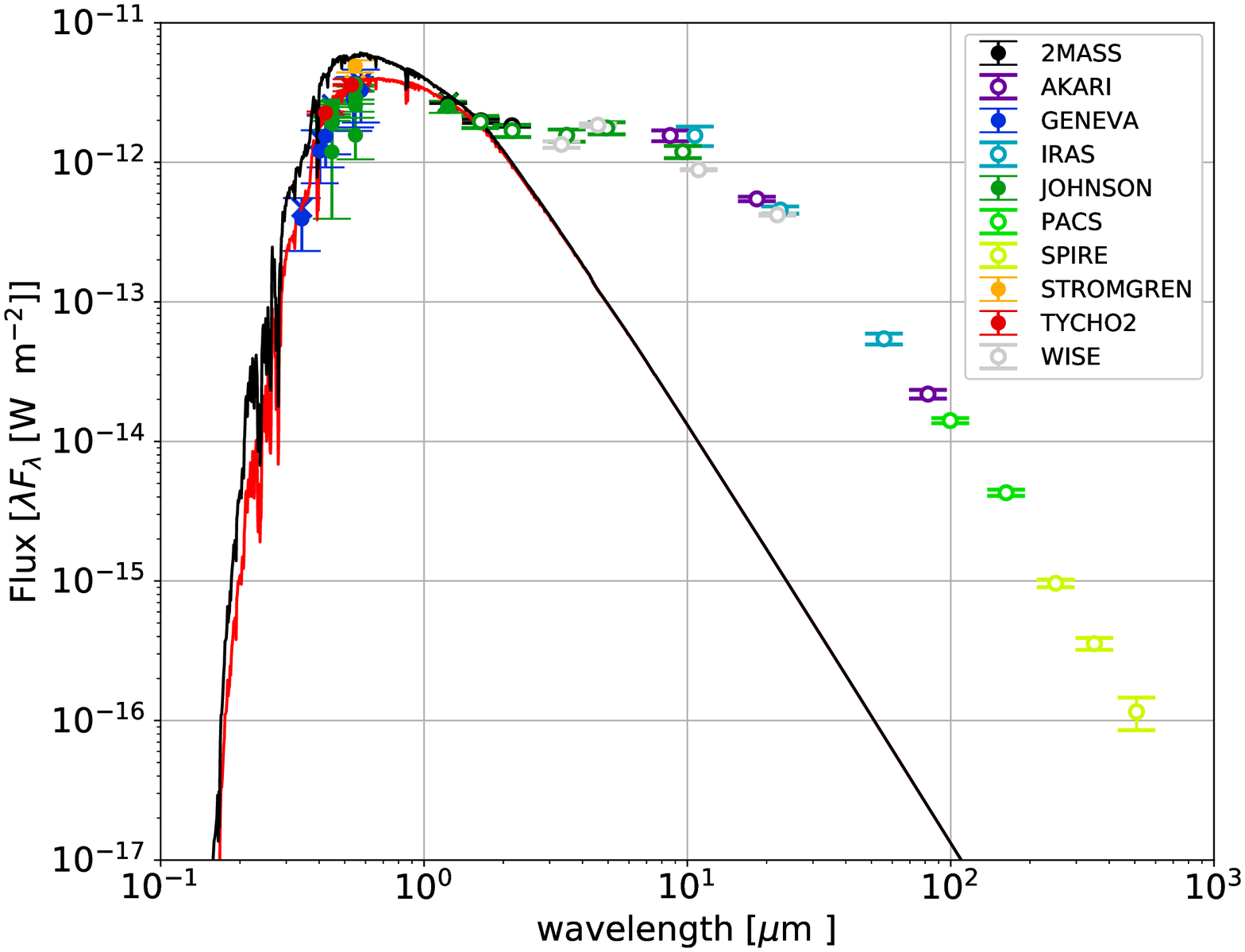}       \label{figure:sed31}
   }
   \subfloat[\#32: TW~Cam]{%
     \includegraphics[width=0.33\textwidth]{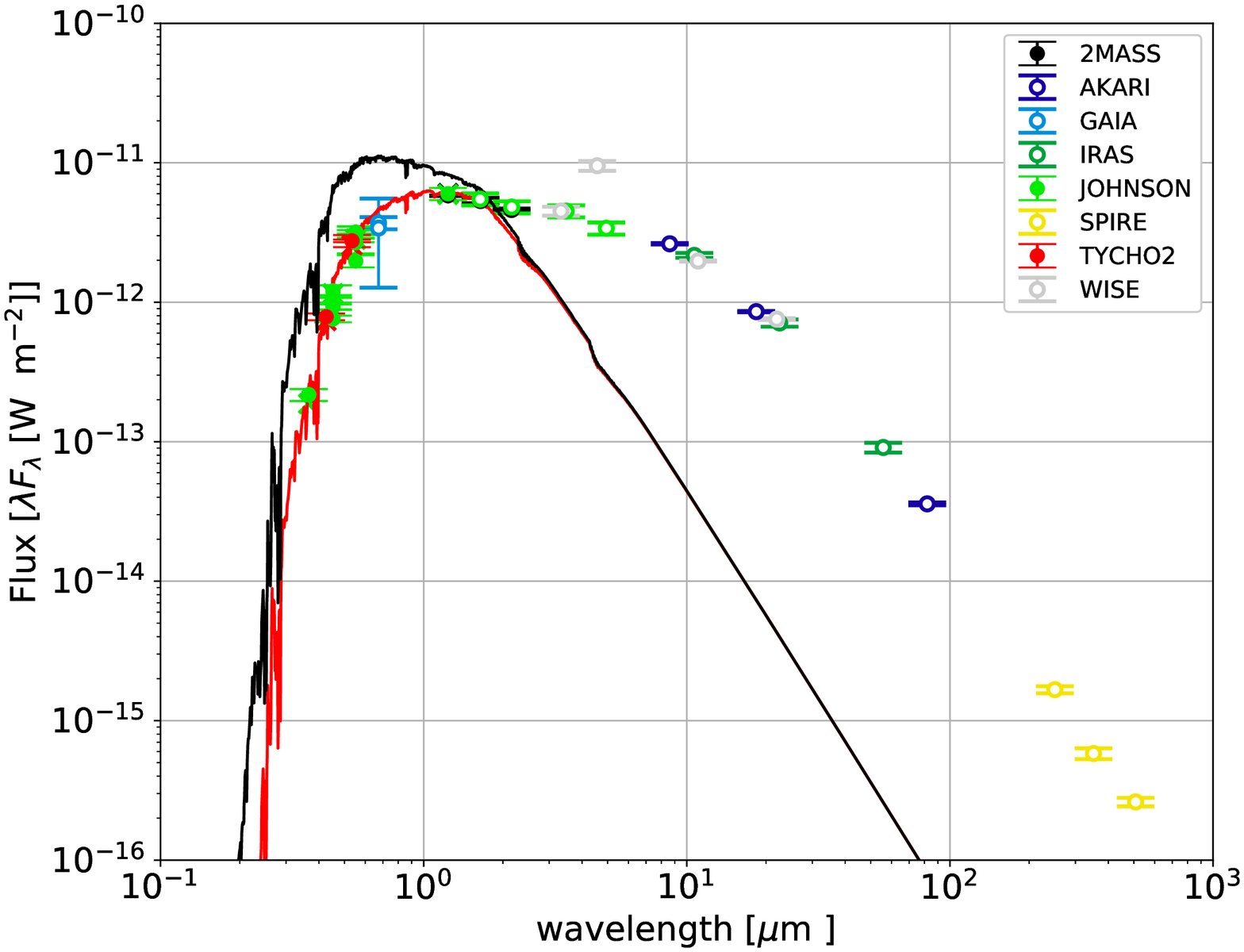}        \label{figure:sed32}
   }
   \subfloat[\#33: U~Mon]{%
     \includegraphics[width=0.33\textwidth]{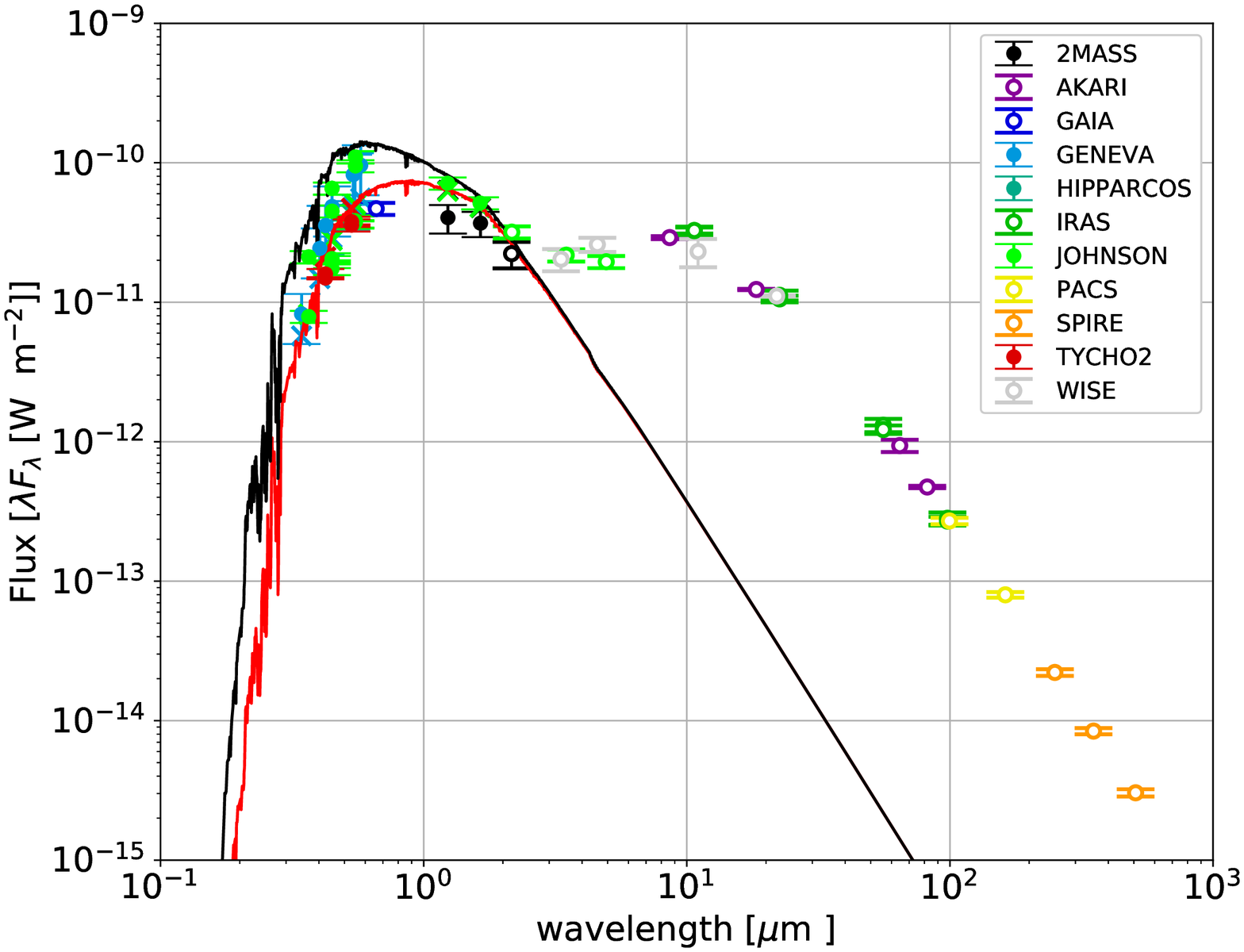}      \label{figure:sed33}
   }\\
   \caption{\textit{continued}}
   %\label{fig:discsedsLMC}
 \end{figure*}

\section{Radial velocity curves} \label{appendix:orbits}
 \captionsetup[subfigure]{labelformat=empty}
This appendix contains the RV curves for all post-AGB binaries in the sample. The fitted curve has been plotted in each of the figures, with their corresponding orbital elements given in Table~\ref{tableorbitalelements}. The error bars are plotted as well, but they are in many cases smaller than the symbol size of the data points.

The radial velocity measurements used to construct and fit the radial velocity curves of all binary post-AGB stars in our sample are available at CDS. These tables contain the following information. Column~1 shows the date at which the radial velocity measurement was taken, Column~2 gives the radial velocity in km/s, and the uncertainty on the measurement is given in Column~3.

 \begin{figure*}
 \centering
   \subfloat[\#1: 89~Her]{%
     \includegraphics[width=0.33\textwidth]{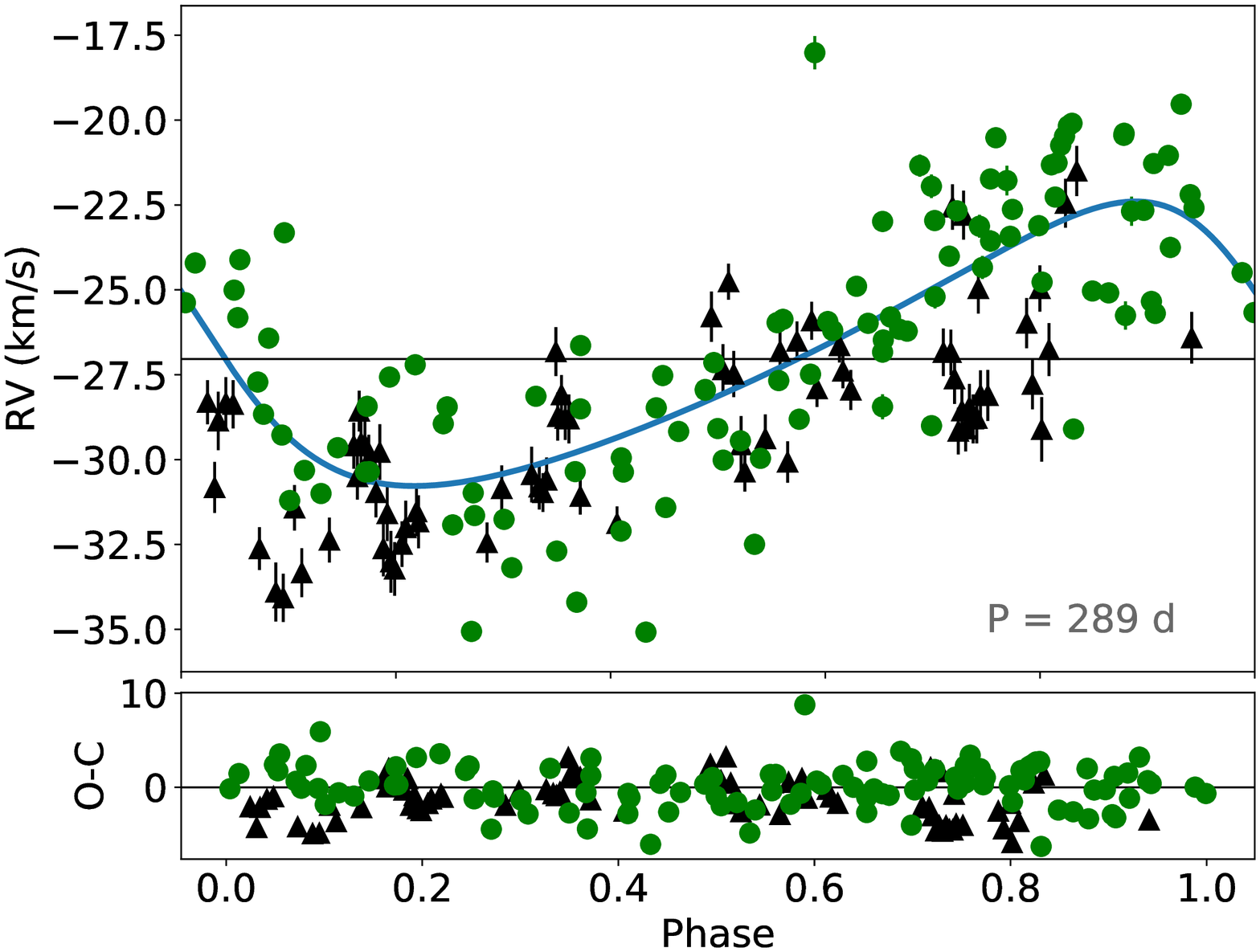}          \label{figure:89her}
   }
   \subfloat[\#2: AC~Her]{%
     \includegraphics[width=0.33\textwidth]{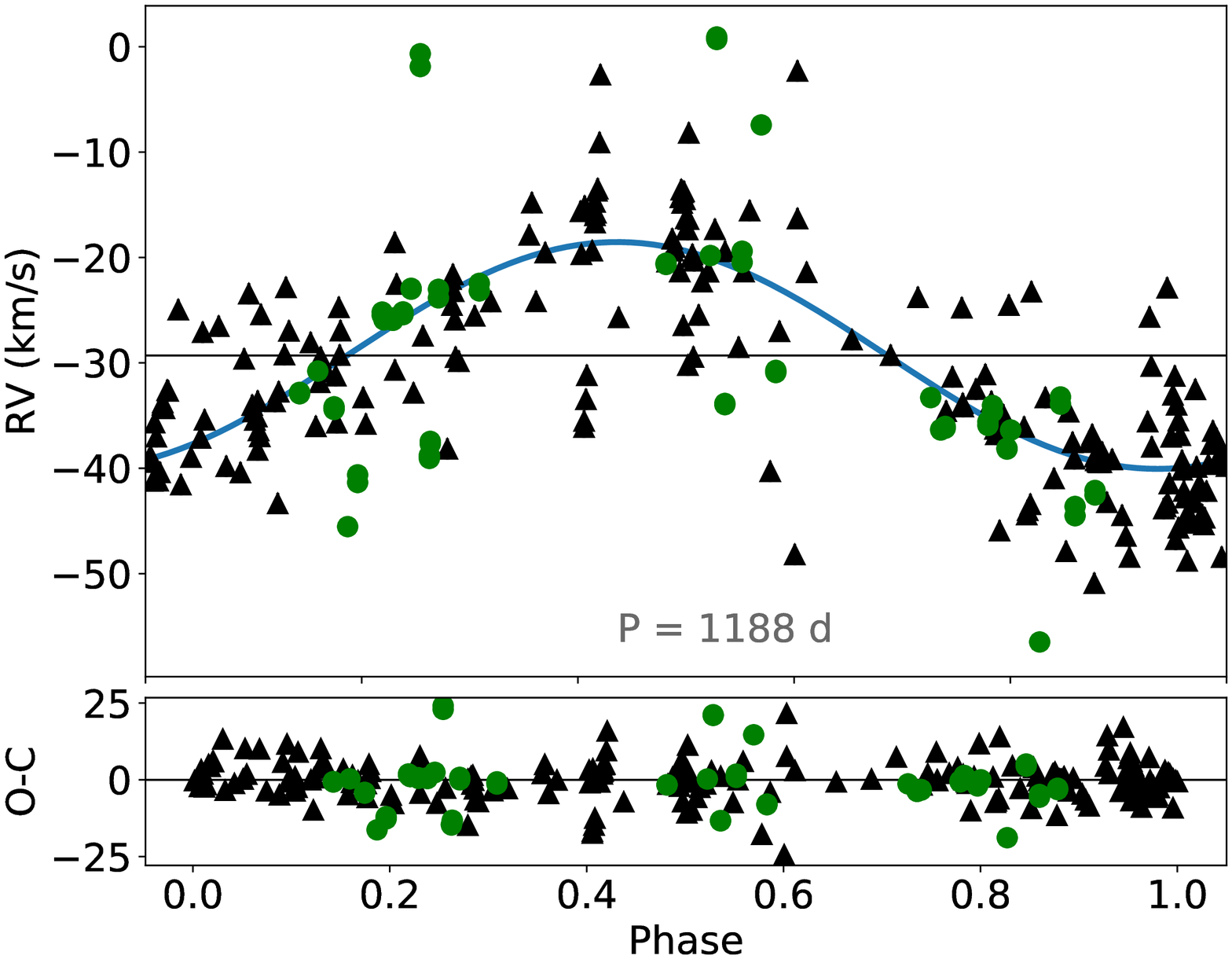}          \label{figure:acher}
   }
   \subfloat[\#3: BD+39~4926]{%
     \includegraphics[width=0.33\textwidth]{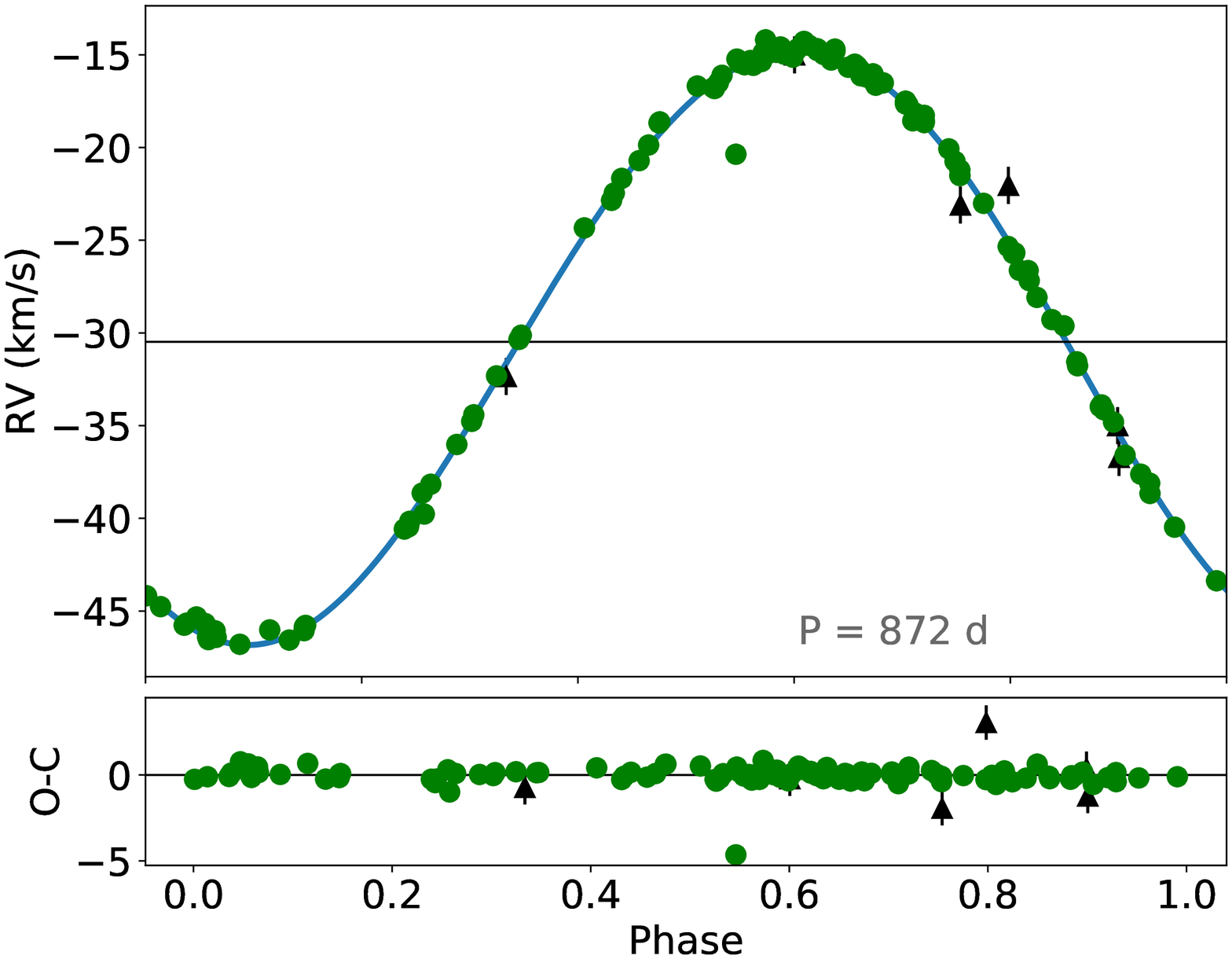}        \label{figure:bd+394926}
   }\\
   
   \subfloat[\#4: BD+46~442]{%
     \includegraphics[width=0.33\textwidth]{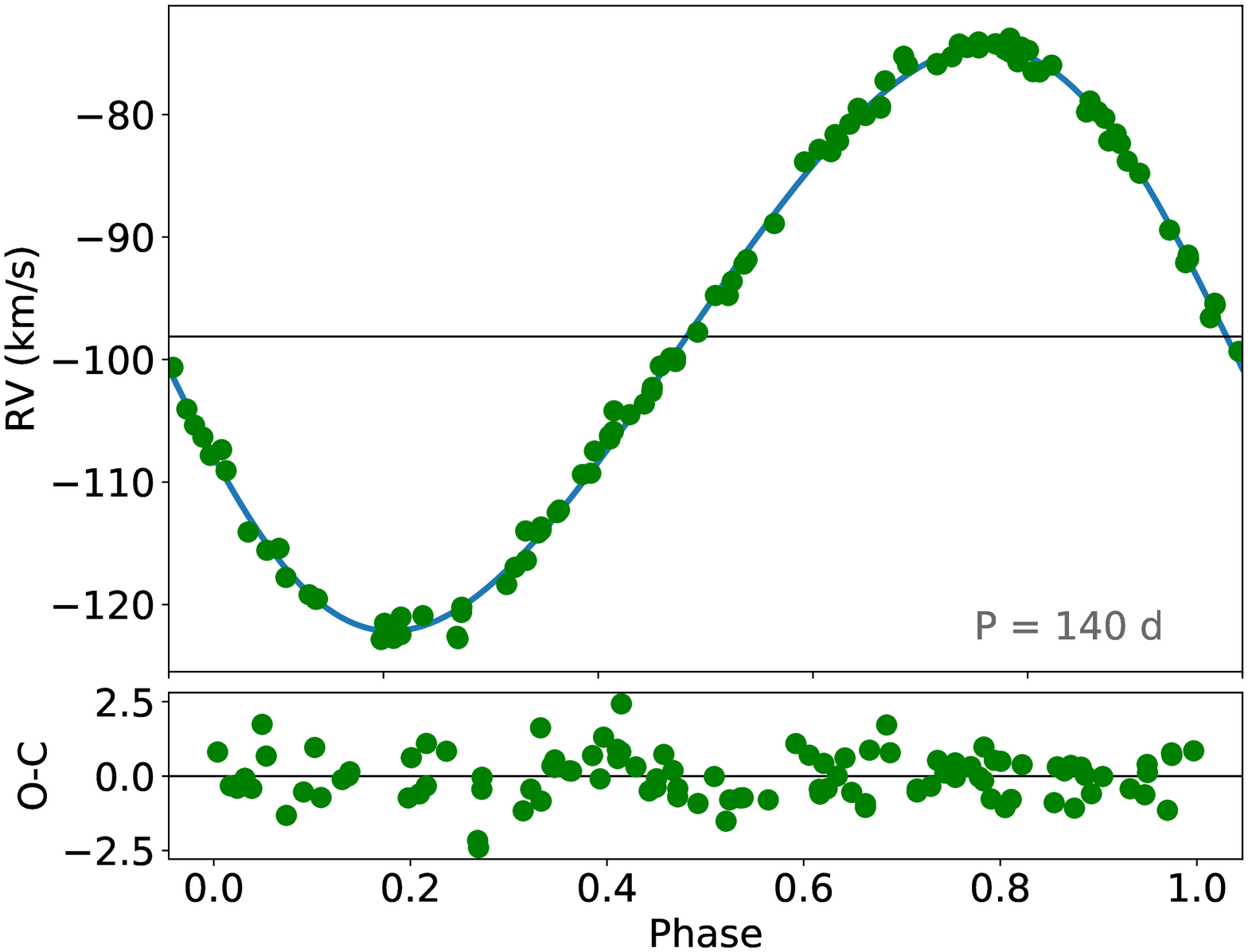}       \label{figure:bd+46442}
   }
   \subfloat[\#5: DY~Ori]{%
     \includegraphics[width=0.33\textwidth]{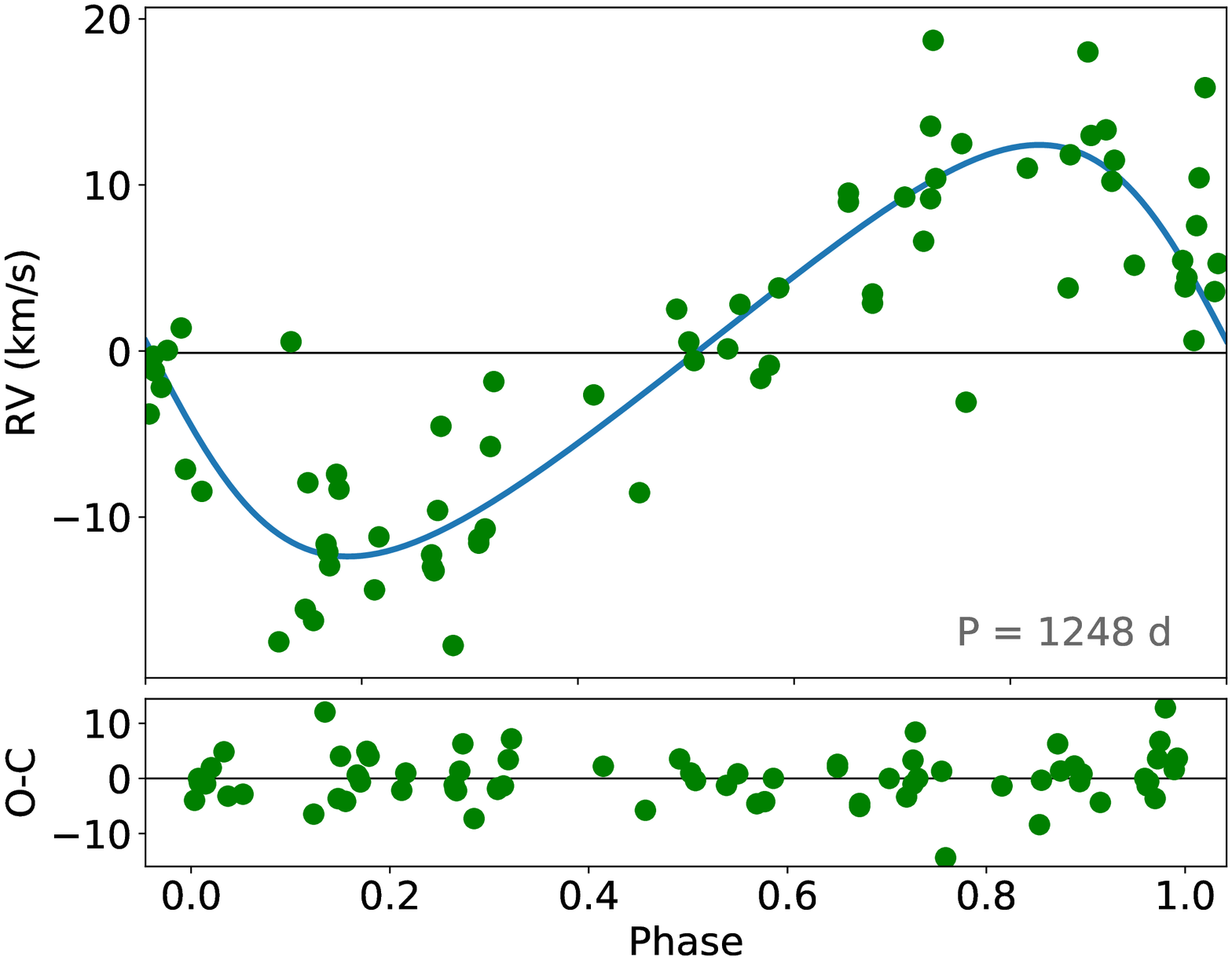}        \label{figure:dyori}
   }
   \subfloat[\#6: EP~Lyr]{%
     \includegraphics[width=0.33\textwidth]{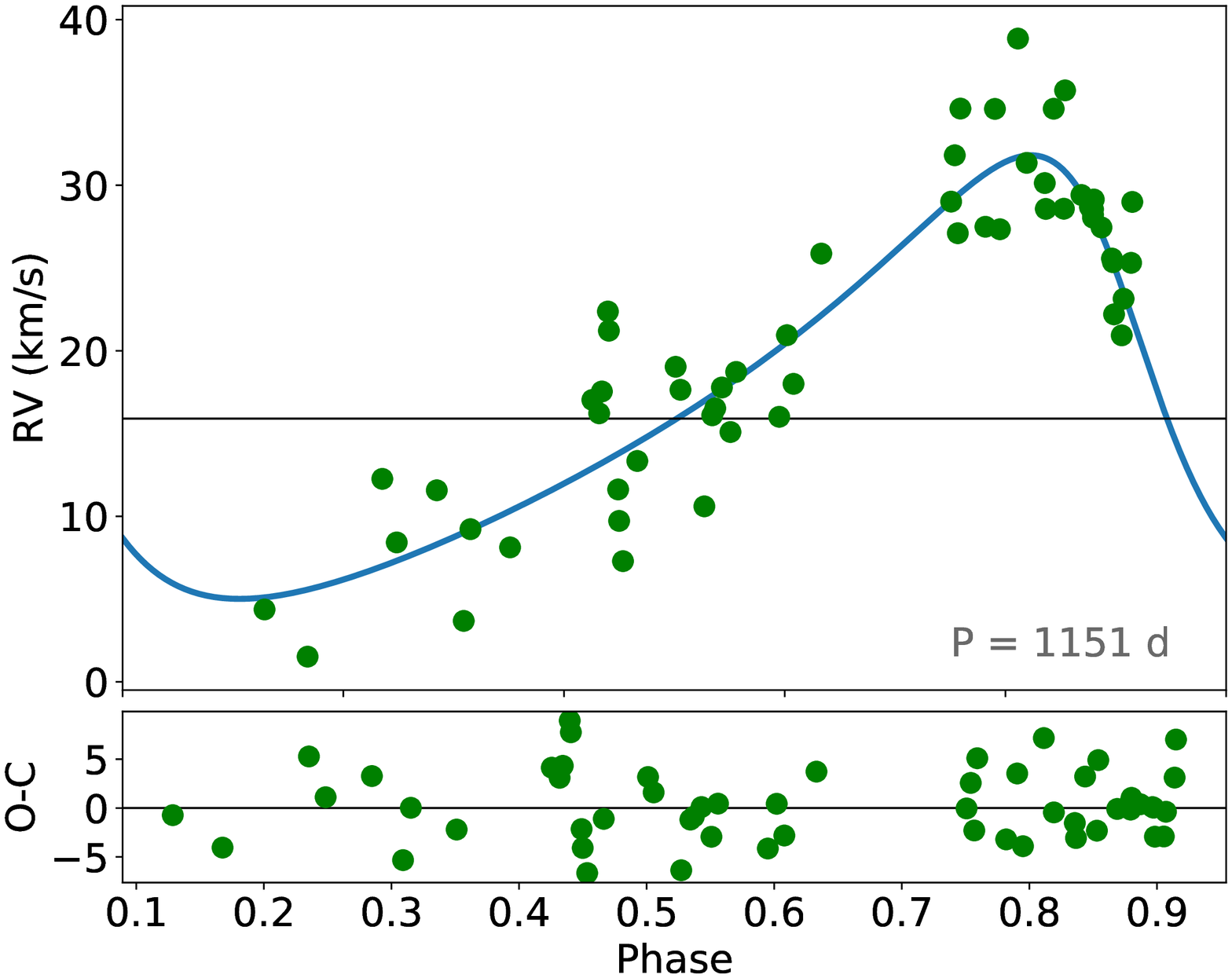}       \label{figure:eplyr}
   }\\
   
   \subfloat[\#7: HD~44179]{%
     \includegraphics[width=0.33\textwidth]{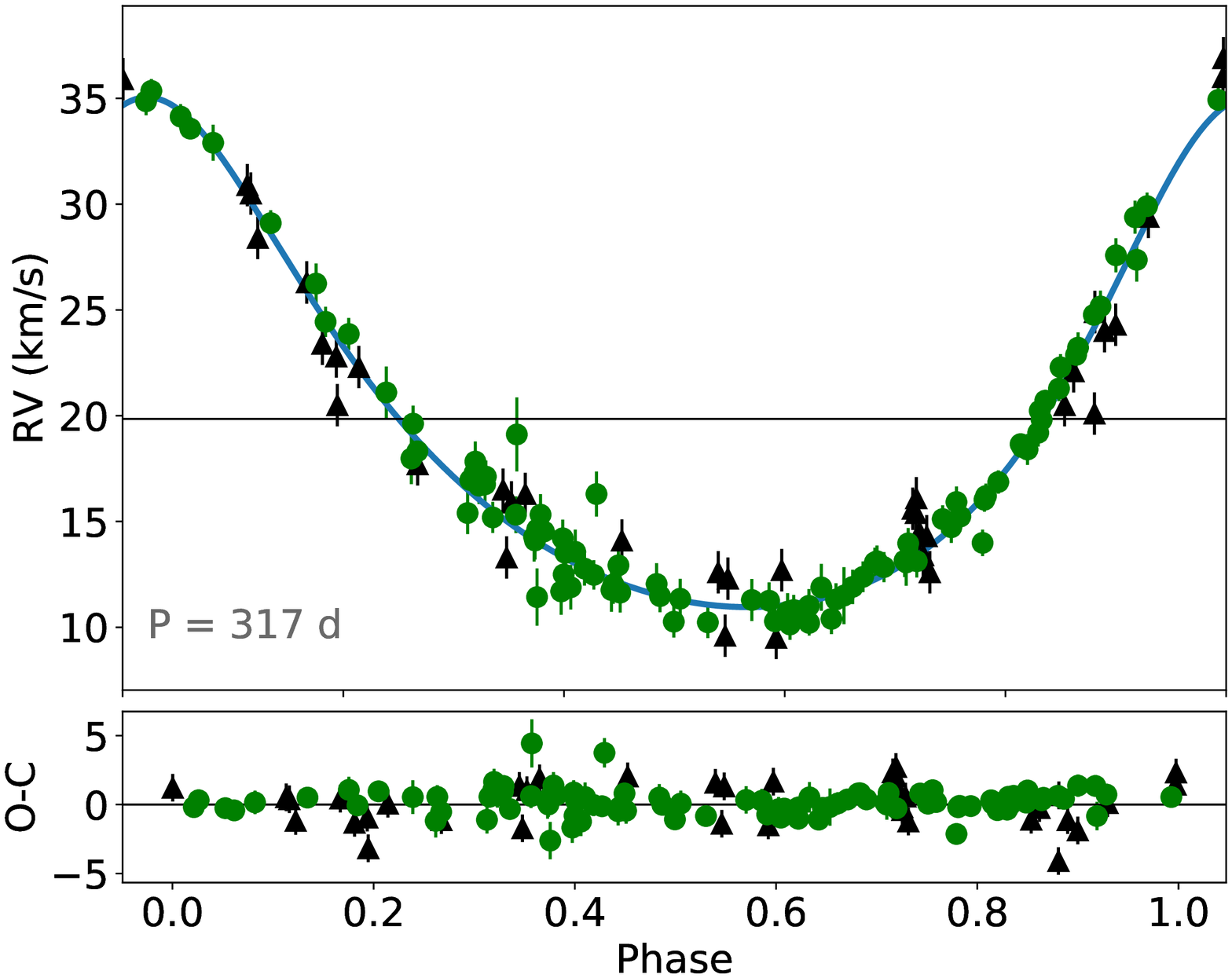}        \label{figure:hd44179}
   }
   \subfloat[\#8: HD~46703]{%
     \includegraphics[width=0.33\textwidth]{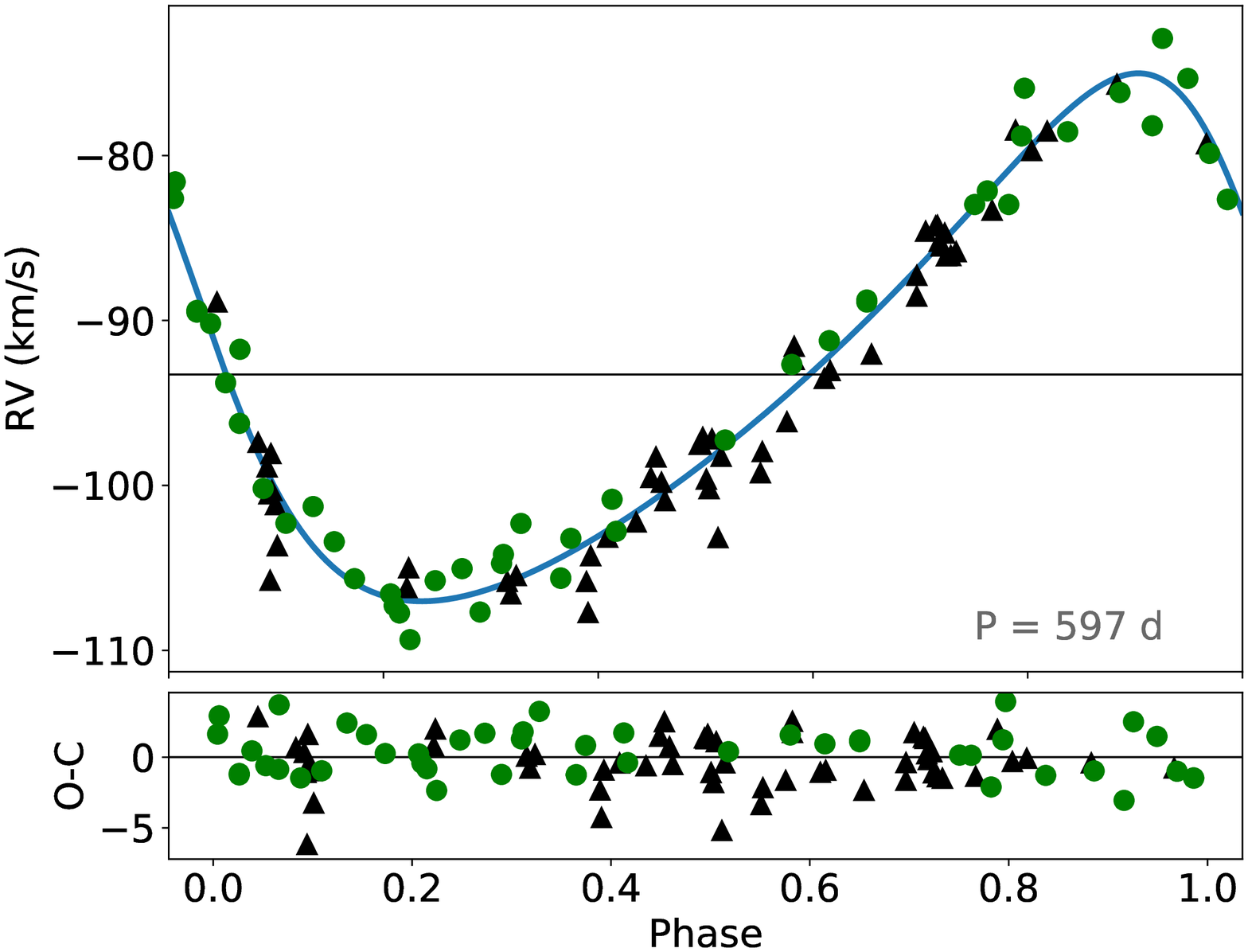}      \label{figure:hd46703}
   }
   \subfloat[\#9: HD~52961]{%
     \includegraphics[width=0.33\textwidth]{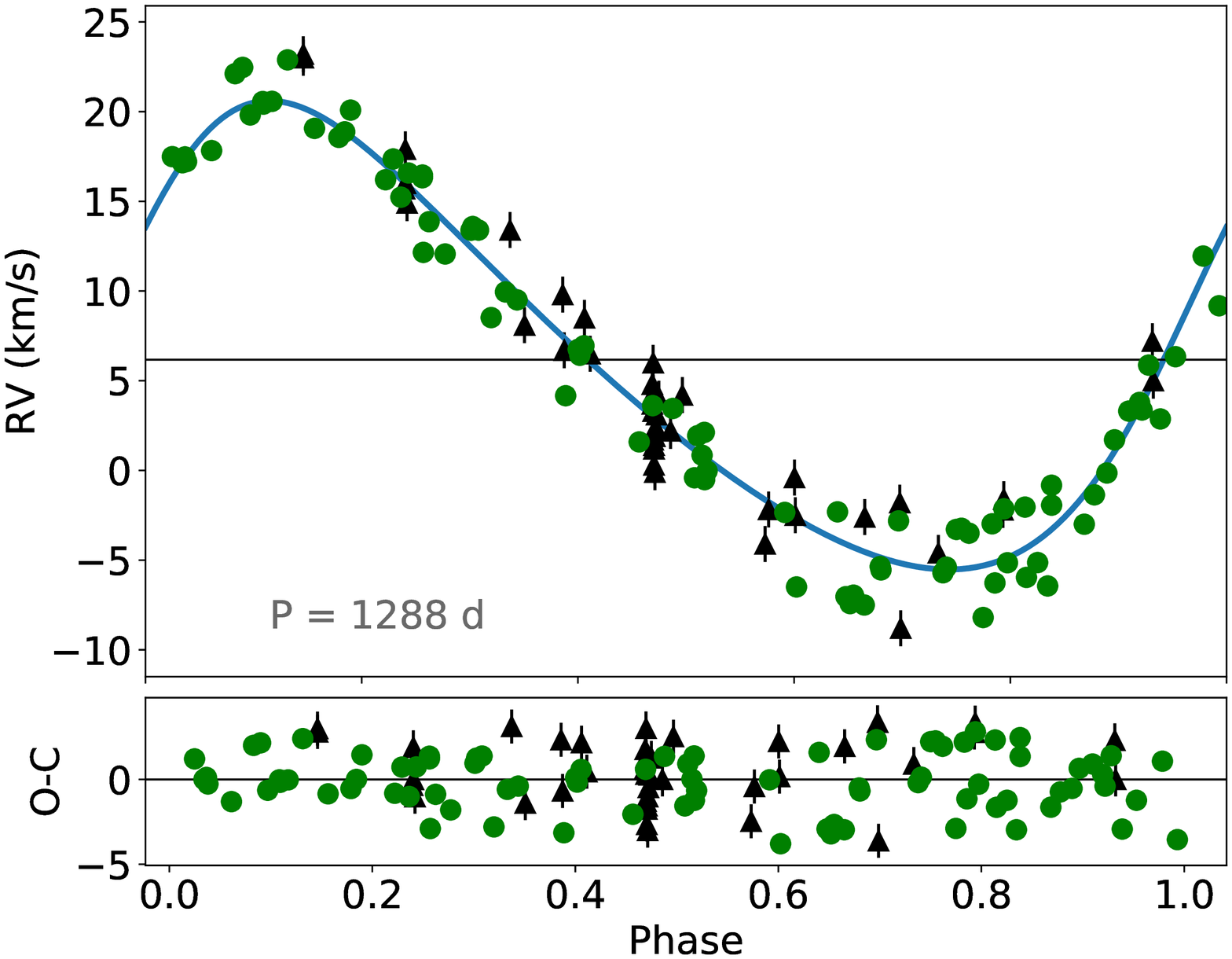}      \label{figure:hd52961}
   }\\
   
   \subfloat[\#10: HD~95767]{%
     \includegraphics[width=0.33\textwidth]{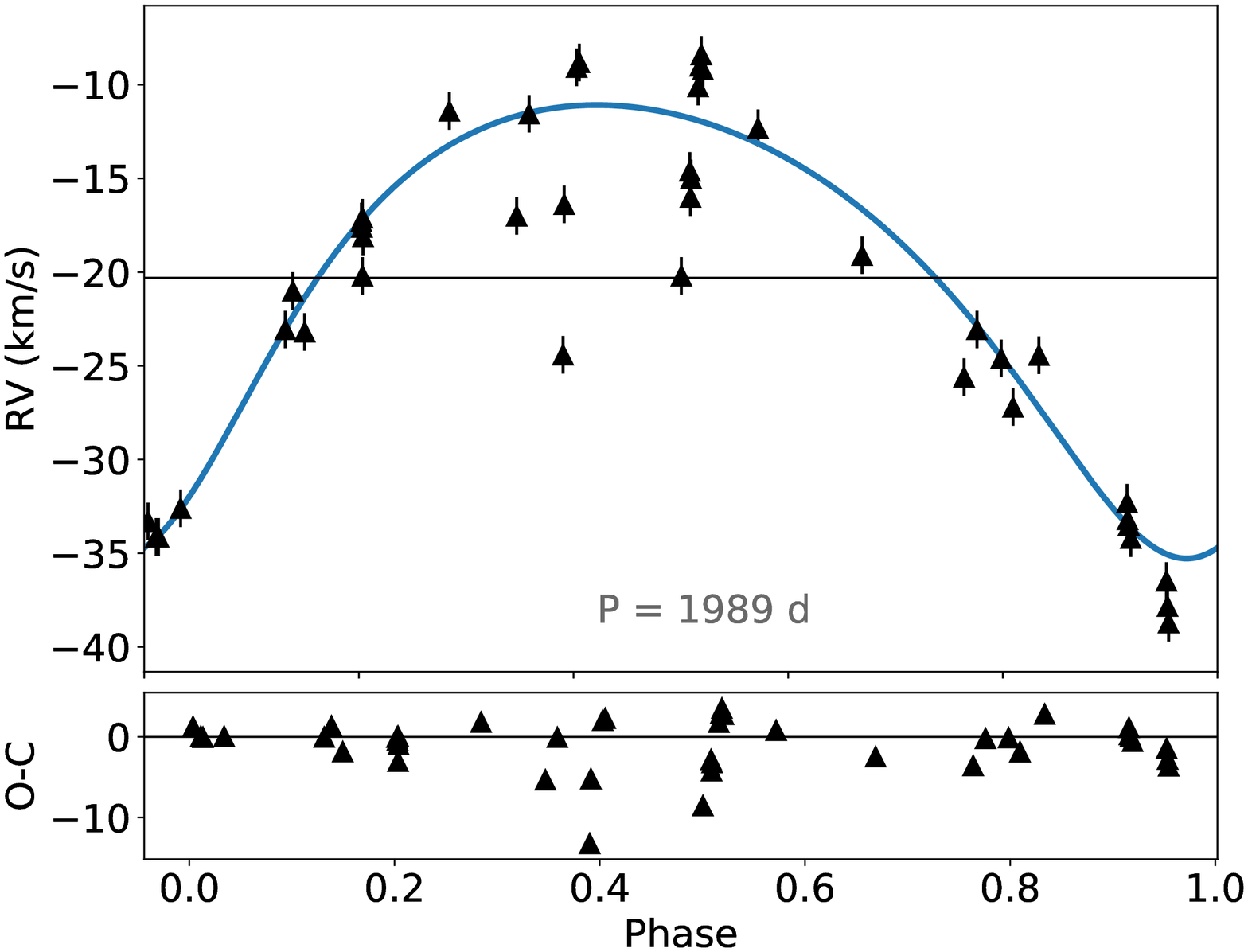}        \label{figure:hd95767}
   }
   \subfloat[\#11: HD~108015]{%
     \includegraphics[width=0.33\textwidth]{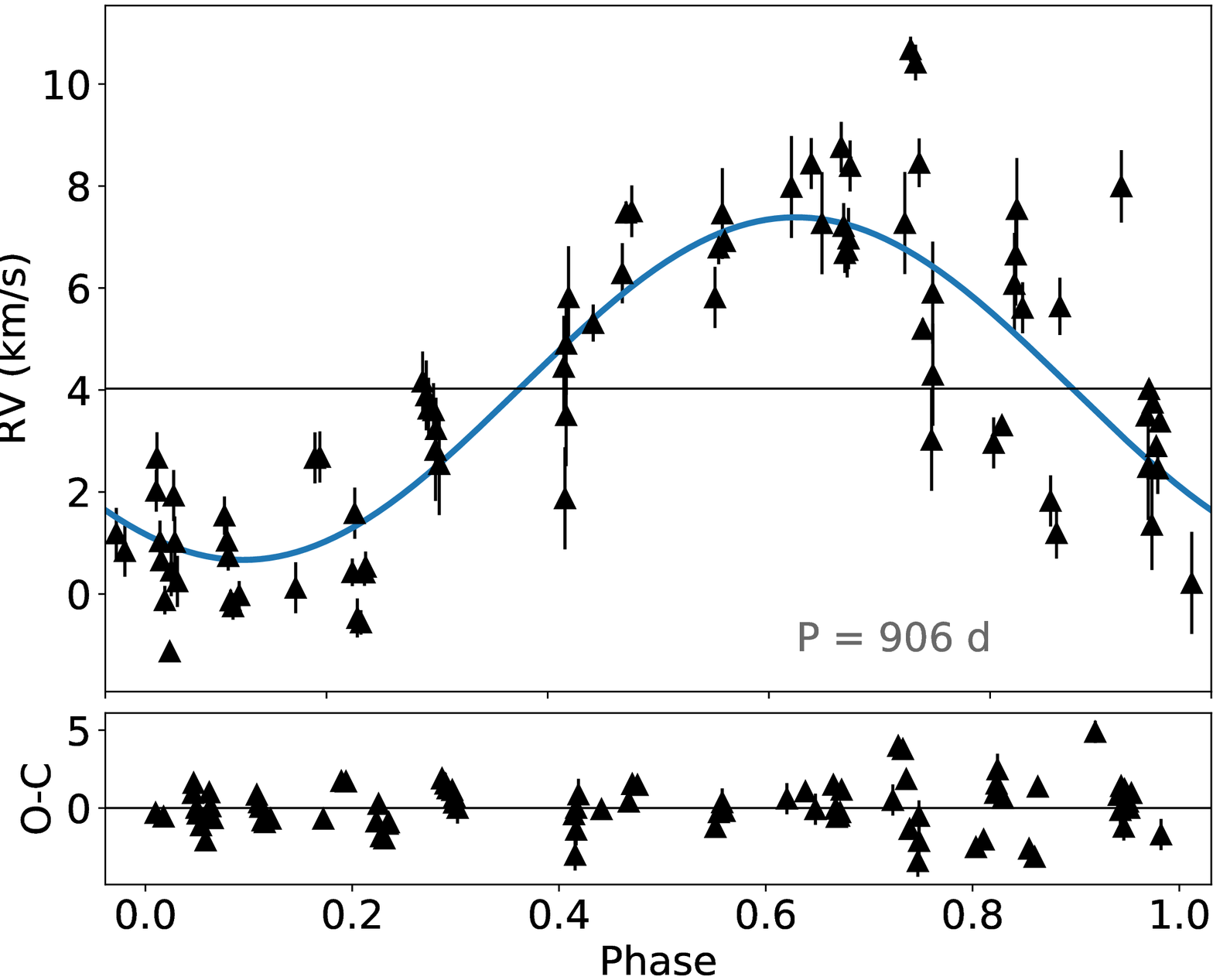}       \label{figure:hd108015}
   }
   \subfloat[\#12: HD~131356]{%
     \includegraphics[width=0.33\textwidth]{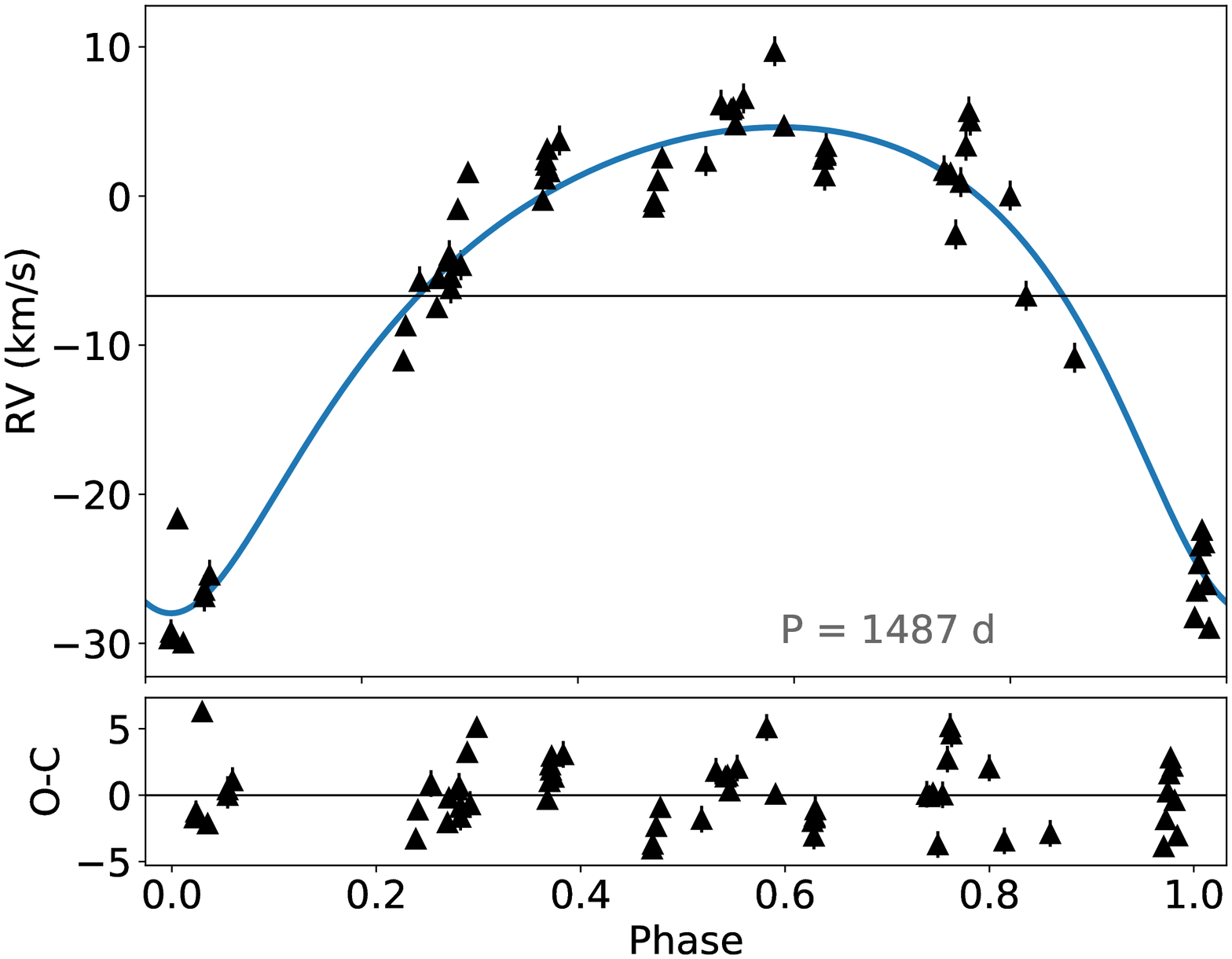}        \label{figure:hd131356}
   }\\
   
   \caption{Phase-folded orbits of post-AGB binaries in our sample. Green circles show data points taken with the HERMES spectrograph, while black triangles are older data. The systemic velocity is given by the horizontal black line. The residuals are shown below each radial-velocity curve.}
   \label{fig:orbits}
 \end{figure*}

 \begin{figure*}
 \ContinuedFloat
 \centering
   \subfloat[\#13: HD~158616]{%
     \includegraphics[width=0.33\textwidth]{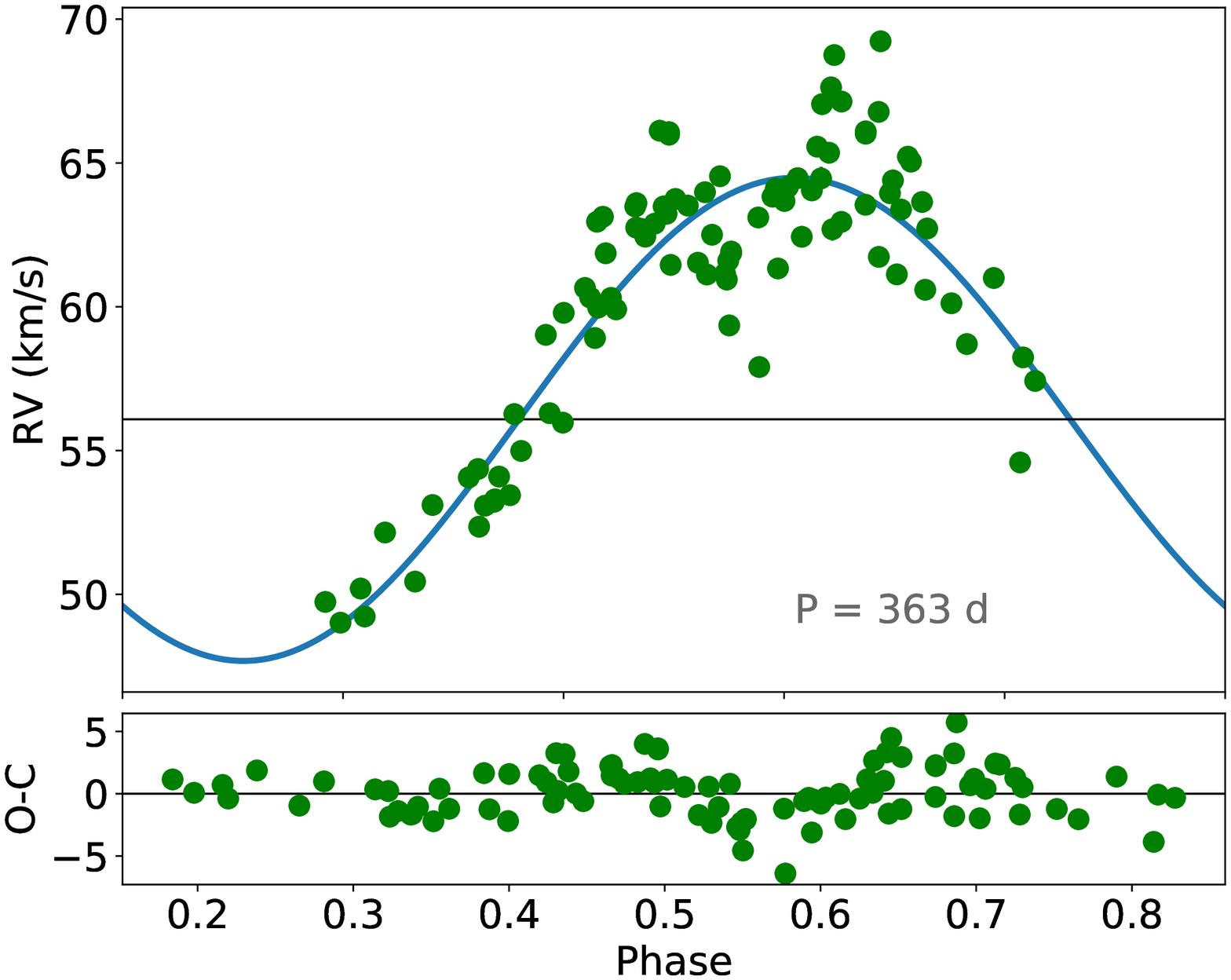}     \label{figure:hd158616}
   }
   \subfloat[\#14: HD~213985]{%
     \includegraphics[width=0.33\textwidth]{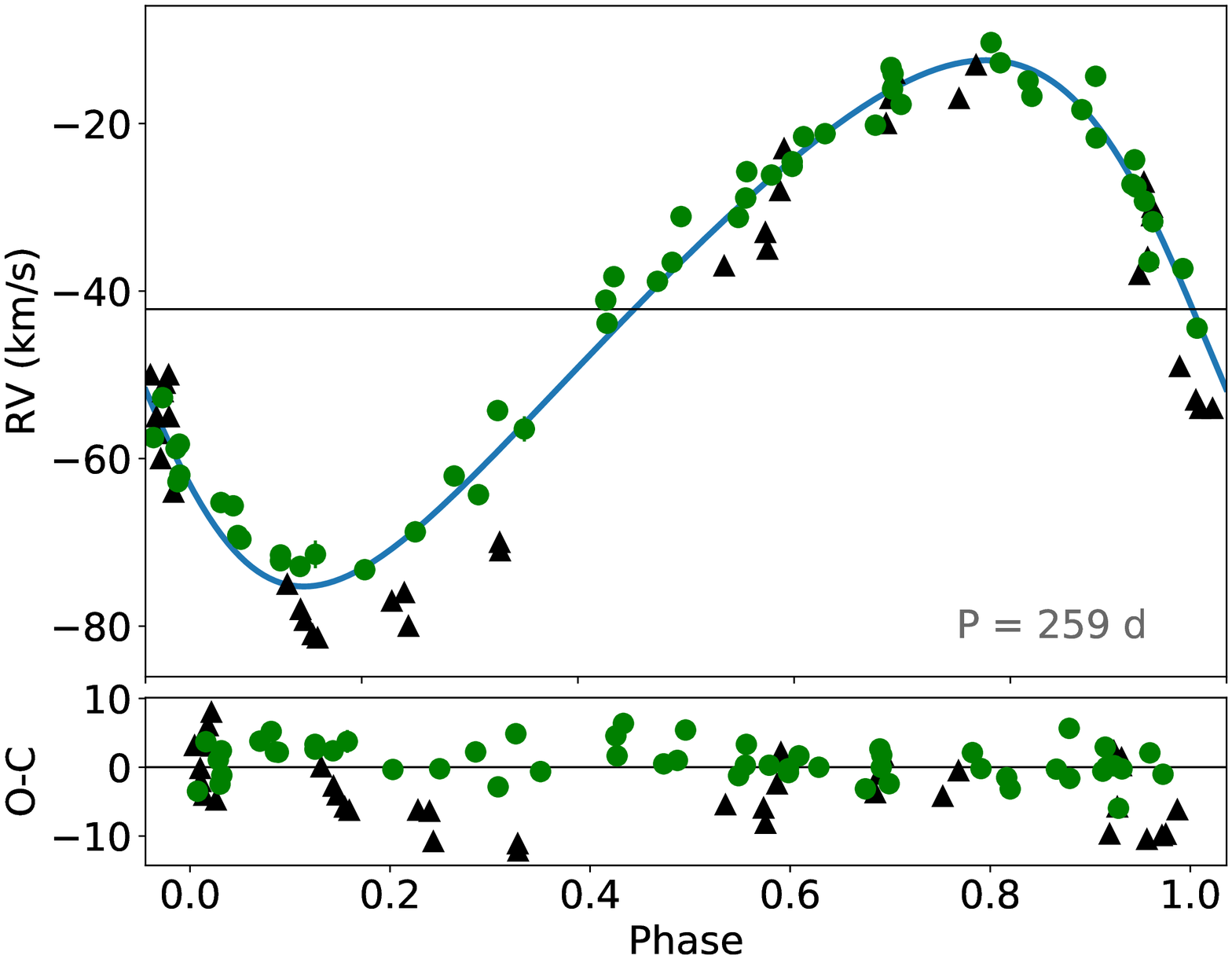}      \label{figure:hd213985}
   }
   \subfloat[\#15: HP~Lyr]{%
     \includegraphics[width=0.33\textwidth]{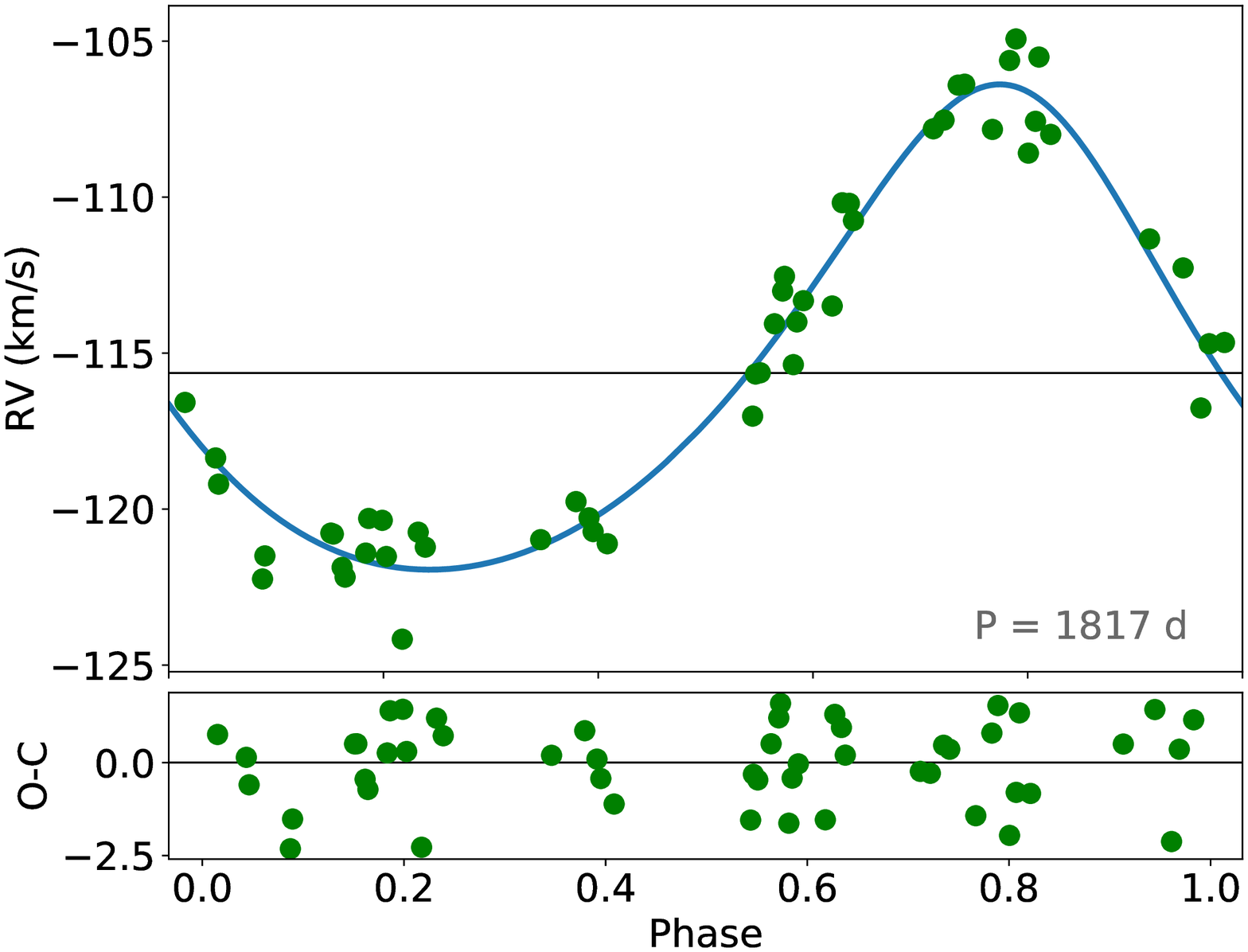}      \label{figure:hplyr}
   }\\
   
   \subfloat[\#16: HR~4049]{%
     \includegraphics[width=0.33\textwidth]{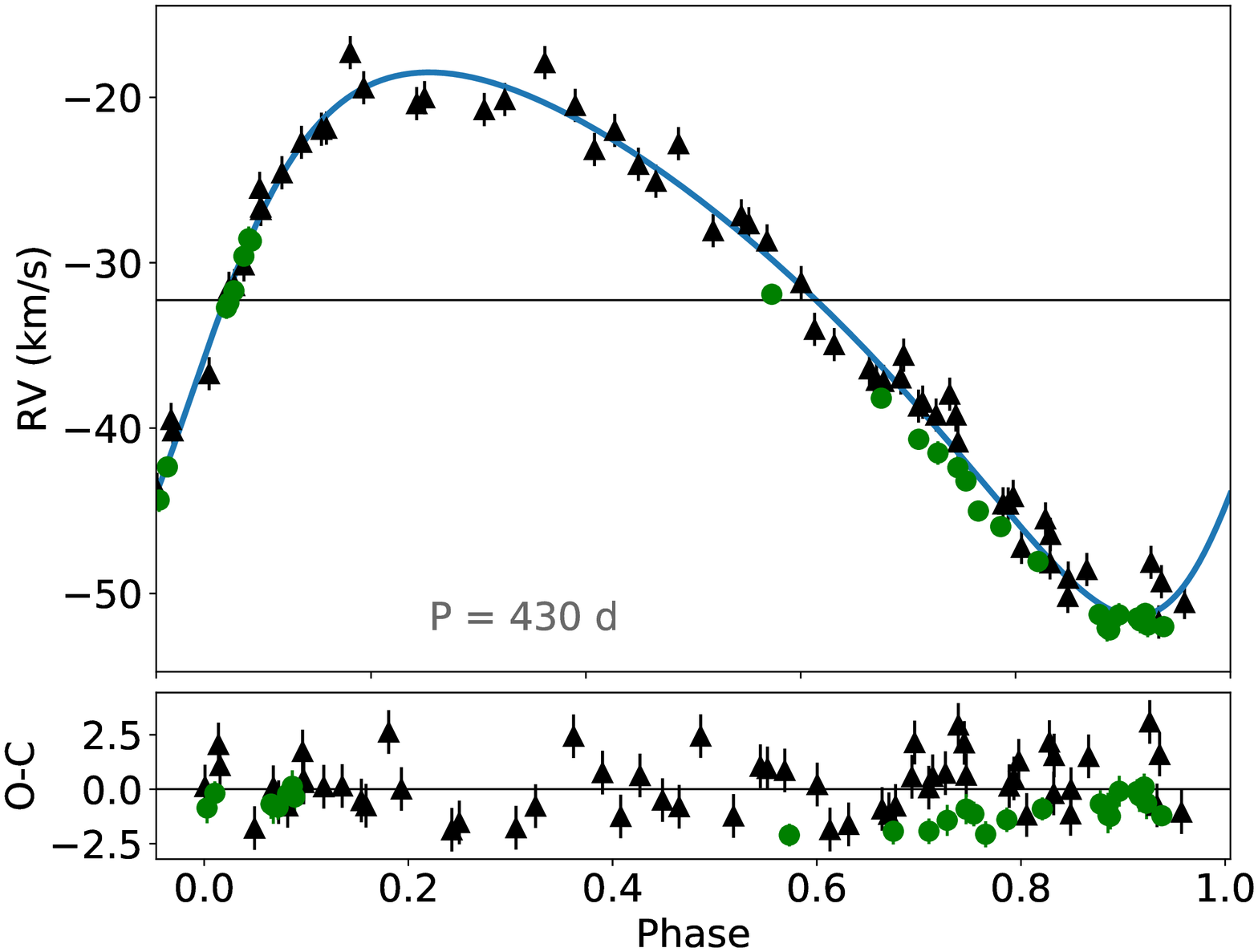}     \label{figure:hr4049}
   }
   \subfloat[\#17: IRAS~05208-2035]{%
     \includegraphics[width=0.33\textwidth]{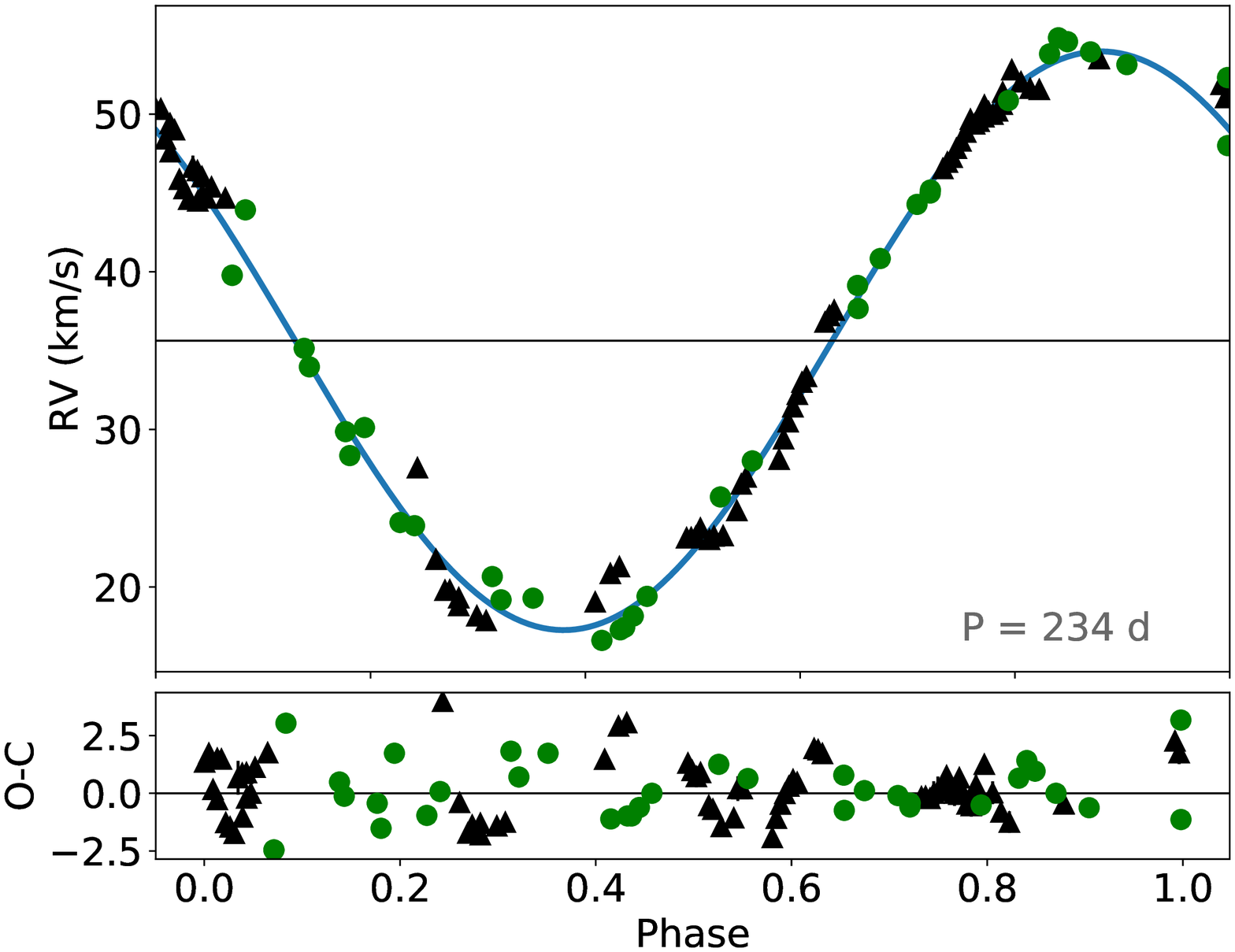}          \label{figure:iras05208}
   }
   \subfloat[\#18: IRAS~06165+3158]{%
     \includegraphics[width=0.33\textwidth]{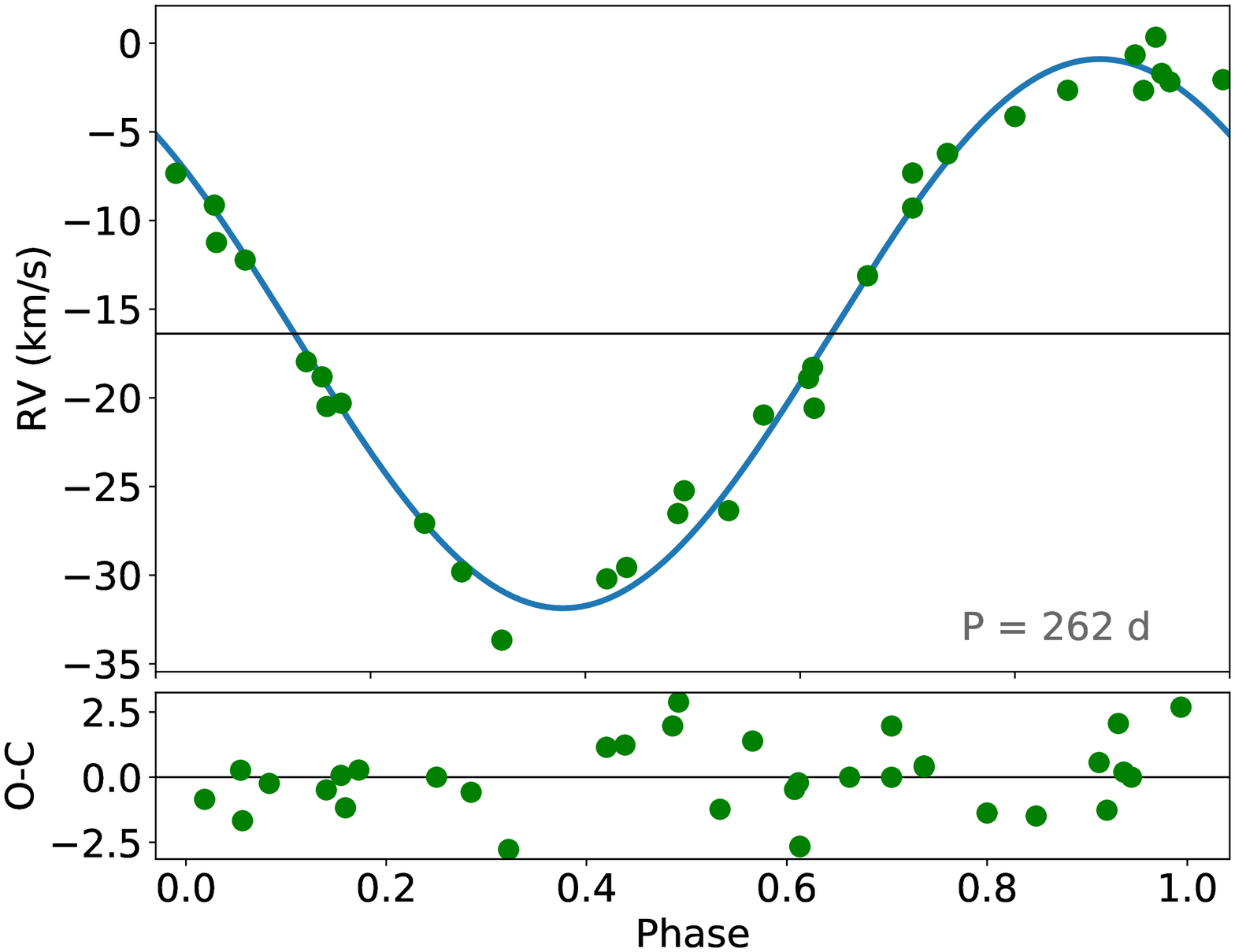}        \label{figure:iras06165}
   }\\
   
   \subfloat[\#19: IRAS~06452-3456]{%
     \includegraphics[width=0.33\textwidth]{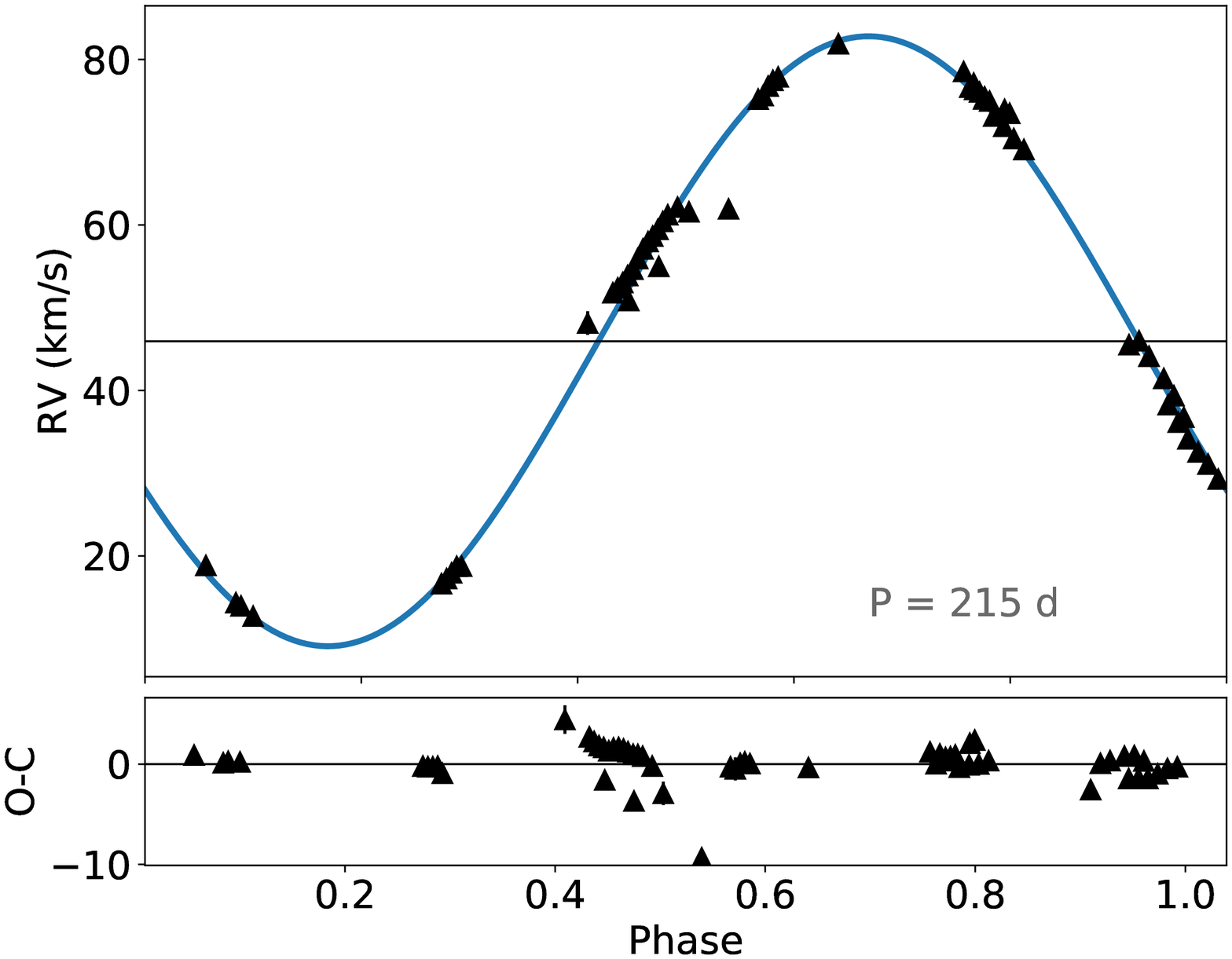}       \label{figure:iras06452}
   }
   \subfloat[\#20: IRAS~08544-4431]{%
     \includegraphics[width=0.33\textwidth]{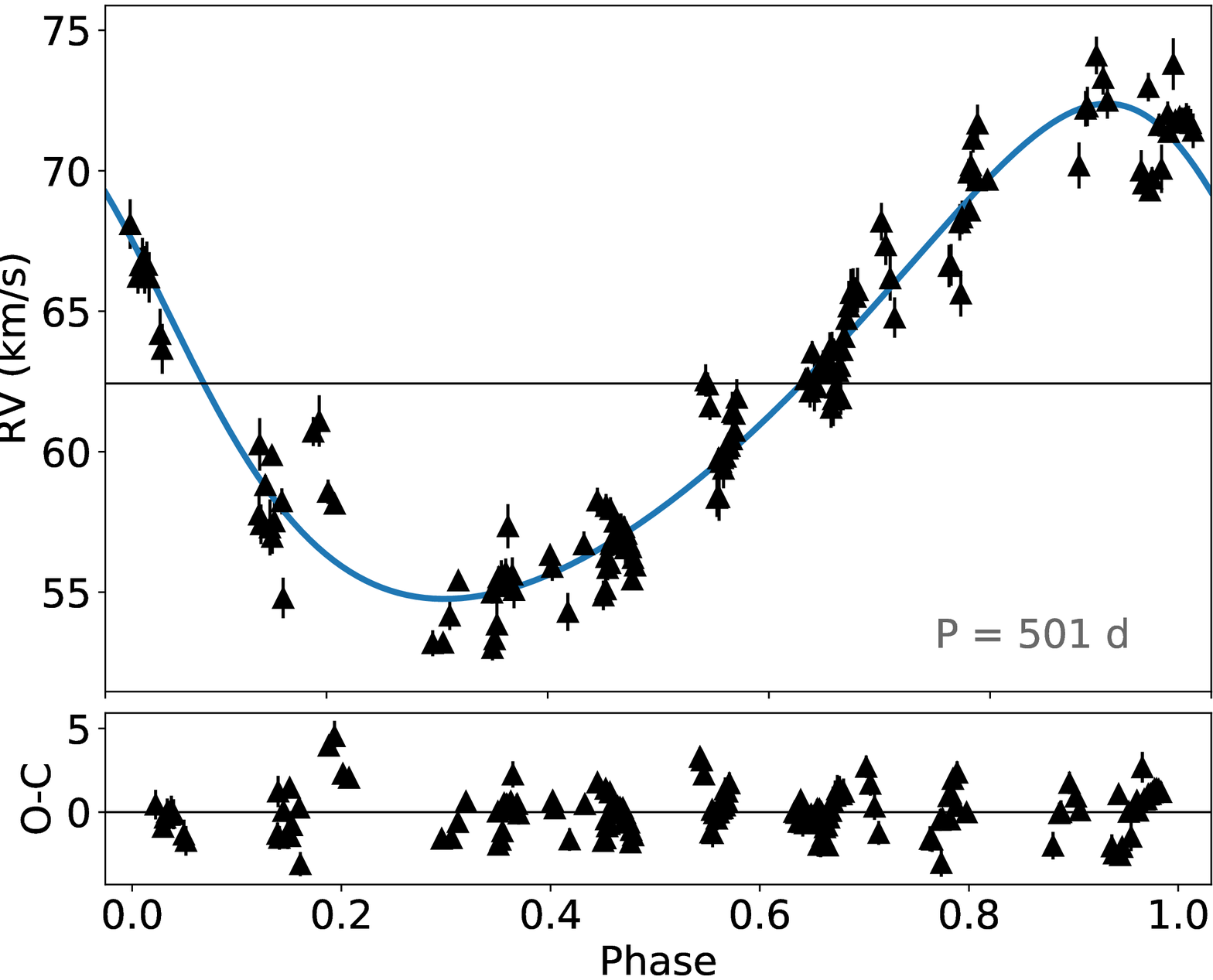}        \label{figure:iras08544}
   }
   \subfloat[\#21: IRAS~09144-4933]{%
     \includegraphics[width=0.33\textwidth]{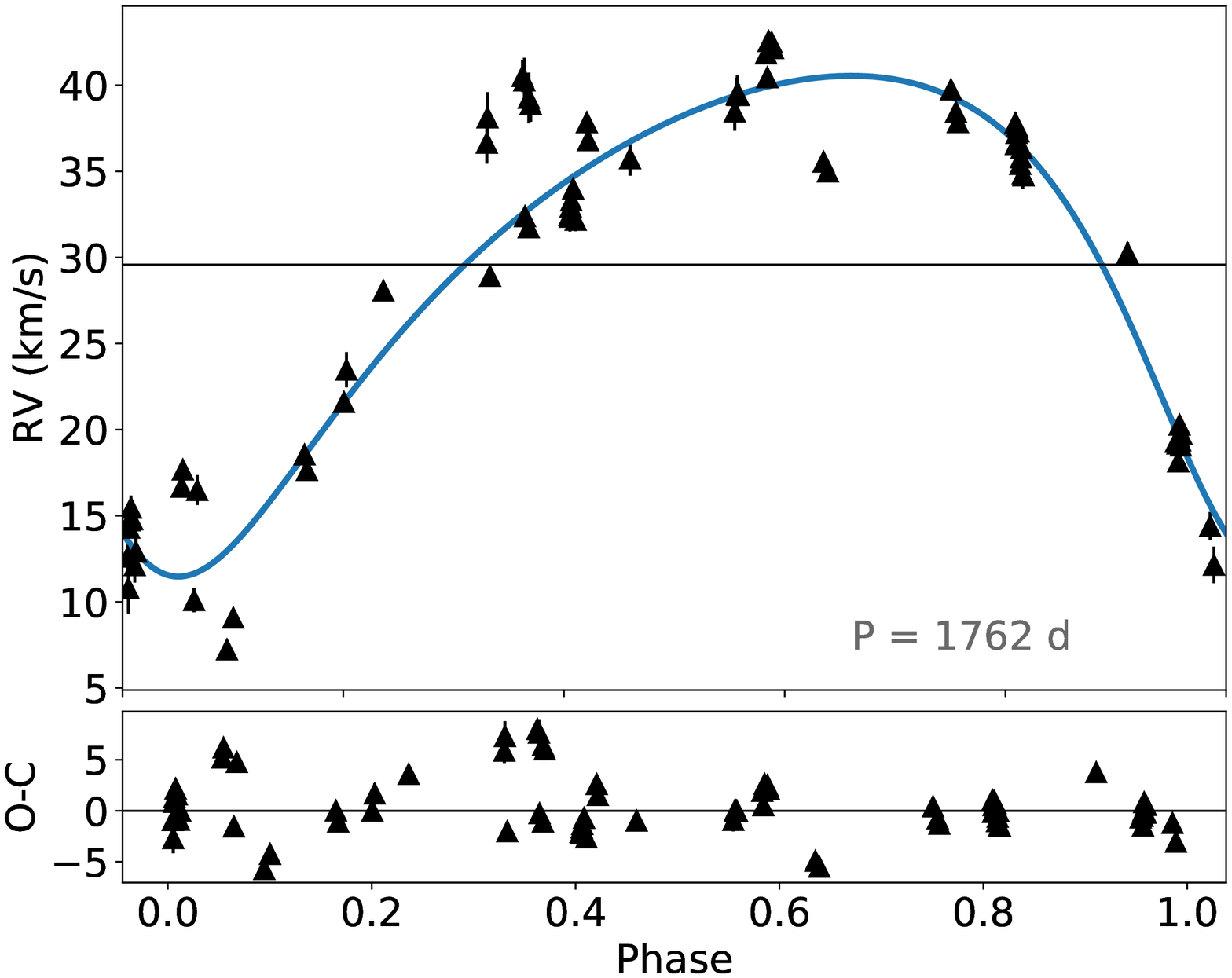}       \label{figure:iras09144}
   }\\
   
   \subfloat[\#22: IRAS~15469-5311]{%
     \includegraphics[width=0.33\textwidth]{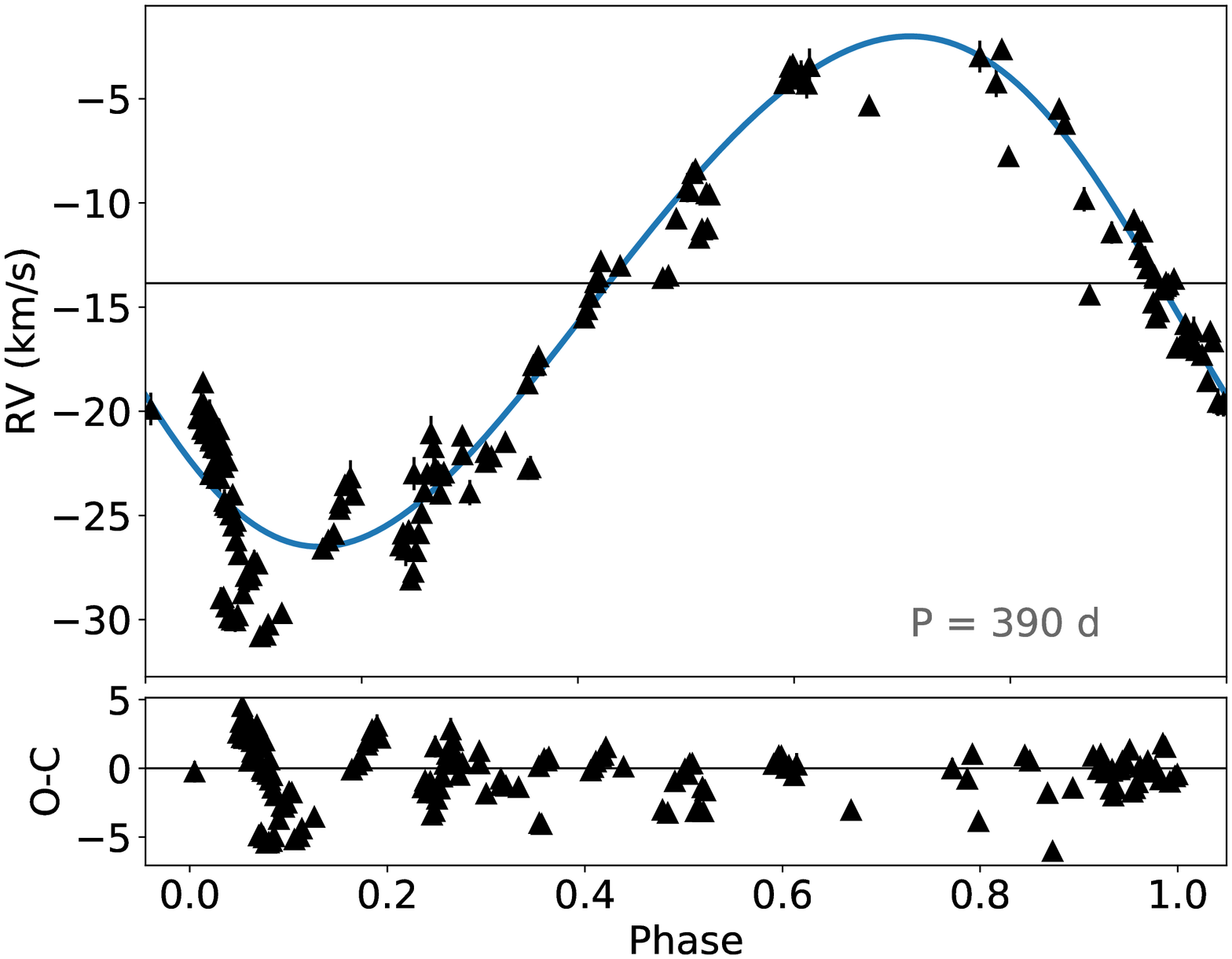}      \label{figure:iras15469}
   }
   \subfloat[\#23: IRAS~16230-3410]{%
     \includegraphics[width=0.33\textwidth]{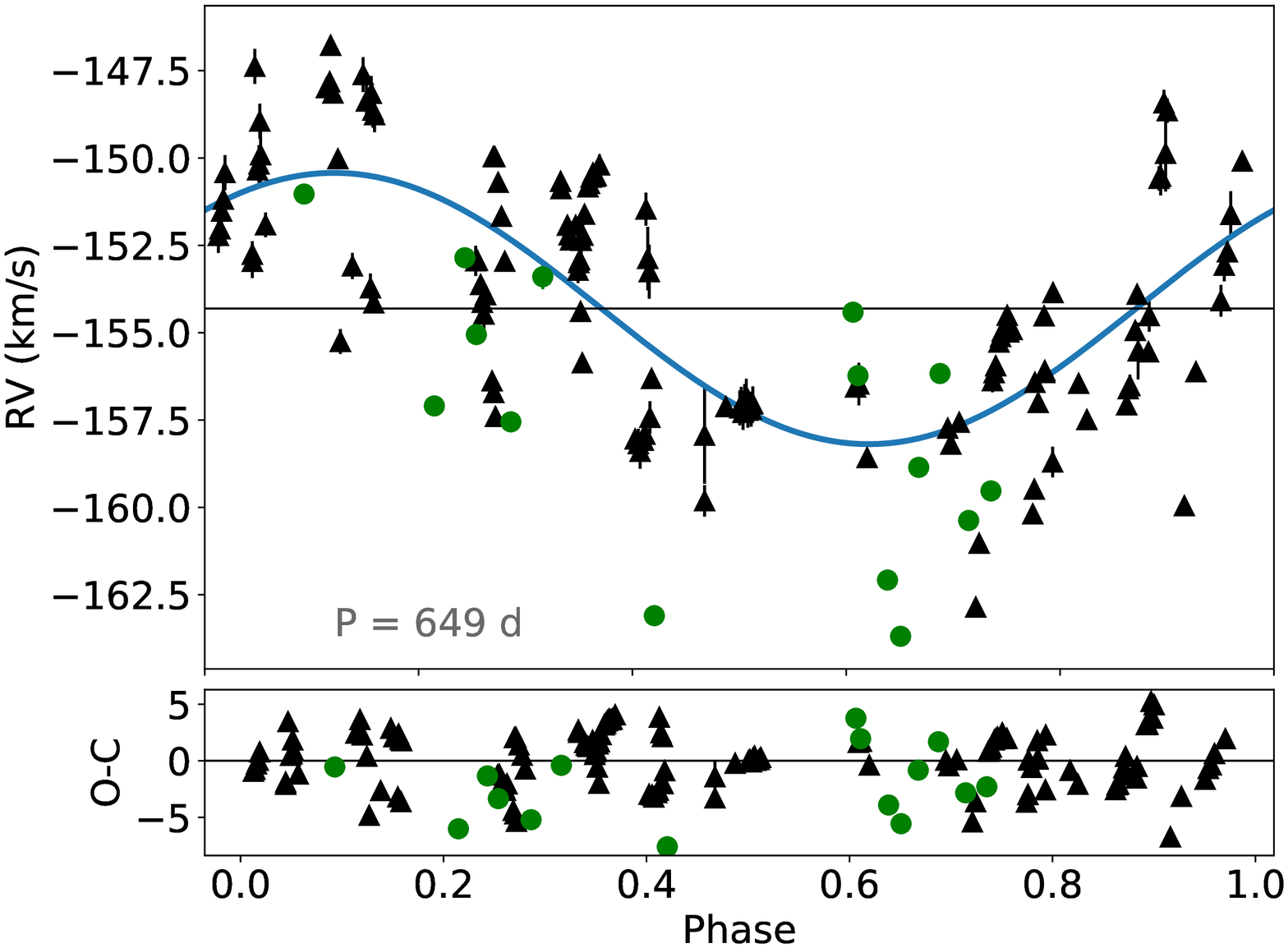}      \label{figure:iras16230}
   }
   \subfloat[\#24: IRAS~17038-4815]{%
     \includegraphics[width=0.33\textwidth]{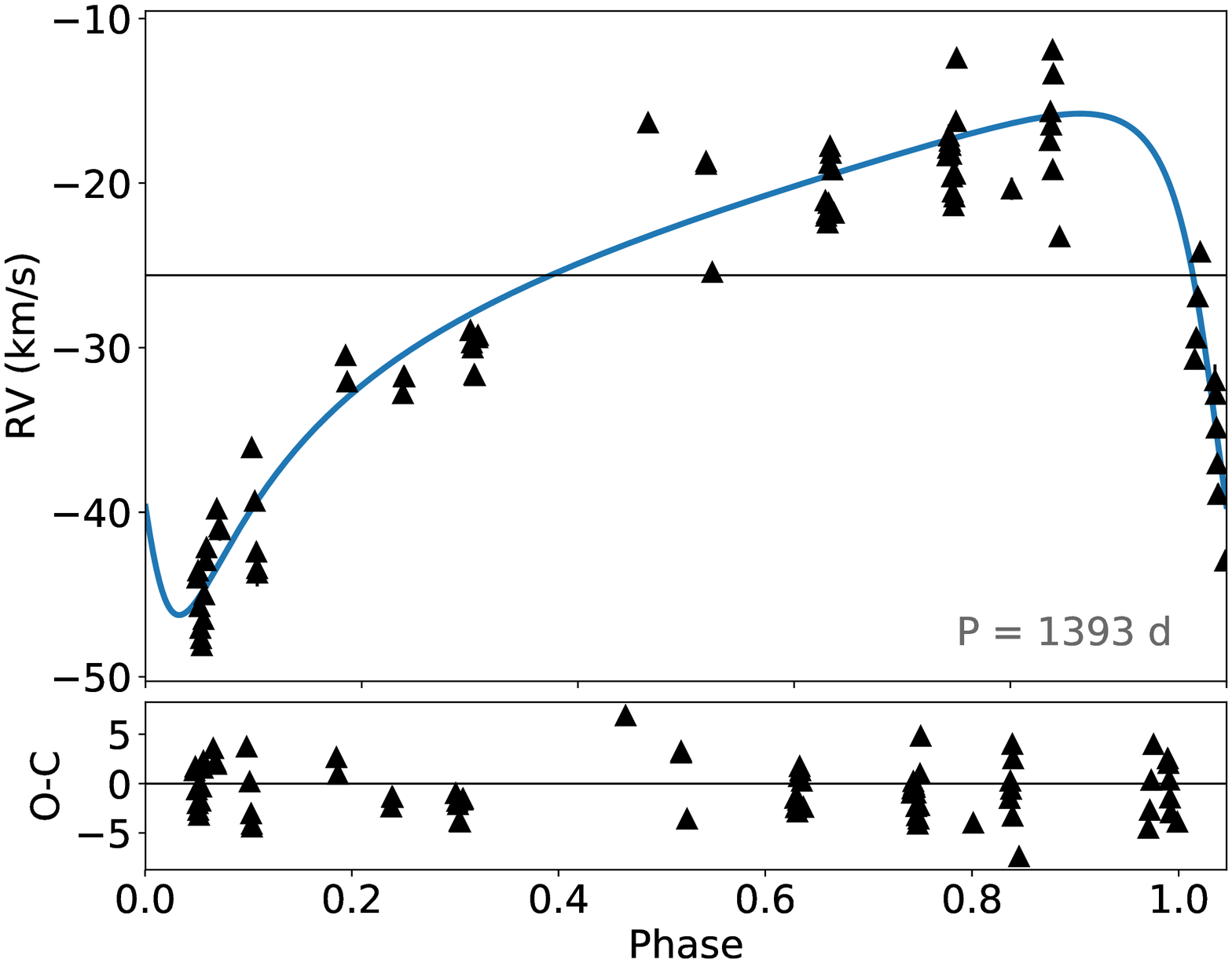}        \label{figure:iras17038}
   }\\
   \caption{\textit{continued}}
   %\label{fig:orbits}
 \end{figure*}

 \begin{figure*}
 \ContinuedFloat
 \centering
   \subfloat[\#25: IRAS~19125+0343]{%
     \includegraphics[width=0.33\textwidth]{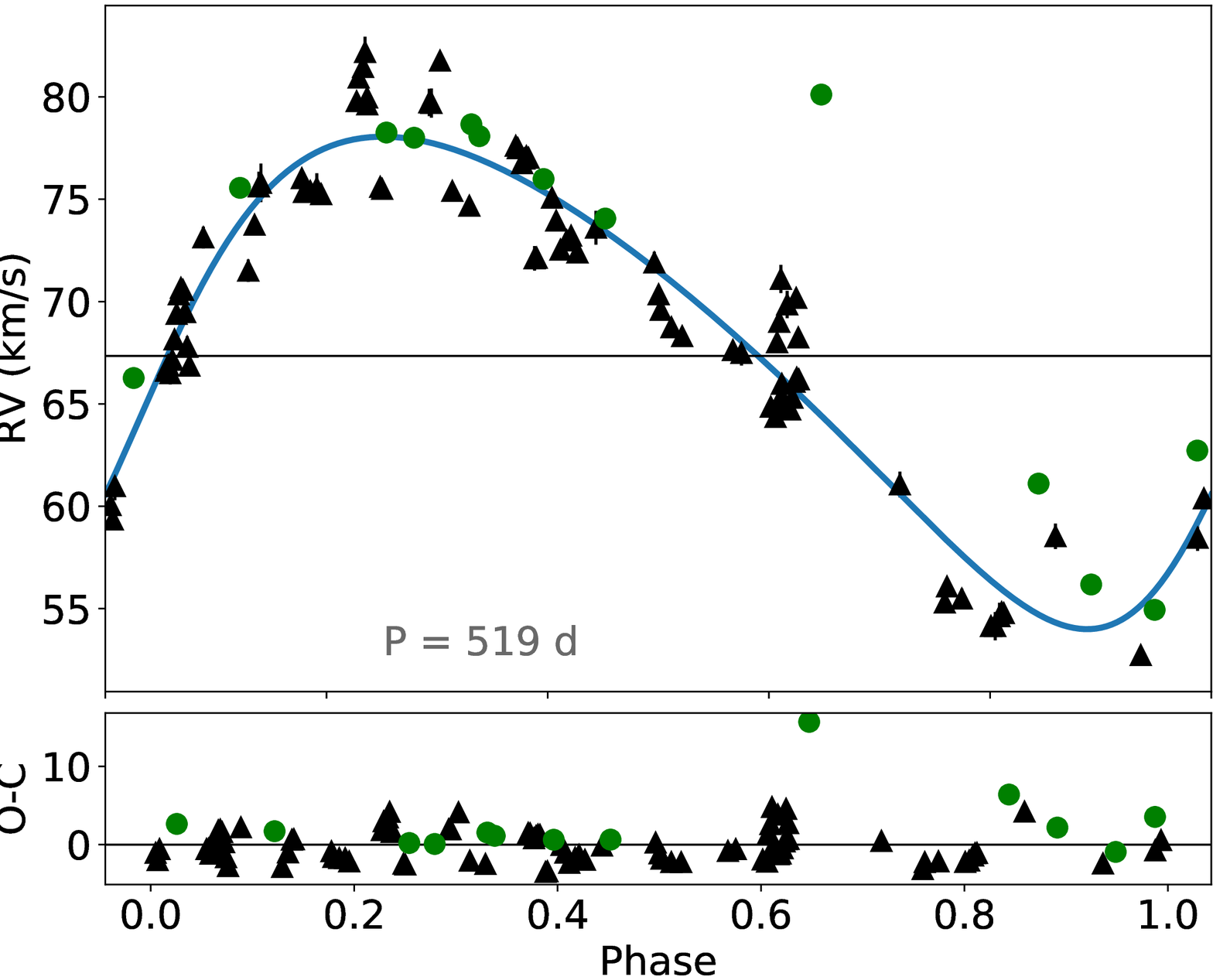}       \label{figure:iras19125}
   }
   \subfloat[\#26: IRAS~19135+3937]{%
     \includegraphics[width=0.33\textwidth]{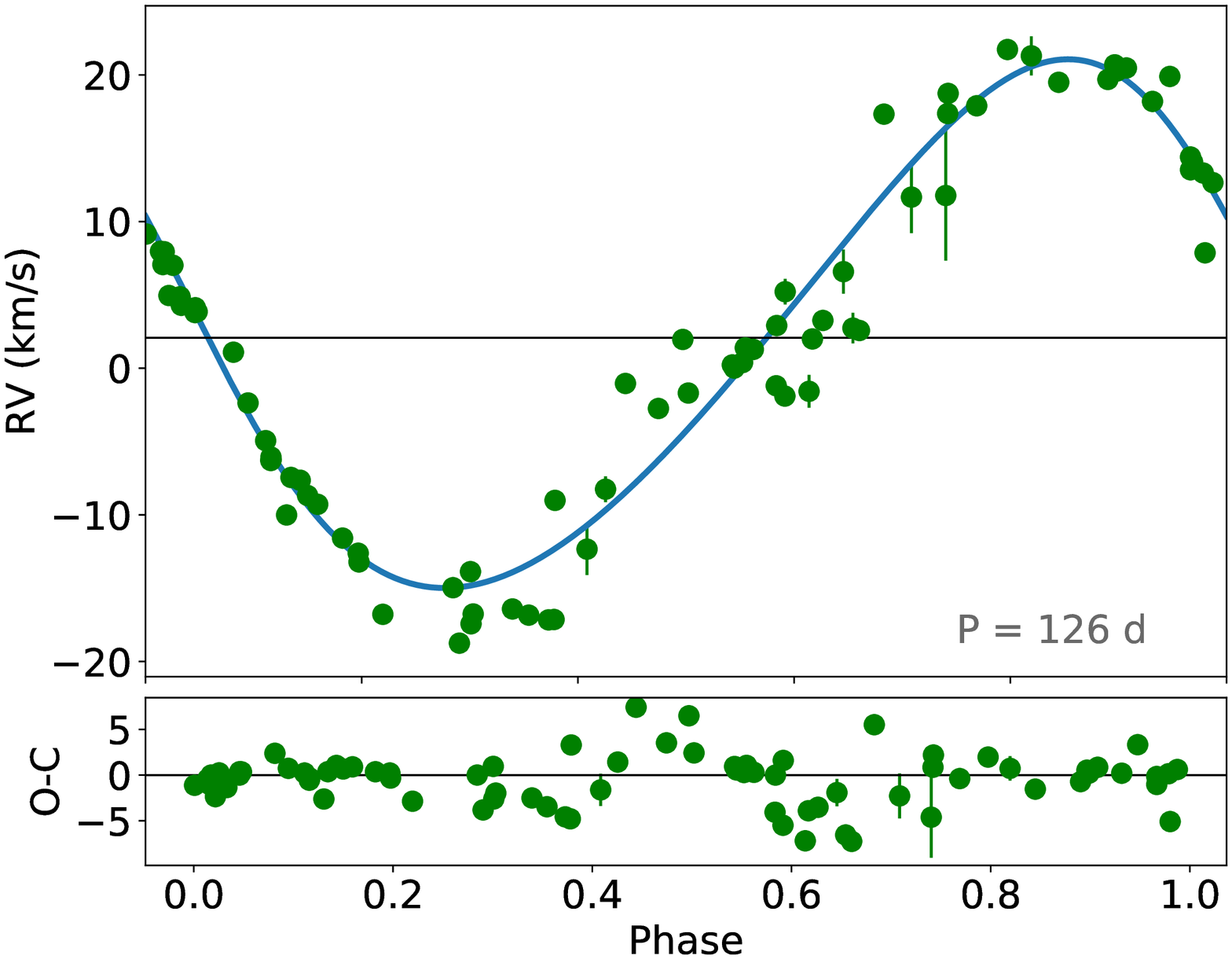}        \label{figure:iras19135}
   }
   \subfloat[\#27: IRAS~19157-0247]{%
     \includegraphics[width=0.33\textwidth]{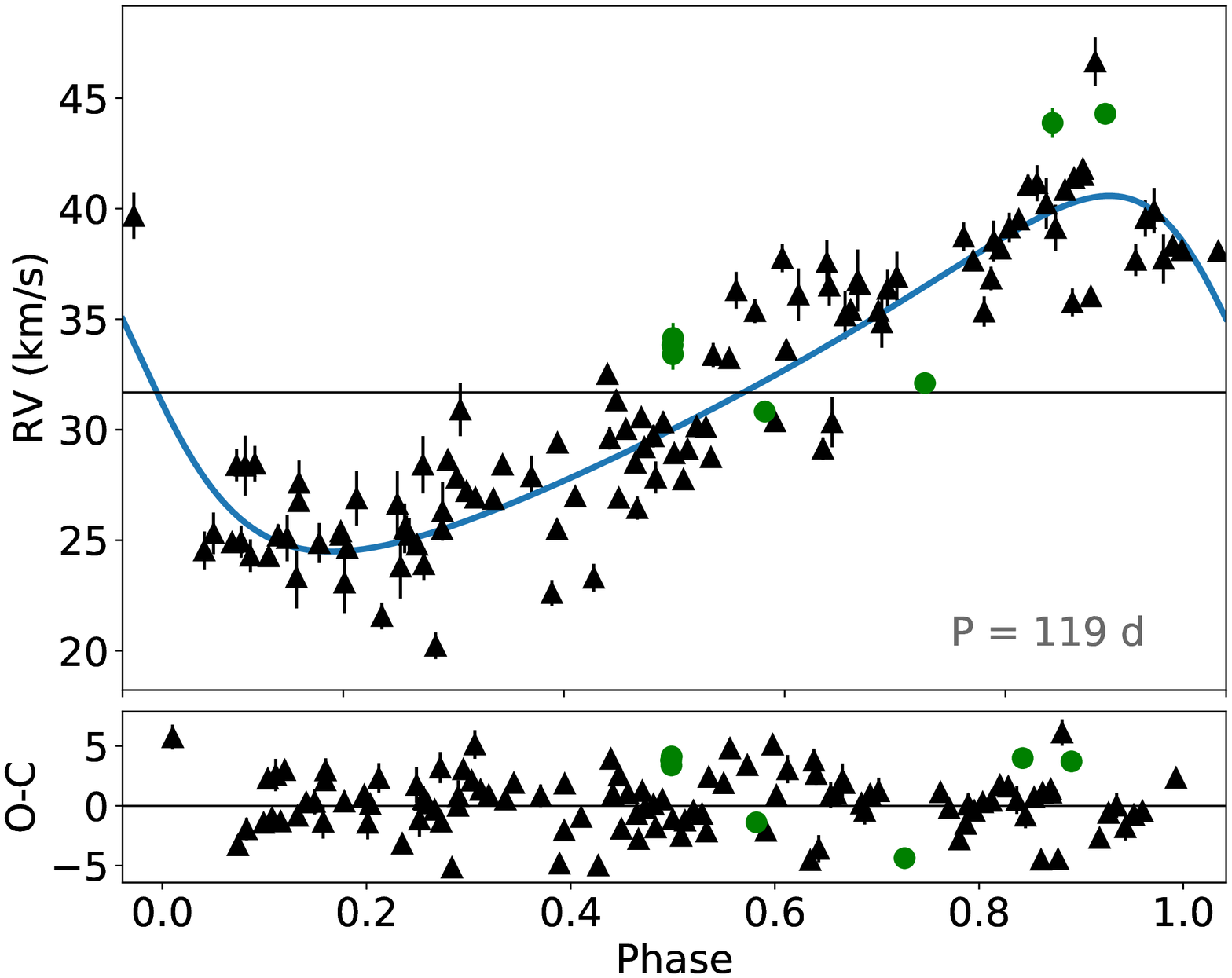}     \label{figure:iras19157}
   }\\
   \subfloat[\#28: RU~Cen]{%
     \includegraphics[width=0.33\textwidth]{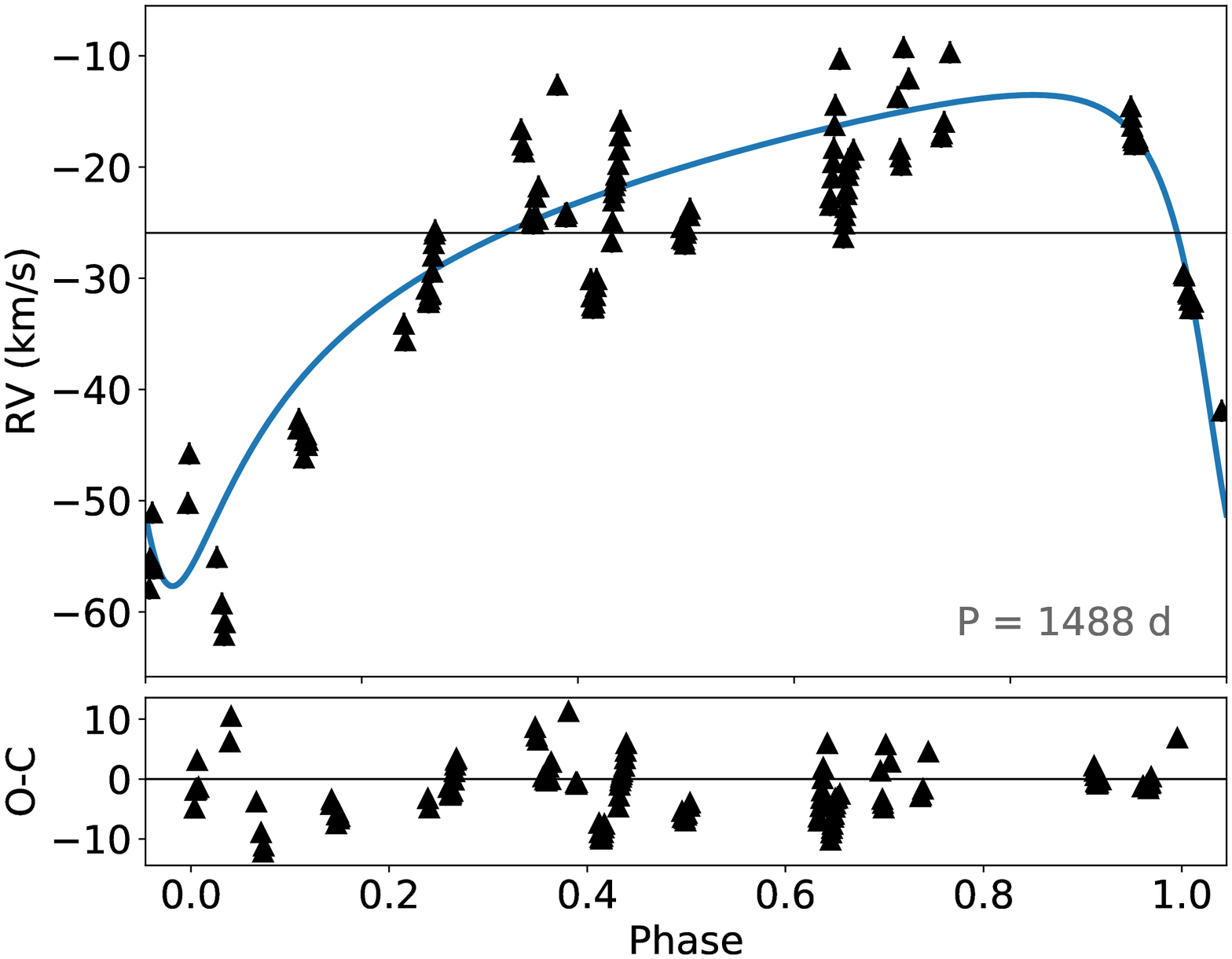}      \label{figure:rucen}
   }
   \subfloat[\#29: SAO~173329]{%
     \includegraphics[width=0.33\textwidth]{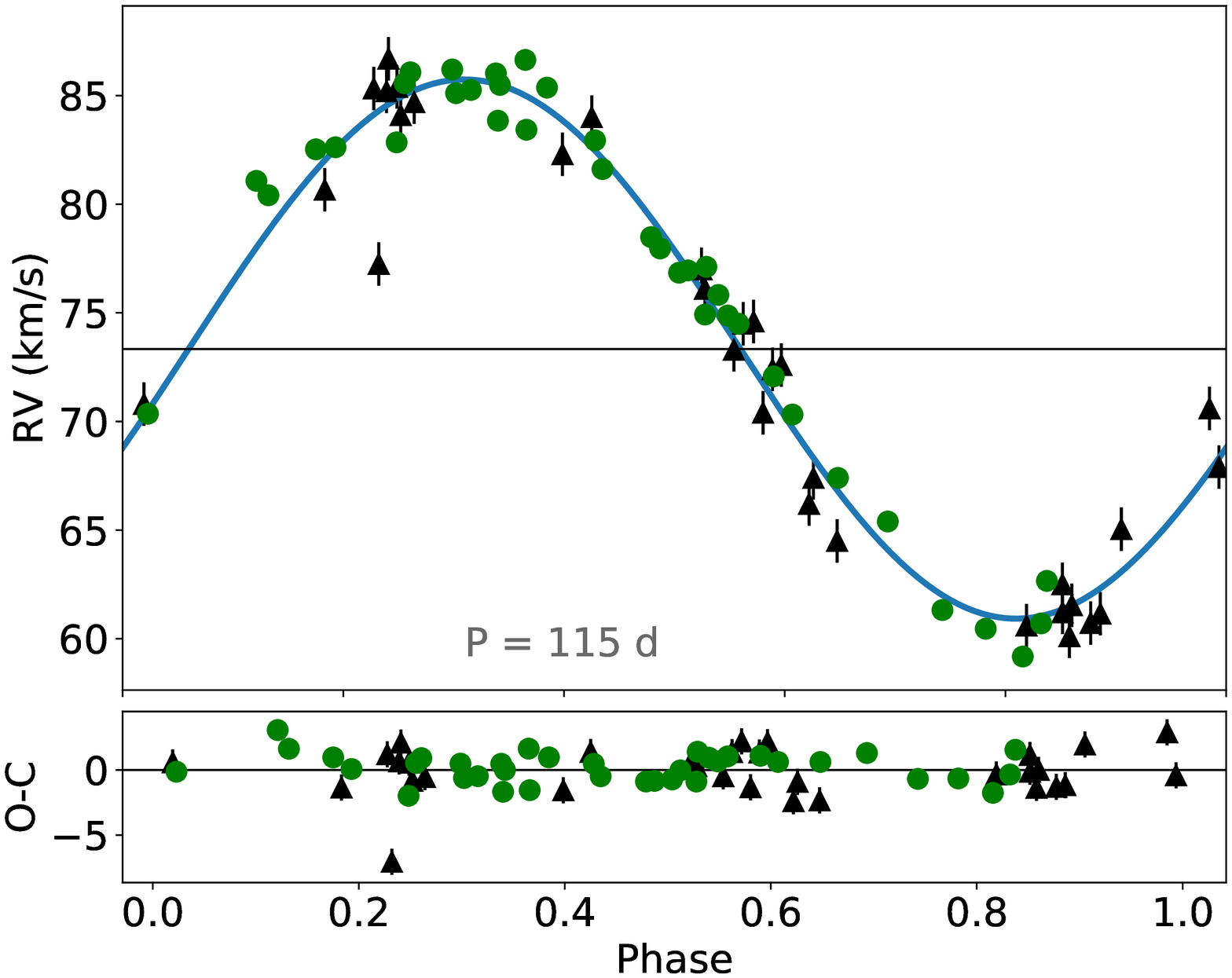}      \label{figure:sao173329}
   }
   \subfloat[\#30: ST~Pup]{%
     \includegraphics[width=0.33\textwidth]{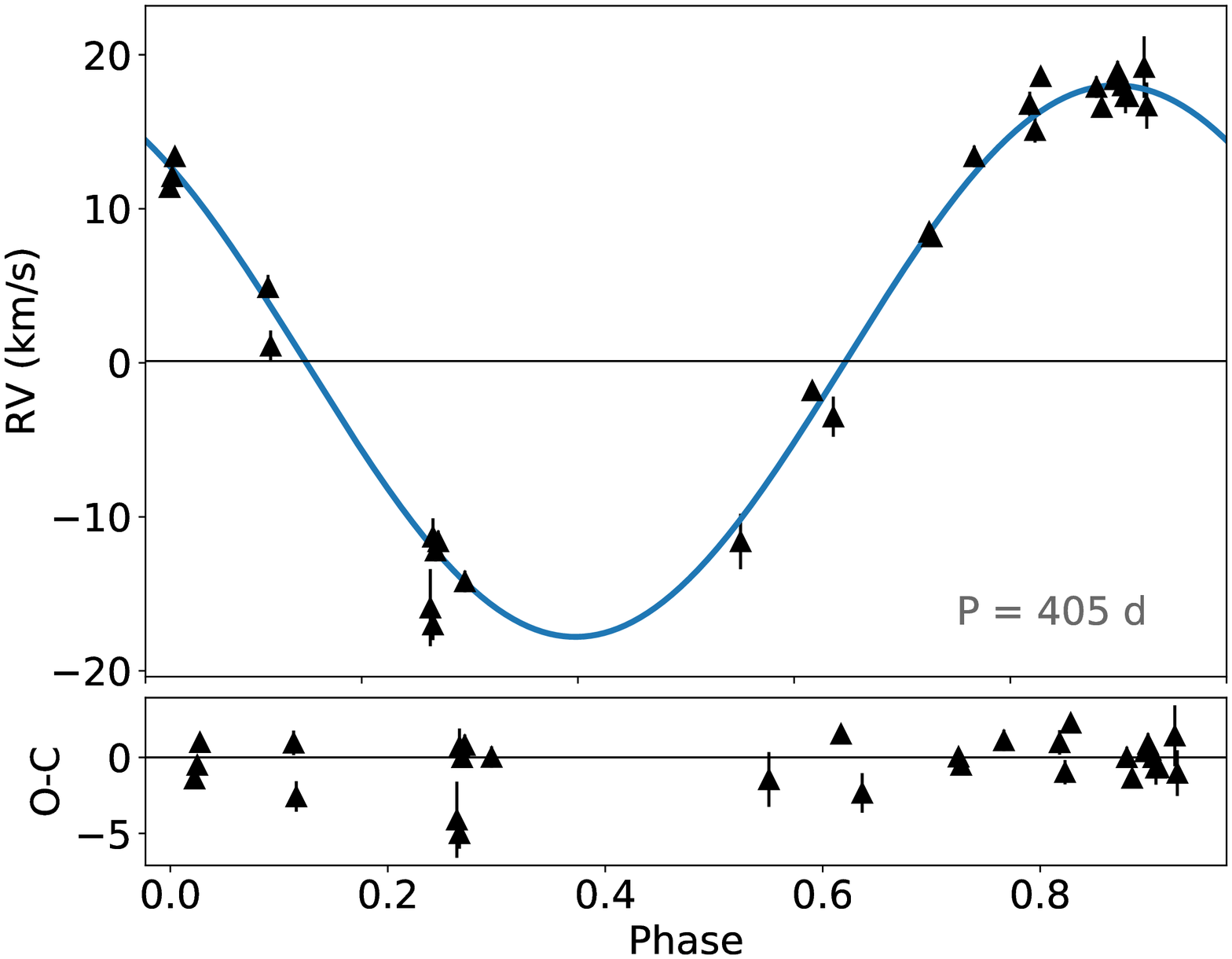}      \label{figure:stpup}
   }\\
   
   \subfloat[\#31: SX~Cen]{%
     \includegraphics[width=0.33\textwidth]{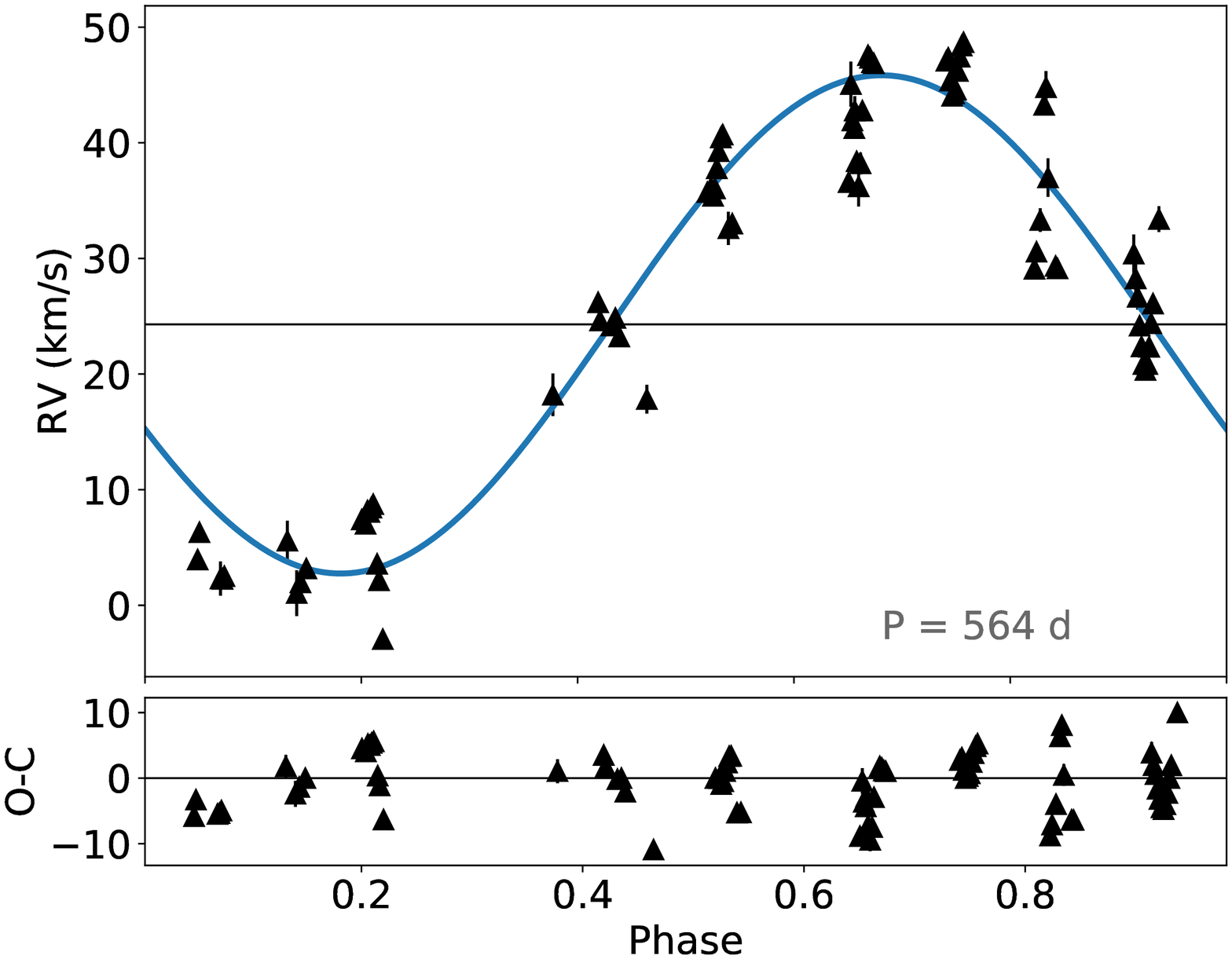}     \label{figure:sxcen}
   }
   \subfloat[\#32: TW~Cam]{%
     \includegraphics[width=0.33\textwidth]{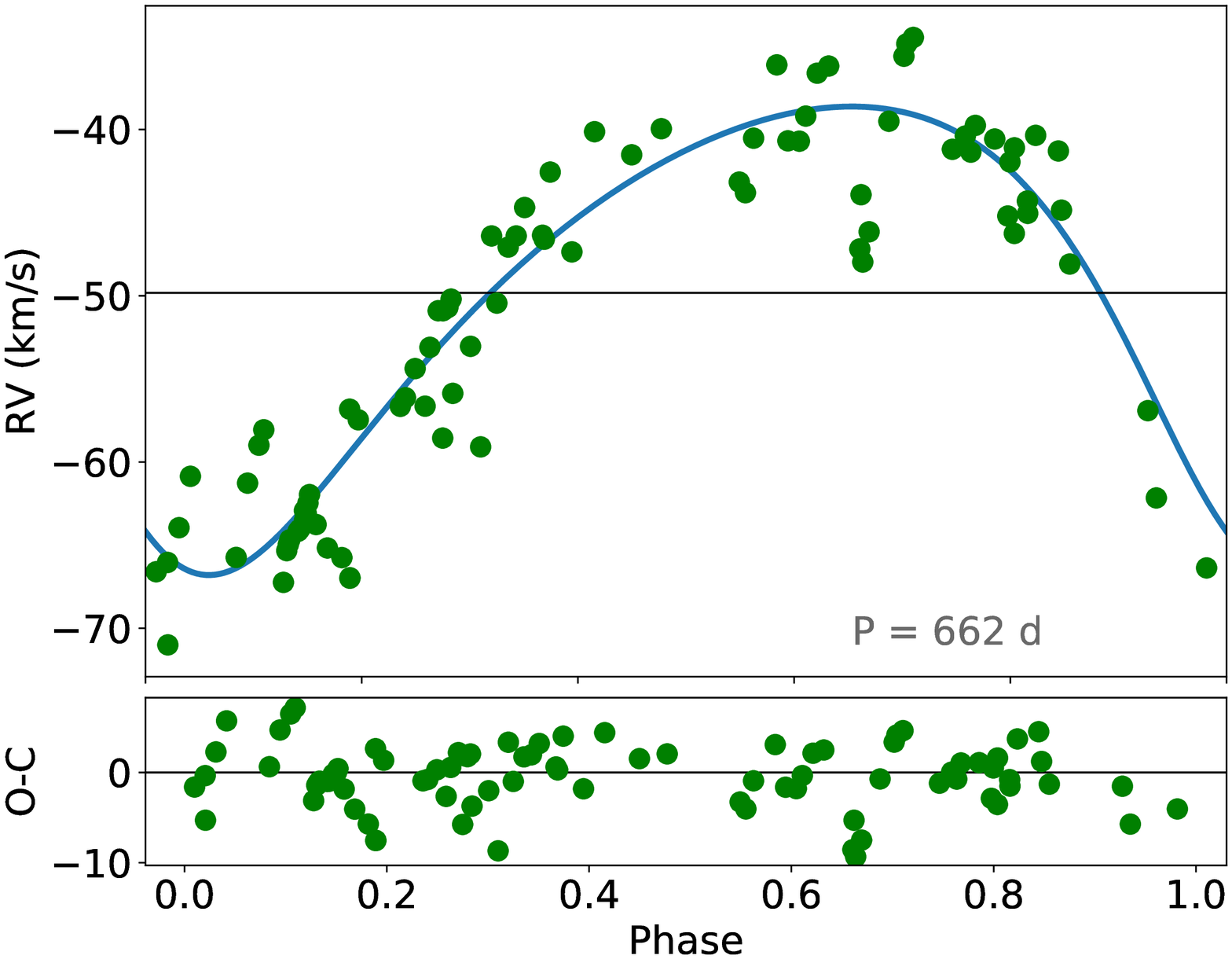}      \label{figure:twcam}
   }
   \subfloat[\#33: U~Mon]{%
     \includegraphics[width=0.33\textwidth]{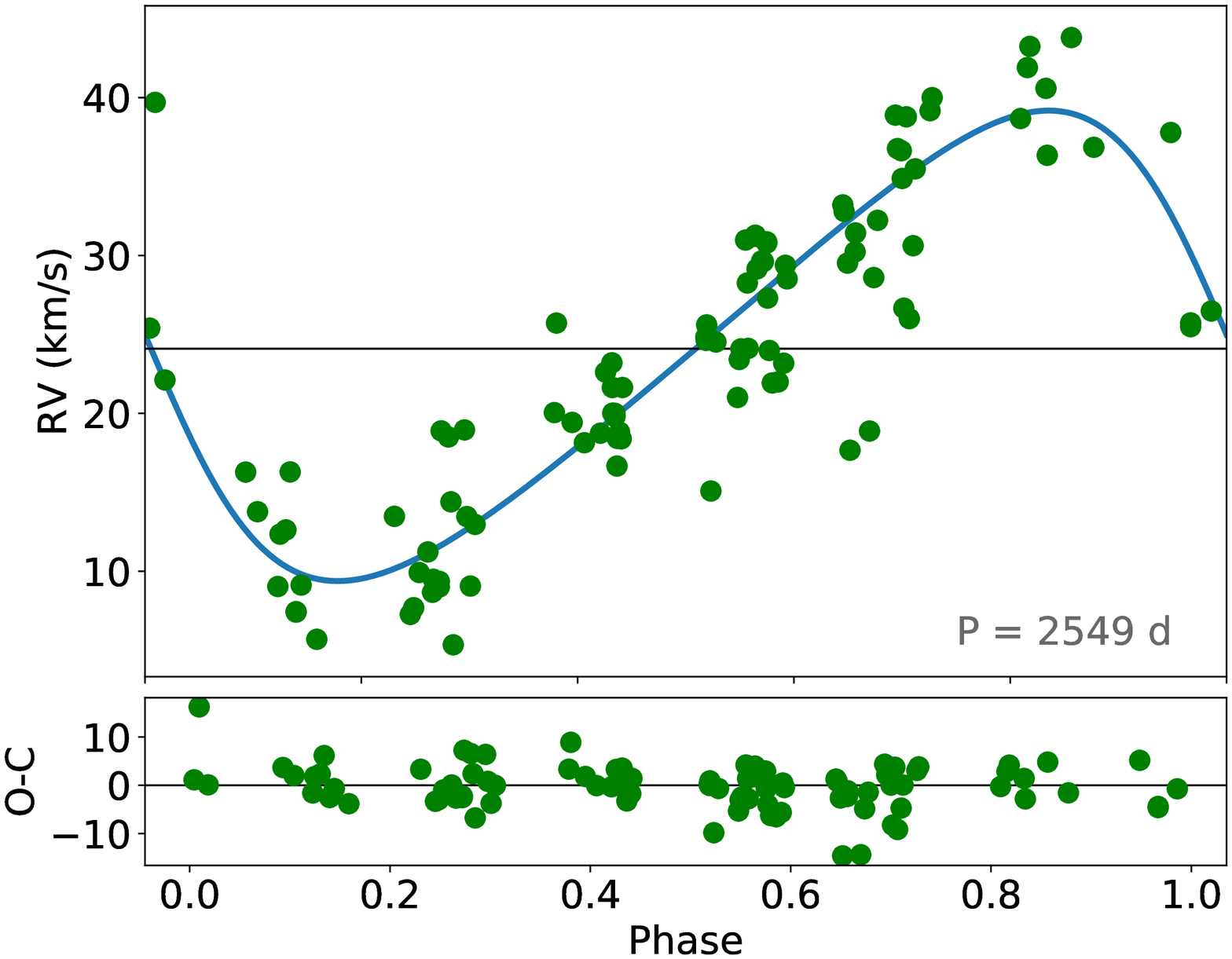}      \label{figure:umon}
   }
   \\
   \caption{\textit{continued}}
 \end{figure*}

\section{MESA models} \label{appendix:MESA}
In Sect.~\ref{sect:depletion discussion}, we use stellar evolution models of the post-AGB phase to argue that the effective temperature of a post-AGB star strongly correlates with its envelope mass. These evolutionary tracks are computed with version 10398 of the \texttt{MESA} code \citep{MESApaper1,MESApaper4}. The binary post-AGB stars in our sample are the result of a strong interaction on the AGB and hence do not follow a single-star evolutionary track. However, we assume that after the interaction, the star continues to evolve as a single star.

Since there is great uncertainty in the evolution prior to the post-AGB phase, we use a simplified procedure. We adopt a single-star model of 2.5~$M_\sun$ and solar metallicity, and evolve it towards the AGB phase. Once the core reaches a mass of 0.45~$M_\sun$,\footnote{Note that a core mass of 0.45~$M_\sun$ is reached before the tip of the RGB. This means that this star never started helium burning, hence becomes a post-RGB star.} 0.55~$M_\sun$, 0.60~$M_\sun$, and 0.65~$M_\sun$, we artificially increase the mass-loss rate to 10$^{-4}$~$M_\sun$/yr until the mass in the envelope is reduced to 0.02~$M_\sun$. 

Once the envelope is removed, we continue the evolution as post-RGB/AGB star. The adopted mass-loss mechanism is similar to the post-AGB models of \citet{millerbertolami16}. We did not include convective overshooting due to numerical stability issues in the post-AGB phase as some convective regions become highly superadiabatic. However, modest values of the overshooting in all the convective regions have only a small impact on the effective temperature.

\begin{figure}
\resizebox{\hsize}{!}{\includegraphics{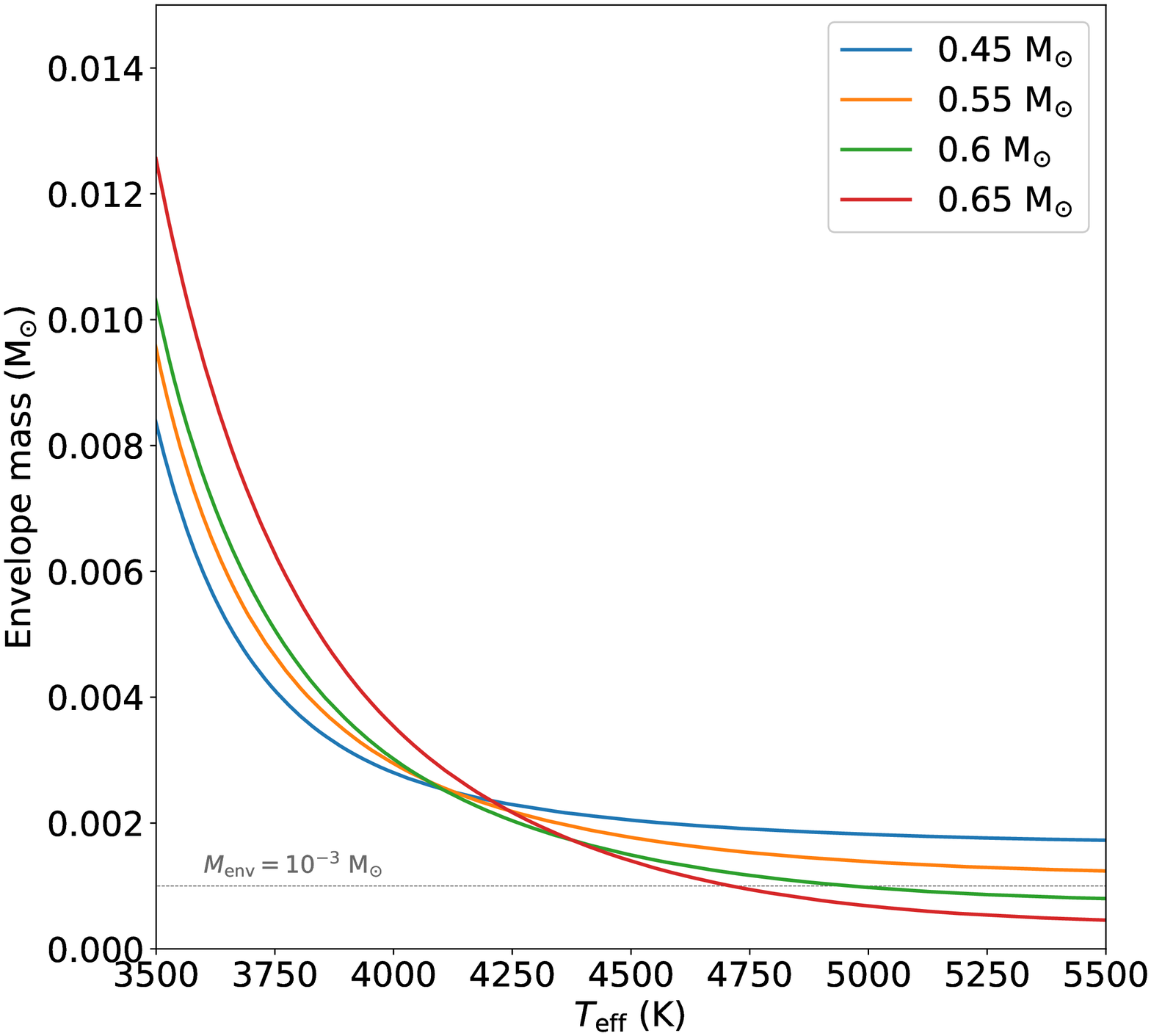}}
\caption{Relation between envelope mass and effective temperature for 4 different masses. Models with a temperature higher than 5000~K are characterised by a low envelope mass. The mass in the envelope decreases to about $10^{-3}$~$M_\sun$ by the time the star reaches 5500~K.}
\label{fig:mesamodels}
\end{figure}

Figure~\ref{fig:mesamodels} shows the evolution of the envelope mass against the effective temperature for 4 different masses. The envelope mass decreases by approximately a factor 10 as the temperature increases from 3500~K to 5500~K. When the envelope is still massive enough, the change in effective temperature is small compared to the change in envelope mass. However, as the envelope mass becomes really small ($\sim10^{-3}$~$M_\sun$, depending on the post-AGB mass), a small decrease in the envelope mass will result in a large decrease in radius and a large increase in effective temperature.

\end{appendix}
\end{document}